\documentclass[twocolumn,iop,twocolappendix,numberedappendix]{emulateapj}

\usepackage{graphics,color}
\graphicspath{{Figures/}{./}}
\usepackage{amsmath}

\usepackage{txfonts}
\usepackage[breaklinks=false]{hyperref} 
\usepackage{color}
\usepackage{sistyle}

\SIthousandsep{,}

\newcommand{\hMpc}{h^{-1}\,\mathrm{Mpc}}
\newcommand{\hGpc}{h^{-1}\,\mathrm{Gpc}}
\newcommand{\kms}{\mathrm{km\,s^{-1}}}
\newcommand{\lya}{Ly$\alpha$}
\newcommand{\lyb}{Ly$\beta$}
\newcommand{\lyalya}{Ly$\alpha$(Ly$\alpha$)}
\newcommand{\lyalyb}{Ly$\alpha$(Ly$\beta$)}
\newcommand{\lyalyalyalya}{\lyalya$\,\times\,$\lyalya}
\newcommand{\lyalyalyalyb}{\lyalya$\,\times\,$\lyalyb}
\newcommand{\lyalyaq}{\lyalya$\,\times\,$quasar}
\newcommand{\lyalybq}{\lyalyb$\,\times\,$quasar}

\newcommand{\comquest}[1]{}
\newcommand{\mmp}[1]{}

\newcommand{\zeffshort}{2.33}
\newcommand{\zeff}{2.334}
\newcommand{\nbLyaForest}{210005}
\newcommand{\nbLybForest}{69656}
\newcommand{\nbQSO}{341468}
\newcommand{\picca}{\texttt{picca}}
\newcommand{\zlyawg}{\texttt{Z\_LYAWG}}
\newcommand{\zpca}{\texttt{Z\_PCA}}
\newcommand{\zciv}{\texttt{Z\_CIV}}
\newcommand{\zciii}{\texttt{Z\_CIII}}
\newcommand{\zvi}{\texttt{Z\_VI}}
\newcommand{\zpipe}{\texttt{Z\_PIPE}}
\newcommand{\zcomp}{\texttt{Z}}

\newcommand{\sigmav}{\sigma_v}
\newcommand{\lrf}{\lambda_{\rm RF}}
\newcommand{\rvec}{{\bf r}}
\newcommand{\kvec}{{\bf k}}
\newcommand{\kpar}{k_{\parallel}}
\newcommand{\apar}{\alpha_{\parallel}}
\newcommand{\aperp}{\alpha_{\perp}}
\newcommand{\rpar}{r_{\parallel}}
\newcommand{\rperp}{r_{\perp}}
\newcommand{\blya}{b_{\rm Ly\alpha}}
\newcommand{\betalya}{\beta_{\rm Ly\alpha}}

\newcommand{\bhcd}{b_{\rm HCD}}
\newcommand{\betahcd}{\beta_{\rm HCD}}
\newcommand{\Fhcd}{F_{\rm HCD}}
\newcommand{\Lhcd}{L_{\rm HCD}}
\newcommand{\imin}{i_{\rm min}}
\newcommand{\imax}{i_{\rm max}}

\newcommand{\jmax}{j_{\rm max}}
\newcommand{\xioned}{\xi_{\rm 1d}}
\newcommand{\DHub}{D_{H}}
\newcommand{\DM}{D_{M}}

\newcommand{\omegak}{\Omega_k}

\newcommand{\lcdm}{$\Lambda$CDM}

\newcommand{\Blomqvist}{B19}
\newcommand{\deSainteAgathe}{dSA19}
\newcommand{\Bautista}{B17}
\newcommand{\duMasdesBourboux}{dMdB17}
\newcommand{\Planck}{Planck (2016)}

\begin{document}

\title{
  The Completed SDSS-IV extended Baryon Oscillation Spectroscopic Survey:
  Baryon acoustic oscillations with Lyman-$\alpha$ forests
}

\slugcomment{Submitted to ApJ}

\author{
H\'elion~du~Mas~des~Bourboux\altaffilmark{1},
James~Rich\altaffilmark{2},
Andreu~Font-Ribera\altaffilmark{3,4},
Victoria~de~Sainte~Agathe\altaffilmark{5},
James~Farr\altaffilmark{3},
Thomas~Etourneau\altaffilmark{2},
Jean-Marc~Le Goff\altaffilmark{2},
Andrei~Cuceu\altaffilmark{3},
Christophe~Balland\altaffilmark{5},
Julian~E.~Bautista\altaffilmark{6},
Michael~Blomqvist\altaffilmark{19},
Jonathan~Brinkmann\altaffilmark{7},
Joel~R.~Brownstein\altaffilmark{1},
Sol\`ene~Chabanier\altaffilmark{2},
Edmond~Chaussidon\altaffilmark{8},
Kyle~Dawson\altaffilmark{1},
Alma~X.~Gonz\'alez-Morales\altaffilmark{9,}\altaffilmark{10},
Julien~Guy\altaffilmark{8},
Brad~W.~Lyke\altaffilmark{11},
Axel~de~la~Macorra\altaffilmark{12},
Eva-Maria~Mueller\altaffilmark{13},
Adam~D.~Myers\altaffilmark{11},
Christian~Nitschelm\altaffilmark{14},
Andrea~Mu\~noz~Guti\'errez\altaffilmark{12},
Nathalie~Palanque-Delabrouille\altaffilmark{2},
James~Parker\altaffilmark{7},
Will~J.~Percival\altaffilmark{15,16,17},
Ignasi~P{\'e}rez-R{\`a}fols\altaffilmark{5},
Patrick~Petitjean\altaffilmark{18}, 
Matthew~M.~Pieri\altaffilmark{19},
Corentin~Ravoux\altaffilmark{2},
Graziano~Rossi\altaffilmark{20},
Donald~P.~Schneider\altaffilmark{21,}\altaffilmark{22},
Hee-Jong~Seo\altaffilmark{23},
An\v{z}e~Slosar\altaffilmark{24},
Julianna~Stermer\altaffilmark{5},
M.~Vivek\altaffilmark{25},
Christophe~Y{\`e}che\altaffilmark{2},
Samantha~Youles\altaffilmark{6}
}

\altaffiltext{1}{
Department of Physics and Astronomy, University of Utah, 115 S 1400 E, Salt Lake City, UT 84112, USA
}
\altaffiltext{2}{
IRFU, CEA, Universit\'e Paris-Saclay, F91191 Gif-sur-Yvette, France
}
\altaffiltext{3}{
Department of Physics and Astronomy, University College London, London, UK
}
\altaffiltext{4}{
Institut de F\'isica d'Altes Energies (IFAE), 
The Barcelona Institute of Science and Technology, 
08193 Bellaterra (Barcelona), Spain
}
\altaffiltext{5}{
Sorbonne Universit\'e, Universit\'e Paris Diderot, CNRS/IN2P3, Laboratoire de Physique Nucl\'eaire et de Hautes Energies, LPNHE, 4 Place Jussieu, F-75252 Paris, France
}
\altaffiltext{6}{
Institute of Cosmology \& Gravitation, University of Portsmouth, Dennis Sciama Building, Portsmouth, PO1 3FX, UK
}
\altaffiltext{7}{
Apache Point Observatory, P.O. Box 59, Sunspot, NM 88349
}
\altaffiltext{8}{
Lawrence Berkeley National Laboratory, 1 Cyclotron Road, Berkeley, CA 94720, U.S.A
}
\altaffiltext{9}{
Departamento de F\'isica, DCI, Campus Le\'on, Universidad de
Guanajuato, 37150, Le\'on, Guanajuato, M\'exico
}
\altaffiltext{10}{
Consejo Nacional de Ciencia y Tecnolog\'ia, Av. Insurgentes Sur 1582. Colonia Cr\'edito Constructor, Del. Benito Ju\'arez, C.P. 03940, M\'exico D.F. M\'exico
}
\altaffiltext{11}{Department of Physics and Astronomy, University of Wyoming, Laramie, WY 82071, USA
}
\altaffiltext{12}{
Instituto de F\'{i}sica,  Universidad Nacional Aut\'{o}noma de M\'{e}xico, Ciudad de M\'{e}xico, 04510 M\'{e}xico.
}
\altaffiltext{13}{
Sub-department of Astrophysics, Department of Physics, University of Oxford, Denys Wilkinson Building, Keble Road, Oxford OX1 3RH, UK
}
\altaffiltext{14}{
Centro de Astronomía, Universidad de Antofagasta, Avenida Angamos 601, Antofagasta 1270300, Chile
}
\altaffiltext{15}{
Waterloo Centre for Astrophysics, University of Waterloo, 200 University Ave W, Waterloo, ON N2L 3G1, Canada
}
\altaffiltext{16}{
Department of Physics and Astronomy, University of Waterloo, 200 University Ave W, Waterloo, ON N2L 3G1, Canada
}
\altaffiltext{17}{
Perimeter Institute for Theoretical Physics, 31 Caroline St. North, Waterloo, ON N2L 2Y5, Canada
}
\altaffiltext{18}{
IAP, Université Paris 6 et CNRS, 98bis blvd. Ara go, 75014 Paris, France
}
\altaffiltext{19}{
Aix Marseille Univ, CNRS, CNES, LAM, Marseille, France
}
\altaffiltext{20}{
Department of Astronomy and Space Science, Sejong University, 209, Neungdong-ro, Gwangjin-gu, Seoul, South Korea
}
\altaffiltext{21}{
Department of Astronomy and Astrophysics, The Pennsylvania State University, University Park, PA 16802
}
\altaffiltext{22}{
Institute for Gravitation and the Cosmos, The Pennsylvania State University, University Park, PA 16802
}
\altaffiltext{23}{
Department of Physics and Astronomy, Ohio University, Clippinger Labs, Athens, OH 45701
}
\altaffiltext{24}{
Brookhaven National Laboratory, Physics Department, Upton, NY 11973, U.S.A
}
\altaffiltext{25}{
Indian Institute of Astrophysics, Koramangala, Bangalore 560034, India
}

\date{\today}

\email{h.du.mas.des.bourboux at gmail.com, james.rich at cea.fr}

\shorttitle{\texorpdfstring{Ly$\alpha$}{Lya} BAO in DR16}

\begin{abstract}
We present a measurement of baryonic acoustic oscillations (BAO)
from Lyman-$\alpha$ (\lya) absorption and quasars
at an effective redshift $z=\zeffshort$ using the complete
extended Baryonic Oscillation Spectroscopic Survey (eBOSS).
The sixteenth and final eBOSS data release (SDSS DR16) contains
all data from eBOSS and its predecessor, 
the Baryonic Oscillation Spectroscopic Survey (BOSS),
providing $\num{\nbLyaForest{}}$ quasars with $z_{q}>2.10$
that are used to measure Ly$\alpha$ absorption.
We measure the BAO scale both in the auto-correlation
of Ly$\alpha$ absorption and in its cross correlation
with  $\num{\nbQSO{}}$ quasars with redshift 
$z_{q}>1.77$.
Apart from the statistical gain from new quasars and deeper observations,
the main improvements over previous work come from
more accurate modeling of physical and instrumental correlations and 
the use of new sets of mock data.
Combining the BAO measurement from the auto- and cross-correlation yields
the constraints of the two ratios
$D_{H}(z=\zeffshort)/r_{d} = 8.99 \pm 0.19$
and
$D_{M}(z=\zeffshort)/r_{d} = 37.5 \pm 1.1$,
where the error bars are statistical.
These results are within $1.5\sigma$ of the prediction of the
flat-$\Lambda$CDM cosmology of Planck~(2016).
The analysis code, \picca, the catalog of the
  flux-transmission field measurements,
and the $\Delta \chi^{2}$ surfaces are publicly available.
\end{abstract}

\hypersetup{pdftitle={
The Completed SDSS-IV extended Baryon Oscillation Spectroscopic Survey:
  Baryon acoustic oscillations with Lyman-\texorpdfstring{Ly$\alpha$}{Lya} forests
}}
\hypersetup{pdfsubject=Cosmology}
\hypersetup{pdfauthor={H. du Mas des Bourboux et al.}}
\hypersetup{pdfkeywords={cosmology, dark energy, large-scale structure, baryon acoustic oscillations, BAO, quasar, \texorpdfstring{Ly$\alpha$}{Lya} forest}}

%
%
\section{Introduction}
\label{section::Introduction}
\setcounter{footnote}{0}

One of the surprising characteristics of the widely accepted
\lcdm~cosmological model is that the expansion of the
universe is presently accelerating.
The most direct evidence for this acceleration
comes from the redshift dependence of
distances  and expansion rates.
Luminosity distances to type Ia supernovae (SNe~Ia) provided the first
evidence for accelerated expansion
\citep{1998AJ....116.1009R,1999ApJ...517..565P}.
More complete information comes from the baryonic acoustic oscillations (BAO)
feature in the matter correlation function whose position
yields both  distances and expansion rates normalized to the
sound horizon, $r_d$.
Constraints on cosmological parameters
from BAO (e.g. \citealt{2015PhRvD..92l3516A})
are consistent with those
from SNe~Ia \citep{Scolnic18,Jones2019} and with 
indirect probes of acceleration
like the spectrum of
CMB anisotropies \citep{{2016A&A...594A..13P},2020A&A...641A...6P}.

The baryonic acoustic oscillations in the pre-recombination universe
\citep{1970ApJ...162..815P,1970Ap&SS...7....3S}
left their imprints as a
peak in the matter correlation function at the sound horizon.
The spectrum of CMB anisotropies allows one to
set the comoving scale for the BAO peak: $r_d=147.3\pm0.5$~Mpc \citep{2016A&A...594A..13P}.
This scale is used as a ``comoving standard ruler'',
i.e. the ruler expands with the expansion of the universe.
This BAO scale is  arguably simpler than the use of SNe~Ia as
standard candles since mean SN luminosities may depend on 
astrophysical conditions.

BAO surveys at a redshift $z$ yield measurements of $\DM(z)/r_d$
and $\DHub(z)/r_d$, where $\DM(z)=(1+z)D_A(z)$ is the comoving
angular-diameter distance to $z$ and $\DHub(z)=c/H(z)$ is
the Hubble distance corresponding to the expansion rate $H(z)$.
The first measurement of BAO was performed using the auto-correlation determined from galaxy positions
\citep{2005ApJ...633..560E} at $z\sim0.35$  and the galaxy power-spectrum \citep{2005MNRAS.362..505C}
at $z\sim0.1$.
At redshifts $z\lesssim2$ the BAO scale has been studied
using discrete tracers such as galaxies
\citep{
2007MNRAS.381.1053P,
2010MNRAS.401.2148P,
2011MNRAS.415.2892B,
2011MNRAS.416.3017B,
2012MNRAS.426..226C,
2012MNRAS.427.2132P,
2012MNRAS.427.2168M,
2013MNRAS.431.2834X,
2012MNRAS.427.3435A,
2014MNRAS.439...83A,
2014MNRAS.441...24A,
2015MNRAS.449..835R,
2017MNRAS.470.2617A,
2018ApJ...863..110B},
galaxy clusters \citep{2016ApJ...826..154H},
and quasars 
\citep{2018MNRAS.473.4773A}.

At redshifts $\gtrsim2$, the number density
of observable discrete tracers is insufficient for
high precision clustering measurements, so
BAO studies have been performed
using instead opacity fluctuations from
the \mbox{Lyman-$\alpha$} (Ly$\alpha$) transition in 
neutral hydrogen towards background quasars (QSOs).
This transition has a restframe wavelength of
$\lambda_{\mathrm{Ly}\alpha} = 121.567 \, \mathrm{nm}$
and traces density fluctuations of neutral hydrogen
in the intergalactic medium (IGM).
From the ground it can be observed 
along the line-of-sight to
 objects with $z\gtrsim2$,
where the transition is redshifted to observed wavelengths
$\lambda_{\mathrm{Obs.}} \gtrsim 360 \, \mathrm{nm}$.
Even though this continuous tracer of the matter density fluctuations
has a lower bias than discrete tracers, the statistical power from the
large wavelength range 
where \lya~absorption can be measured  allows the observation of BAO,
as suggested by \citet{2003ApJ...585...34M} and \citet{2003dmci.confE..18W}.
The BAO scale was first measured in the 
Ly$\alpha$ auto-correlation function
\citep{
2013A&A...552A..96B,
2013JCAP...04..026S,
2013JCAP...03..024K,
2015A&A...574A..59D,
2017A&A...603A..12B,
2019A&A...629A..85D},
and then in the Ly$\alpha$-quasar cross-correlation function
\citep{
2014JCAP...05..027F,
2017A&A...608A.130D,
2019A&A...629A..86B}.

Other continuous tracers of the matter density field have been employed
but have not yet provided competitive BAO measurements.
At redshifts $z \lesssim 2$, \citet{2018JCAP...05..029B} used triply-ionized carbon absorption
and \citet{2019ApJ...878...47D} used singly-ionized magnesium absorption correlated with the quasar and
galaxy distribution to measure large-scale clustering. Both measure the
biased 3D correlation of the matter density field and find a signal consistent with that of BAO.
At redshifts $z \gtrsim 2$, \citet{2016JCAP...11..060L} measured the 3D auto-correlation of quasars 
and \citet{2018MNRAS.473.3019P} the
3D cross-correlation of damped Ly$\alpha$ absorption (DLA) and Ly$\alpha$,
systems, but neither report a measurement of BAO.

This study presents the measurements of BAO from the Ly$\alpha$
auto-correlation function and the Ly$\alpha$-quasar cross-correlation
function updated to the sixteenth data release
(DR16, \citealt{2020ApJS..249....3A})
of the fourth generation of the Sloan Digital Sky Survey
\citep[SDSS-IV:][]{2000AJ....120.1579Y,2017AJ....154...28B}.
This data release contains all of the clustering and \lya\ forest data from the
completed extended Baryonic Oscillation Spectroscopic
Survey \citep[eBOSS:][]{2016AJ....151...44D}.
We use \lya~absorption in two spectral regions illustrated
in the quasar spectrum of Figure~\ref{figure::exemple_data_forest}:
the ``\lya`` region between the quasar \lya~and \lyb-OVI emission peaks,
and in the ``\lyb'' region
between the \lyb-OVI emission peak and the quasar restframe Lyman limit.
We thus measure two \lya~auto-correlation functions:
\begin{itemize}
\item \lyalyalyalya
  \item \lyalyalyalyb
  \end{itemize}
where the ``transition(spectral~region)'' notation means, for example,
that \lyalyb~signifies
\lya~absorption in the \lyb~region.
Similarly, we measure two \lya-quasar cross-correlation functions:
\begin{itemize}
\item \lyalyaq
  \item \lyalybq \hspace*{1mm}.
\end{itemize}
Final BAO constraints are derived using
various combinations of these four correlation functions.
  We do not use \lyb~absorption because its oscillator strength
  is $\approx20\%$ that of \lya, making its contribution to the
  BAO measurement insignificant.
  Neither do we use
  \lya(\lyb)$\times$\lya(\lyb) since the BAO measurement would
  be complicated by its superposition with the  \lyb(\lyb)$\times$\lyb(\lyb)
  correlation.

This work builds upon previous studies
of the \lya~auto- and \lya-quasar
cross-correlations from SDSS.  Those previous studies are
the auto-correlation
in SDSS DR12 (BOSS) 
by \citet[][hereafter ``\Bautista '']{2017A&A...603A..12B}
and in SDSS DR14 (eBOSS)
by \citet[][hereafter ``\deSainteAgathe'']{2019A&A...629A..85D}
and the cross-correlation
in DR12 by \citet[][hereafter ``\duMasdesBourboux'']{2017A&A...608A.130D}
and in DR14 by \citet[][hereafter ``\Blomqvist '']{2019A&A...629A..86B}.

Compared to the studies of \deSainteAgathe~and \Blomqvist~on 
DR14, this analysis using DR16 adds the following improvements:
\begin{itemize}
    
    \item Compared to DR14, DR16 provides
    $\num{67606}$ ($25$\%) more quasars in the
    redshift range $z_{q}>1.77$ as tracers
    and $\num{24944}$ ($12$\%) more in the
    range $z_{q}>2.10$, with spectra allowing measurement of 
     Ly$\alpha$ absorption
    (see Section~\ref{subsection::Host_and_tracer_quasar_samples}).
    Compared to DR12, the DR16 sample has $\num{117940}$ ($52$\%) more quasars at $z_{q}>1.77$,
    and $\num{44257}$ ($23$\%) more quasars at $z_{q}>2.10$.

    DR16 provides additional observations	for greater depth, leading to 
    spectra with improved mean signal-to-noise ratio (SNR) by comparison to 
    the fewer observations of DR12.
    Furthermore \Bautista~and \duMasdesBourboux~used only 
    the best observation of each object, while this study, following
    as \Blomqvist~and  \deSainteAgathe, co-add 
    acceptable observations together.
    
      The larger sample of deeper spectra leads to a decrease in the
      variance of both the auto- and cross-correlation functions
      by a factor 0.5 relative to DR12
    
    \item In addition to the quasar redshift estimators used in
    DR12 and DR14 analyses,
    we develop a new estimator referred to as \zlyawg.
    This estimator is similar to the estimator of the standard pipeline \citep{2012AJ....144..144B},
    but we do not use wavelengths in the vicinity of the
    Ly$\alpha$ emission line or at shorter wavelengths.
    The accuracy of the redshifts with this estimator depends less of
    redshift than that of previously used estimators (see
    Section~\ref{subsection::quasar_redshifts}
    and Appendix \ref{section::quasar_redshifts}).

  \item Compared to \deSainteAgathe, the so-called
    ``Ly$\beta$ spectral region''
    has been extended from the range $[97.4,102]\,\mathrm{nm}$ to
    $[92,102]\,\mathrm{nm}$. This extends the spectral region beyond
    the Ly$\gamma$ absorption line
    ($\lambda_{\mathrm{Ly}\gamma} = 97.2537 \, \mathrm{nm}$)
    but not as far as  the Lyman-limit
    ($91.18 \, \mathrm{nm}$).
    Doing so increases the  number of \lya~absorption
    measurements in the  \lyb~spectral region by a factor of two
    ($\sim 1.5$ in weighted number; see
    Section~\ref{subsection::Measurement_of_the_fluctuations_of_the_transmission_field}).
    
  \item
          We measure any spurious correlations due
      the to sky-subtraction procedure and
      model their effect on the measured correlation function
      (see Sections~\ref{subsection::The_Lya_auto_correlation}
       and \ref{section::Model_of_the_correlations}).
       This results in a better fit of the correlation function
       away from the BAO peak.

    \item Two new sets of three-dimensional (3D) Gaussian random field simulations have been
    developed to characterize our combined measurement of the auto- and
    cross-correlation (see Section~\ref{section::Validation_of_the_analysis_with_mocks}).
    These mocks allow us to test the steps of the data analysis,
     thus study
    potential sources of systematic errors and test the estimation of
    statistical errors.
    Unlike our previous mock spectra, the \lyb~spectral region is also
    simulated and studied.

\end{itemize}
The analysis code developed by our team, \picca:
``Package for Igm Cosmological-Correlations Analyses'',
is publicly available on
\texttt{GitHub}\footnote{\url{https://github.com/igmhub/picca}}.
Modules include the algorithms to  
estimate the \lya~forest signal,
to compute the auto- and cross-correlations and estimate
their covariance, and to fit the measured correlations using an input
matter power spectrum.
The best-fit BAO results and likelihood, and a tutorial, are also given in this
repository\footnote{\url{https://github.com/igmhub/picca/tree/master/data}}.
We also publicly release the catalog of 
estimated \lya~forest signal 
(Appendix~\ref{section::catalog_of_fluctuations_of_transmitted_flux_fraction}).
The correlation functions, their covariance matrix, and other files used in fits
are available upon request.

  The publication of this study is coordinated with the
  release of the final eBOSS measurements
  of BAO and redshift-space distortions (RSD) in the clustering 
  of luminous red galaxies (LRG) in the redshift range $0.6<z<1.0$
  \citep{2020MNRAS.tmp.2651B,2020MNRAS.498.2492G},
  of emission line galaxies (ELG) in the range $0.6<z<1.1$
  \citep{raichoor20a,tamone20a,demattia20a},
  and of quasars in the range $0.8<z<2.2$
  \citep{hou20a,2020MNRAS.tmp.2610N}.
The cosmological interpretation of all DR16 eBOSS and BOSS results,
in combination with  other probes,
is found in
  \citet{eBOSS_Cosmology}
.\footnote{
  A summary of all SDSS BAO and RSD measurements with accompanying legacy
figures can be found here:
\url{https:/www.sdss.org/science/final-bao-and-rsd-measurements/} .  The full
cosmological interpretation of these measurements can be found here:
\url{https:/www.sdss.org/science/cosmology-results-from-eboss/}.
}

The analysis procedure we have used is reflected in
the organization of this paper and can be summarized as follows.
The first step is the selection of quasar and forests samples
(Sect.~\ref{subsection::Host_and_tracer_quasar_samples})
followed by an estimation of quasar redshifts
(Sect.~\ref{subsection::quasar_redshifts}).
The determination of the fluxes in the selected forests
is described in 
\ref{subsection::Measurement_of_the_fluctuations_of_the_transmission_field}.
The calculation of the fluctuations in the transmitted flux fraction, 
$\delta_q(\lambda)$, is presented in Sect.~\ref{subsection::deltafield}).

The calculation of the correlations of $\delta_q(\lambda)$
with itself (auto-correlation) and with quasars (cross-correlation) is
described in 
Section~\ref{section::Measurement_of_the_auto_and_cross_correlations}.
Correlations
are a-priori a  function of the angular and redshift separations
and of redshift.
It is useful to condense these three dimensions 
to two dimensions by adopting a fiducial cosmology
(Sect. \ref{subsection::From_redshifts_and_angles_to_distances}).
This allows the correlation functions to be calculated simply
as functions of radial reparation ($\rpar$) and
of transverse separation ($\rperp$).

The continuum-fitting procedure uses all flux measurements
in a forest  to determine each individual $\delta_q(\lambda)$.
Thus individual measured $\delta_q(\lambda)$ have small admixtures of
all the true $\delta_q(\lambda)$ in the same forest.
This effect is accounted for by the
``distortion matrix'', as described in
Sect.~\ref{section::the_distortion_matrix}.

In Section~\ref{section::Model_of_the_correlations}, we summarize
the physical model of the correlation functions.
We include the effects of the dominant \lya~absorption, of absorption
by high-column-density systems and of metals. The model
of the cross-correlation includes effects of quasar proper motion
and of the proximity effect due to quasar UV radiation.
Un-suspected correlations not included in the model can be accounted
for by the addition of terms that are polynomial functions of $(\rperp,\rpar)$.

The validation of the analysis using synthetic data
is presented in Section~\ref{section::Validation_of_the_analysis_with_mocks}.
The fits of the correlation functions and the 
best fit results for the BAO parameters
are presented in Section~\ref{section::Fit_to_the_data}.

We present a comparison with previous \lya\ BAO analyses in 
Section~\ref{section::Comparison}, and we summarize our study in
Section~\ref{section::Summary_and_conclusions}.
The constraints on cosmological parameters derived from this
analysis, from other DR16 analyses, and from other cosmological
probes are presented in a companion paper \citep{eBOSS_Cosmology}.

%
%
\section{Catalog of quasar and \texorpdfstring{Ly$\alpha$}{Lya} tracers}
\label{section::Quasar_samples_and_data_reduction}

In our study, we use two different tracers of matter density
fluctuations:  quasars and \lya~absorption in quasar spectra.
We thus have two overlapping quasar samples.
The first consists of 
quasars used as discrete tracers at $z_q\gtrsim1.77$ and is referred to
hereafter as ``tracer quasars''.
The second sample consists of 
quasars whose spectra are used to measure \lya~absorption,
referred to hereafter as ``background quasars''.
Background quasars have redshifts $z_q>2.10$ and provide measurements (pixels)
of \lya~absorption at $z>1.96$.
This section presents the catalog of both tracers and the details of the pipeline
that is used to collect and process the data.

\subsection{Tracer and background quasar spectra}
\label{subsection::Host_and_tracer_quasar_samples}

This analysis benefits from more than ten
years of cosmological observations from
SDSS \citep{2000AJ....120.1579Y} on the 
$2.5$~m Sloan Foundation telescope
\citep{2006AJ....131.2332G} at the Apache Point Observatory.
Most of the tracer quasar  and the entirety of the background quasar spectra were 
gathered during SDSS-III \citep{2011AJ....142...72E} in the
Baryon Oscillation Spectroscopic Survey
\citep[BOSS:][]{2013AJ....145...10D} and in the
eBOSS component of SDSS-IV.
A small fraction of tracer quasars were observed during SDSS-I and -II and are
found in the seventh data release \citep{2010AJ....139.2360S}.

\begin{figure}
  \centering
    \includegraphics[width=0.98\columnwidth]{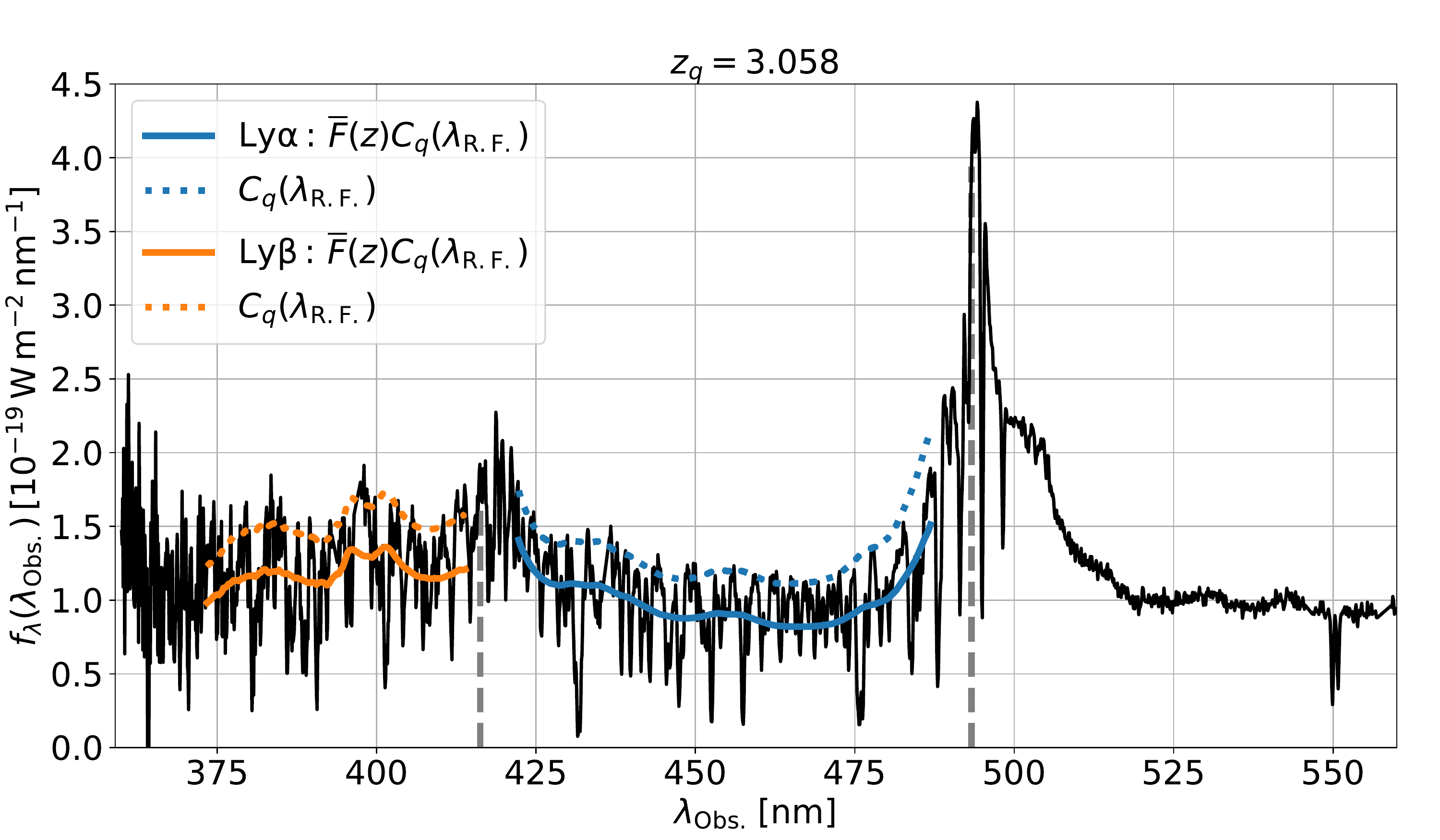}
    \caption{
    A high signal-to-noise  eBOSS quasar spectrum
    as a function of observed wavelength.
    The two spectral regions are
    \lya~(blue), $104<\lrf<120 \, \mathrm{nm}$,
    and \lyb~(orange), $92<\lrf<102 \, \mathrm{nm}$.
    The two solid curves give
    the two independent best fit models
    for $\overline{F}(z)C_{q}(\lrf)$,
    and the dashed curves the unabsorbed continuum
    $C_{q}(\lrf)$
      assuming the $\overline{F}(z)$ of \citet{2012MNRAS.422.3019C}.
    The quasar has a redshift $z_{q} = 3.058$ and is defined in the
    catalog DR16Q by
    $\mathrm{Thing\_id} = 498518806$.
    The Ly$\alpha$ ($\approx493~{\rm nm}$) and the overlapping
    Ly$\beta$+OVI emission lines ($\approx420~{\rm nm}$)
    are marked with vertical gray
    dashed lines.
    }
    \label{figure::exemple_data_forest}
\end{figure}

\begin{figure*}[tb]
    \centering
    \includegraphics[width=.90\textwidth]{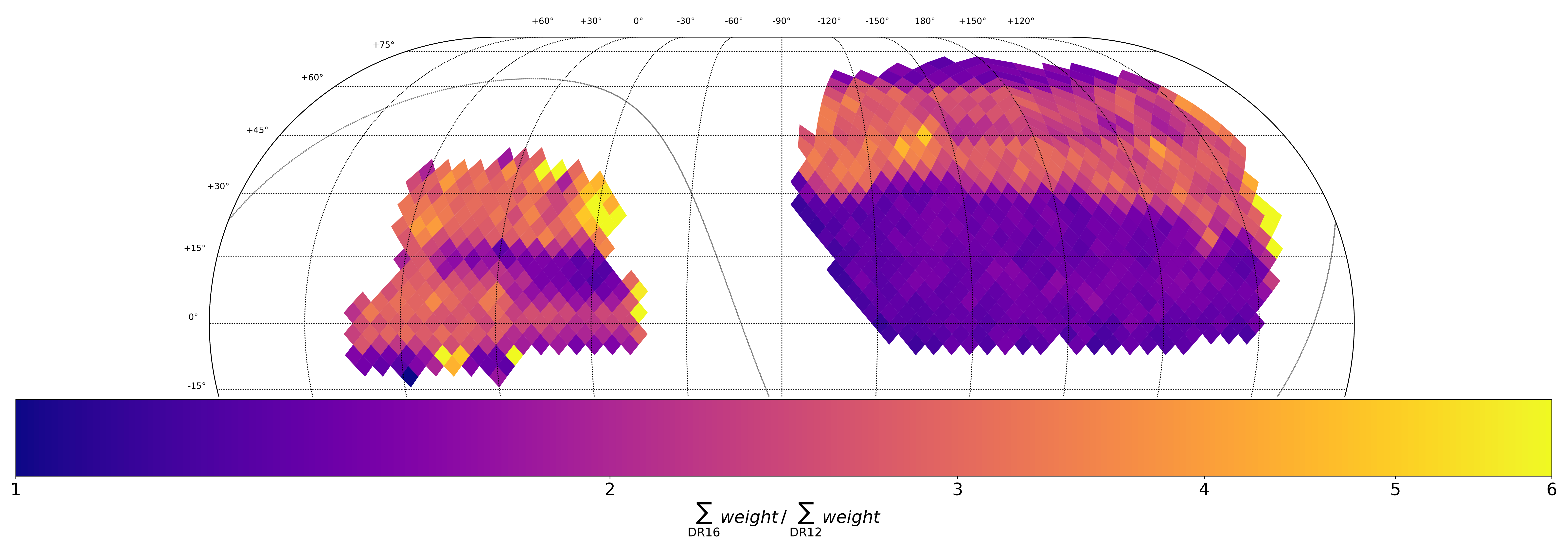}
    \caption{
    Footprint of the eBOSS DR16 survey in a Mollweide projection.
    The South Galactic Cap (SGC) is on the left and the North Galactic Cap
    (NGC) is on the right.
    The gray curve shows the position of the Galactic plane.
    The color scale gives the ratio of the weighted number of pairs for the
    auto-correlation between eBOSS DR16 and BOSS DR12.
    }
    \label{figure::footprint}
\end{figure*}

The DR16 data used in this paper
\citep{2020ApJS..249....3A}  contains the complete
five-year BOSS sample (Fall 2009 - Spring 2014) and the
five-year eBOSS sample (July 2014 - March 2019) including
its pilot program SEQUELS \citep{2015ApJS..221...27M}
which began in the
final year of SDSS-III and concluded in the first year of SDSS-IV.
The quasar target selection algorithms for BOSS are summarized in
\citet{2012ApJS..199....3R} and for eBOSS are summarized in
\citet{2015ApJS..221...27M}.
An additional sample of targets were observed as part of the
Time-Domain Spectroscopic Survey
(TDSS: \citealt{2015ApJ...806..244M,2016ApJ...825..137R})
and the SPectroscopic IDentification of ERosita Sources survey 
(SPIDERS: \citealt{2017MNRAS.469.1065D}).
An example high signal-to-noise quasar spectrum is shown in
Figure~\ref{figure::exemple_data_forest}.
The DR16 footprint is presented in
Figure~\ref{figure::footprint}, where the color map reflects
the statistical improvement 
between DR12 and DR16 for the auto-correlation of Ly$\alpha$ pixels.

The tracer and background quasars are taken from 
the DR16 quasar catalog
(DR16Q: \citealt{2020ApJS..250....8L})
which includes the DR7 quasar catalog \citep{2010AJ....139.2360S}.
Most of the DR7 quasars at $z>2.1$ were
re-observed in BOSS or eBOSS, they are all included in DR16Q.
For the few quasars that were not re-observed,
we use them as tracer quasars, but not
as background quasars, since their spectra were processed with
another pipeline.

Each spectrum from BOSS or eBOSS was acquired using one of the two
spectrographs with wavelength coverage ranging from $360\,\mathrm{nm}$
to $1000\,\mathrm{nm}$ \citep{2013AJ....146...32S}.
Each spectrograph views 500 fibers, roughly 450 of which are dedicated to
science targets and the remaining to standard F-stars
for flux calibration and to empty sky locations for sky-background subtraction.

The data were processed
by version \texttt{v5\_13\_0} of the eBOSS pipeline.
The  reduction is organized in two steps.
The pipeline initially extracts the two-dimensional raw data into a
one-dimensional 
flux-calibrated spectrum. During this procedure, the spectra are
wavelength and flux calibrated and the individual exposures of one
object are coadded into a rebinned spectrum with
$\Delta \log (\lambda) = 10^{-4}$. 
The spectra are then classified as \texttt{STAR},  \texttt{GALAXY},
or \texttt{QSO},
and their
redshift is estimated.

  Damped Lyman-$\alpha$ (DLA) systems  and Broad-absorption lines (BAL) were
  identified in the quasar spectra.
  The details of these searches are presented in
\citet{2020ApJS..250....8L}.
  The BAL search used a procedure similar to that
  used by \citet{2019ApJ...879...72G}.
  DLAs were identified using a neural network as
  described in \citet{2018MNRAS.476.1151P}.

The current DR16 pipeline slightly differs from its last public release
\citep[DR14:][]{2018ApJS..235...42A}.
Details of the changes can be found in \citet{2020ApJS..249....3A}.
We focus on two relevant changes for the \lya~analysis here.
The first concerned improvements in the background estimates
on the spectral CCD images.
This allowed for more accurate determination of spectral densities $f(\lambda)$.
The second  improves modeling of the spectra of calibration stars
by using a new set of stellar templates.
The set was produced for the Dark Energy
Spectroscopic Instrument
\citep{2016arXiv161100036D}
pipeline and was provided to eBOSS.
These templates are able to reduce residuals in flux calibration by
improving the modeling of spectral lines in F-stars.

  For the  purposes of measuring \lya~correlations,
we rebin three original pipeline
$\Delta \log_{10}(\lambda) \sim 10^{-4}$ spectral pixels into one
$\Delta  \log_{10}(\lambda) \sim 3\times10^{-4}$ ``analysis pixel''.
Throughout this paper, the use of the word ``pixel'' refers
to these rebinned analysis pixels, unless otherwise stated.

\subsection{Quasar redshifts}
\label{subsection::quasar_redshifts}

  The DR16Q catalog provides up to four estimates of quasar redshift
  based on the position of the broad emission lines.
  The absence of visible narrow spectral lines that would be
  associated with
  the quasar host galaxies means that individual quasar redshifts 
  have statistical uncertainties
  of order $100~{\rm km\,s^{-1}}$ and the different
  estimators have somewhat different systematic differences.
  As described in Appendix~\ref{section::quasar_redshifts},
  we studied in detail these differences using synthetic quasar spectra.
  Based on these studies, we choose to 
  employ an estimator that
  does not use any spectral information in the vicinity of the \lya-emission
  line or at shorter wavelengths.
  This choice of restframe wavelength coverage
  mitigates redshift dependent systematic errors due, for example, to 
  evolution in the mean \lya~opacity.

\begin{figure*}
    \centering
    \includegraphics[width=0.98\columnwidth]{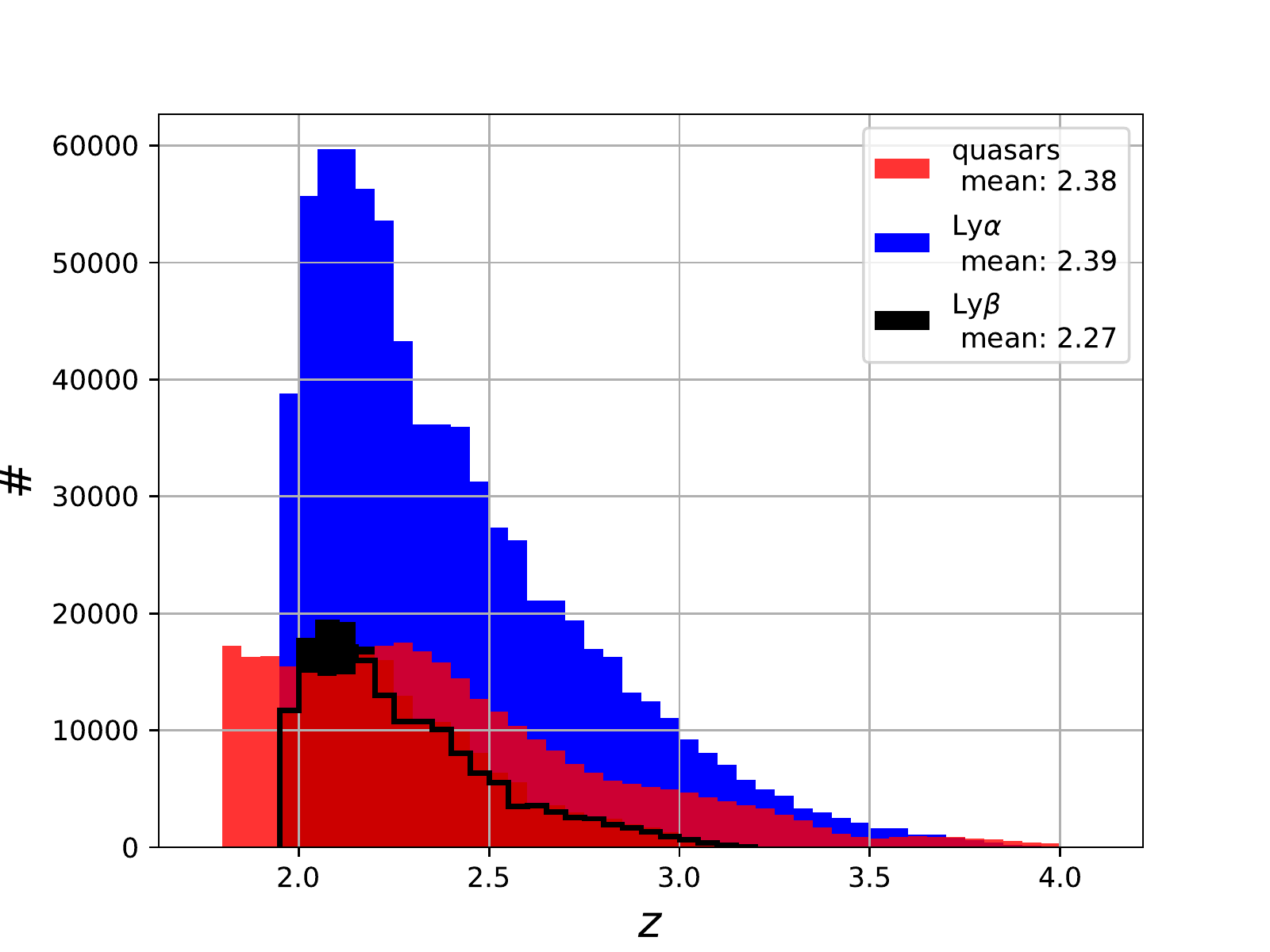}
    \includegraphics[width=0.98\columnwidth]{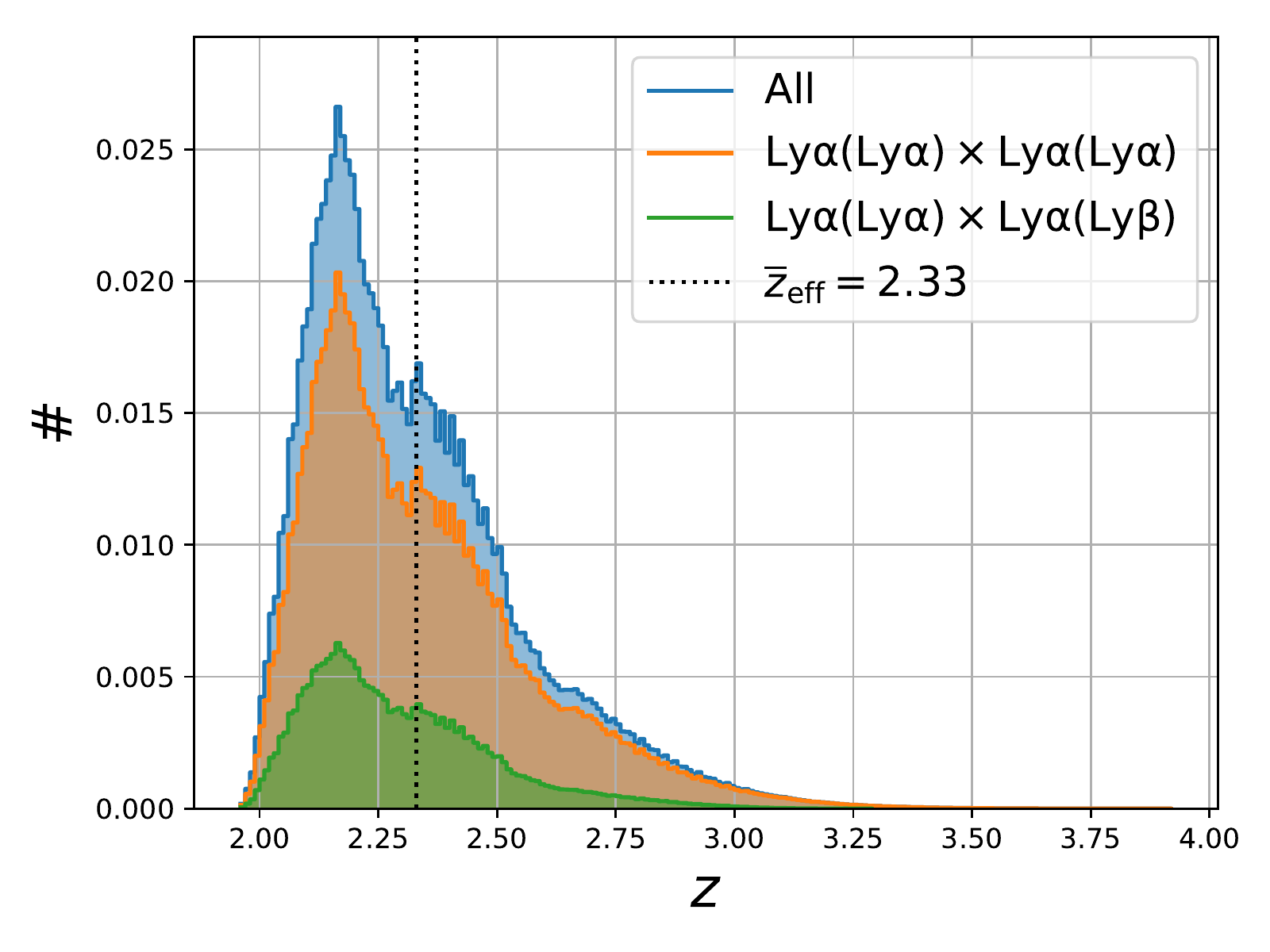}
    \caption{
      {\bf Left:}
      Redshift distribution of \lyalya~and \lyalyb~pixels (numbers divided by 50)
      and of quasars.
    {\bf Right:} Normalized weighted redshift distribution of pixel-pixel pairs
    in the auto-correlation.
    The pairs are in the BAO region of the correlation functions:
    $r \in [80,120] \, h^{-1}\mathrm{Mpc}$.
    The orange histogram gives the contribution
    of correlations
    involving pixels in the Ly$\alpha$ region, i.e. the
    \lyalyalyalya~correlation function.  The green histogram
    for one pixel in the Ly$\beta$ region, the other in the Ly$\alpha$ region,
    i.e. the \lyalyalyalyb~correlation function.
    The blue histogram is the combination of the two.
    The black dotted line shows the effective redshift of the measurements,
      i.e. the pivot redshift for the BAO measurement
      (Section~\ref{section::Fit_to_the_data}).
    }
    \label{figure::histo_pairs_tracer}
\end{figure*}

The redshift range of useful quasars is set by the goal of
measuring the quasar-\lya~cross-correlations up to separations
of  $200\,\hMpc$.
The lowest redshift \lya~pixel has $z=1.96$, meaning that quasars
with redshifts $z>1.77$ can be used to sample the cross-correlations up
to the maximum separation.
We impose a maximum redshift of $z=4$, beyond which
the number of tracers is insufficient for a useful correlation measurement.
Our final sample is thus composed of $\num{341468}$ tracer quasars,
as summarized in Table~\ref{table::definition_forests}.
The redshift distribution is shown in the left panel of
Figure~\ref{figure::histo_pairs_tracer}.

\subsection{Forest selection and residual calibration}
\label{subsection::Measurement_of_the_fluctuations_of_the_transmission_field}

As was done in DR14 (\deSainteAgathe, \Blomqvist ), we use two
different spectral regions shown in Figure~\ref{figure::exemple_data_forest}:
the ``Ly$\alpha$ region'', $\lrf \in [104,120]\,\mathrm{nm}$,
and the ``Ly$\beta$ region'', $\lrf \in [92,102]\,\mathrm{nm}$.
The Ly$\alpha$ region is defined between the Ly$\beta$+OVI quasar emission line
and the Ly$\alpha$ emission line. Defined as such, we ensure that this
region only has Ly$\alpha$ absorption and weak metal absorption (see \citealt{2014MNRAS.441.1718P}).
This is the same definition as in previous analyses.
The Ly$\beta$+OVI and Ly$\alpha$ emission line wings are avoided because of their higher
quasar-to-quasar diversity.

The Ly$\beta$ spectral region covers most of the range  between the Lyman-limit,
$\lambda_{\mathrm{Ly-limit}} = 91.18 \, \mathrm{nm}$, and the
Ly$\beta$+OVI quasar emission line.
This increase in range over the
$[97.4,102]\,\mathrm{nm}$ in DR14 to $[92,102]\,\mathrm{nm}$
increases the number of spectral pixels in the region by a factor two.

This analysis uses pixels in the observed wavelength range
$\lambda \in [360,600]\,\mathrm{nm}$.
The upper limit comes from the requirement $z_{q}<4$, 
which is motivated by the decreasing number of quasars and the increasing
contamination from sky emission lines.
The lower limit is motivated
by  high atmospheric absorption in the UV and the low CCD throughput at the lowest wavelengths.
From the previously defined rest-frame wavelength range of the
two spectral regions, and the observed wavelength range, the minimum
background quasar redshift is $z_{q}>2$ for the Ly$\alpha$ region and
$z_{q}>2.53$ for the Ly$\beta$ region. 
This selection gives us
$\num{265328}$ background quasars for the Ly$\alpha$ region and
$\num{103080}$ background quasars for the Ly$\beta$ region.
We removed BALs with BALPROB $> 0.9$ in the DR16Q Catalog
\citep{2020ApJS..250....8L}.
This reduced the numbers to
$\num{249814}$ and $\num{97124}$.


As in \deSainteAgathe~and \Blomqvist , all good spectral observations
of a given background quasar are stacked to give a higher signal-to-noise ratio
spectrum.
The use of only good BOSS or eBOSS observations
reduces the number of forests
to
$\num{244514}$ and $\num{96110}$
for the Ly$\alpha$ and Ly$\beta$ regions, respectively.

A spectral regions is used only if it contains at least
$50$ pixels,
since short lines-of-sight may be overfitted leading to erasing of structure
in a correlated manner.
This requirement shifts the minimum quasar redshift to $z_{q}>2.10$ for the Ly$\alpha$ region and
$z_{q}>2.65$ for the Ly$\beta$ region, leaving
$\num{214542}$ and $\num{71602}$
background quasars, respectively.
The fit of the continuum
(eqn.~\ref{equation::definition_quasar_continuum}) fails for a few
low SNR or outlier spectral regions.
The final sample has respectively
$\num{\nbLyaForest{}}$ and $\num{\nbLybForest{}}$
background quasars.
We summarize the sample in Table~\ref{table::definition_forests}.

\begin{table}
    \caption{
    Definitions and statistics for the
    matter density tracers.
    }
\centering
    \begin{tabular}{l r r r r r r r}

    Tracer &
    $\lambda^{\min}_{\mathrm{RF}}$ &
    $\lambda^{\max}_{\mathrm{RF}}$ &
    $z_{\mathrm{q,\,\min}}$ &
    $z_{\mathrm{q,\,\max}}$ &
    $N_{q}$ &
    $N_{pix}$ &
    $\overline{SNR}$ \\

    &
    $[\text{nm}]$ &
    $[\text{nm}]$ &
    &
    &
    &
    $[10^{6}]$
    &
    \\

    \noalign{\smallskip}
    \hline \hline
    \noalign{\smallskip}

    quasar                                        &        &        & $1.77$ & $4$ & $\num{341468}$    &   & \\ 
    $\mathrm{Ly\alpha \left(Ly\beta \right)}$     & $92$   & $102$  & $2.65$ & $4$ & $\num{\nbLybForest{}}$  & $8.4$  & $1.88$ \\
    $\mathrm{Ly\alpha \left(Ly\alpha \right)}$    & $104$  & $120$  & $2.10$ & $4$ & $\num{\nbLyaForest{}}$ & $34.3$ & $2.56$ \\

    \end{tabular}
      \tablecomments{
   The rest-frame wavelength intervals are shown in the 2nd and 3rd columns,
    the associated redshift interval in the 4th and 5th
    columns, the number of background or tracer quasars in the 6th column,
    the total number of spectral pixels in the 7th column,
    and, in the last column,  the mean signal-to-noise ratio of the pixels,
    ${\rm SNR}=\langle(\delta+1)*\sqrt{w}\rangle$
    where the weight, $w$ is given by the inverse of
    eqn. \ref{equation::definition_weight}.
        }
    \label{table::definition_forests}
\end{table}

Spectra and noise variance estimates are corrected for Galactic
extinction using the dust map of \citet{1998ApJ...500..525S}.
This had a very small effect in the final results, since extinction
corrections are very smooth and can be absorbed in the continuum fitting.

Following \Bautista~and \citet{2019ApJ...878...47D},
we correct for small residual flux calibration errors from the eBOSS pipeline.
To do so we 
designate a `calibration spectral region' on the red side of the
MgII emission line, $\lrf \in [290,312]\,\mathrm{nm}$.
We compute the mean flux as a function of observed wavelength 
$\overline{f_{\mathrm{calib.}}}(\lambda_{i})$
, correcting
for the shape of the quasar continuum (see below).
The new flux is then defined to be
$f(\lambda_{i}) \rightarrow f(\lambda_{i})/\overline{f_{\mathrm{calib.}}}(\lambda_{i})$,
and its inverse variance
$ivar(\lambda_{i}) \rightarrow ivar(\lambda_{i}) \times \overline{f_{\mathrm{calib.}}}^{2}(\lambda_{i})$.
We find that the correction varies by at most $3$\%.
This process captures residual errors from incomplete modeling of standard
F-stars and sky-emission lines, as well as the CaII H\&K absorption
of the Milky Way.

The next step is to mask out spectral intervals, in observed wavelength,
where the variance  increases sharply due to
unmodeled emission lines from the sky.
However, a balance has to be found between
  pixel quantity and pixel quality.
We use the previously defined
calibration spectral region
to compute the variance as a function of observed wavelength.
Comparing to the intrinsic variance of
large-scale-structure in the Ly$\alpha$ and Ly$\beta$ regions,
we mask the following six intervals: 
$\lambda \in [404.30,405.13]$,
$\lambda \in [435.31,436.51]$,
$\lambda \in [545.68,546.73]$,
$\lambda \in [557.05,559.00]$,
$\lambda \in [588.33,590.33]$, and
$\lambda \in [629.48,631.68]$ $\mathrm{nm}$.
This mask produces less data loss compared to previous studies,
in large part due to the improved sky calibration in the latest eBOSS pipeline.
In addition, we mask both CaII H\&K absorption
of the Milky Way, $\lambda \in [393.01,393.82]$ and $\lambda \in [396.55,397.38]$
$\mathrm{nm}$.  Even though we do not observe a significant excess of variance
in these regions, masking them avoids spurious
large-scale angular correlated absorption between lines-of-sight.

\subsection{The flux-transmission field $\delta_q(\lambda)$}
\label{subsection::deltafield}

For each spectral region in each line-of-sight, $q$,
the flux-transmission field, $\delta_{q}(\lambda)$,
at the observed wavelength, $\lambda$,
is obtained from  the ratio of the observed flux, $f_{q}(\lambda)$,
to the mean expected flux, $\overline{F}(z)C_{q}(\lambda)$:
\begin{equation}
    \delta_{q}(\lambda) =
    \frac{
    f_{q}(\lambda)
    }{
    \overline{F}(\lambda)C_{q}(\lambda)
    } - 1.
    \label{equation::definition_delta}
\end{equation}
Here, $C_q(\lambda)$ is the unabsorbed quasar continuum
and $\overline{F}(\lambda)$ is the mean transmission.
  Their product  is taken to be a universal function of the restframe
  wavelength,
  $\overline{C}(\lrf)$, corrected
by a first degree polynomial in
$\log\lambda$:
\begin{equation}
  \overline{F}(\lambda)C_{q}(\lambda) = \overline{C}(\lrf)
    \left(
    a_{q} + b_{q}\Lambda
    \right).
    \hspace*{5mm}\Lambda\equiv\log\lambda
    \label{equation::definition_quasar_continuum}
\end{equation}
The coefficients $(a_q,b_q)$ are fitted separately for the \lya~and
\lyb~spectral regions
of each line-of-sight 
by maximizing
the likelihood function
\begin{equation}
  2\ln L= - \sum_i
  \frac{[f_i-\overline{F}C_q(\lambda_i,a_q,b_q)]^2}{\sigma_q^2(\lambda_i)}
  -\ln [\sigma_q^2(\lambda_i)] \;,
\label{equation::likelihood}
\end{equation}
where the sum is over all forest pixel for the quasar $q$ and
$\sigma_q^2(\lambda)$ is the estimated variance of the flux, $f_i$.
Our  estimate of the $\sigma_q(\lambda)$  depends on $(a_q,b_q)$,
 so we include this dependence in the likelihood function
 of eqn.\ref{equation::likelihood}.

  The variance $\sigma_q^2$ receives contribution from both
instrumental noise (readout and photo-statistics)
and from large scale structure (LSS), and
we model it as the sum of three terms:
\begin{equation}
\frac{\sigma_q^2(\lambda)}{(\overline{F}C_q(\lambda))^2} =
  \eta(\lambda) \tilde\sigma^{2}_{\mathrm{pip},q}(\lambda)
    + \sigma^{2}_{\mathrm{LSS}}(\lambda)
    + \frac{\epsilon(\lambda)}{\tilde\sigma^{2}_{\mathrm{pip},q}(\lambda)}
    \; .
    \label{equation::definition_weight}
\end{equation}
      The first term is the instrumental
      noise,
      $\tilde\sigma_{\mathrm{pip},q}(\lambda)=\sigma_{\mathrm{pip},q}(\lambda)/\overline{F}C_q(\lambda)$,
      where $\sigma^2_{\mathrm{pip},q}$ is the pipeline estimate of the flux
      variance.
      We multiply this by
      a wavelength-dependent correction, $\eta(\lambda)$, that is found
      to range from $1.04$ at $\lambda=360$~nm to $1.20$ at
      $\lambda=580$~nm.
      The LSS variance, $\sigma^{2}_{\mathrm{LSS}}$, gives a minimum
      value of the variance equal to the
intrinsic variance of the flux-transmission field.  
Finally, we add an ad hoc factor proportional to
$1/\tilde\sigma_{\mathrm{pip},q}^2$.
It describes the observed
increase of variance, at high SNR, most probably due to quasar-to-quasar
spectral diversity.
The three terms in (\ref{equation::definition_weight}) dominate
for different ranges of flux and wavelength:  for our sample of
quasars, the instrumental and LSS terms
dominate, respectively, for wavelengths less than or greater than
$\lambda\approx450$~nm.  The $\epsilon$ term is significant
only for the brightest quasars, corresponding to $\approx10^{-4}$ of the
pixels.



\begin{figure*}
    \includegraphics[width=\columnwidth]{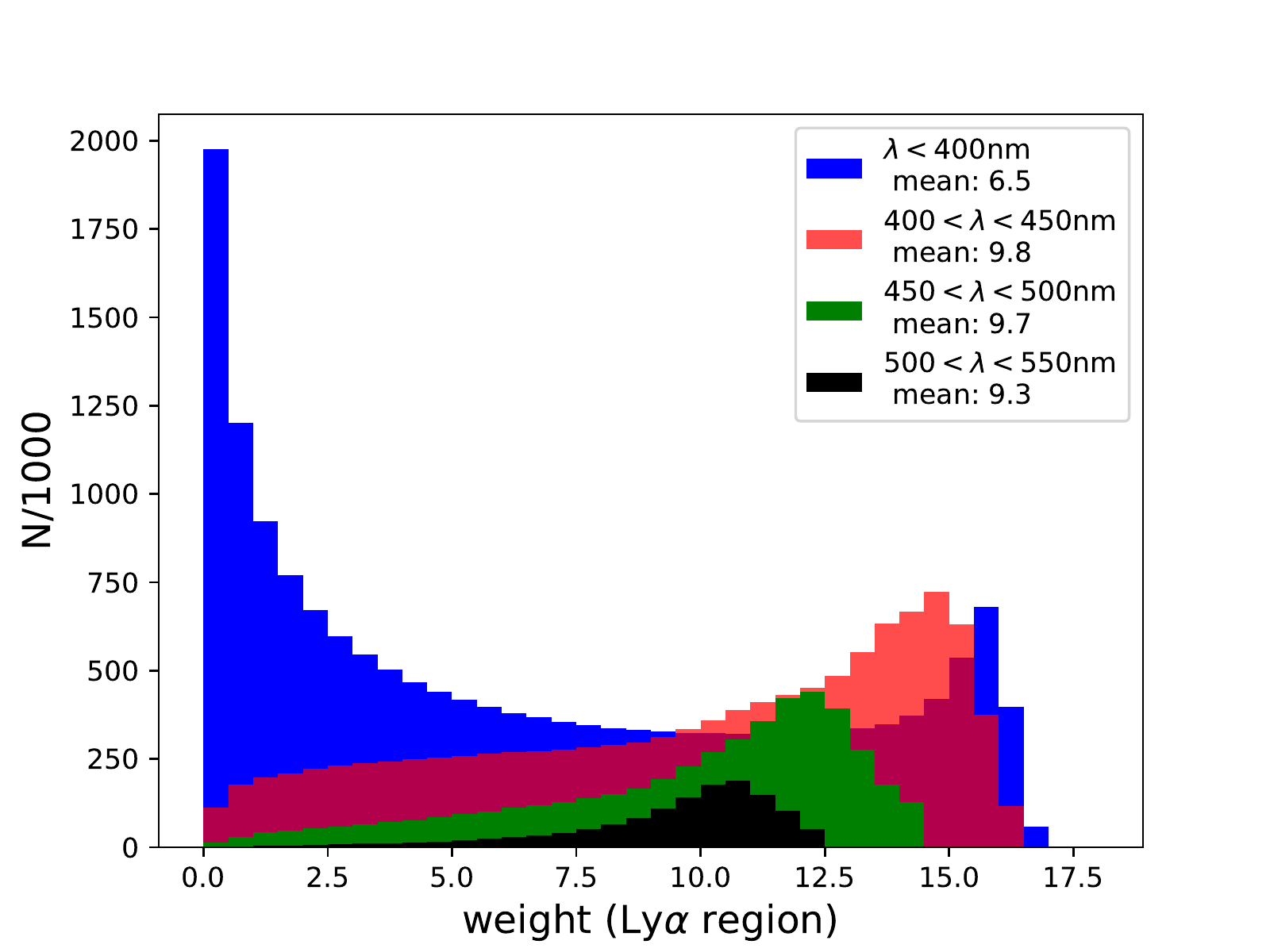}
    \includegraphics[width=\columnwidth]{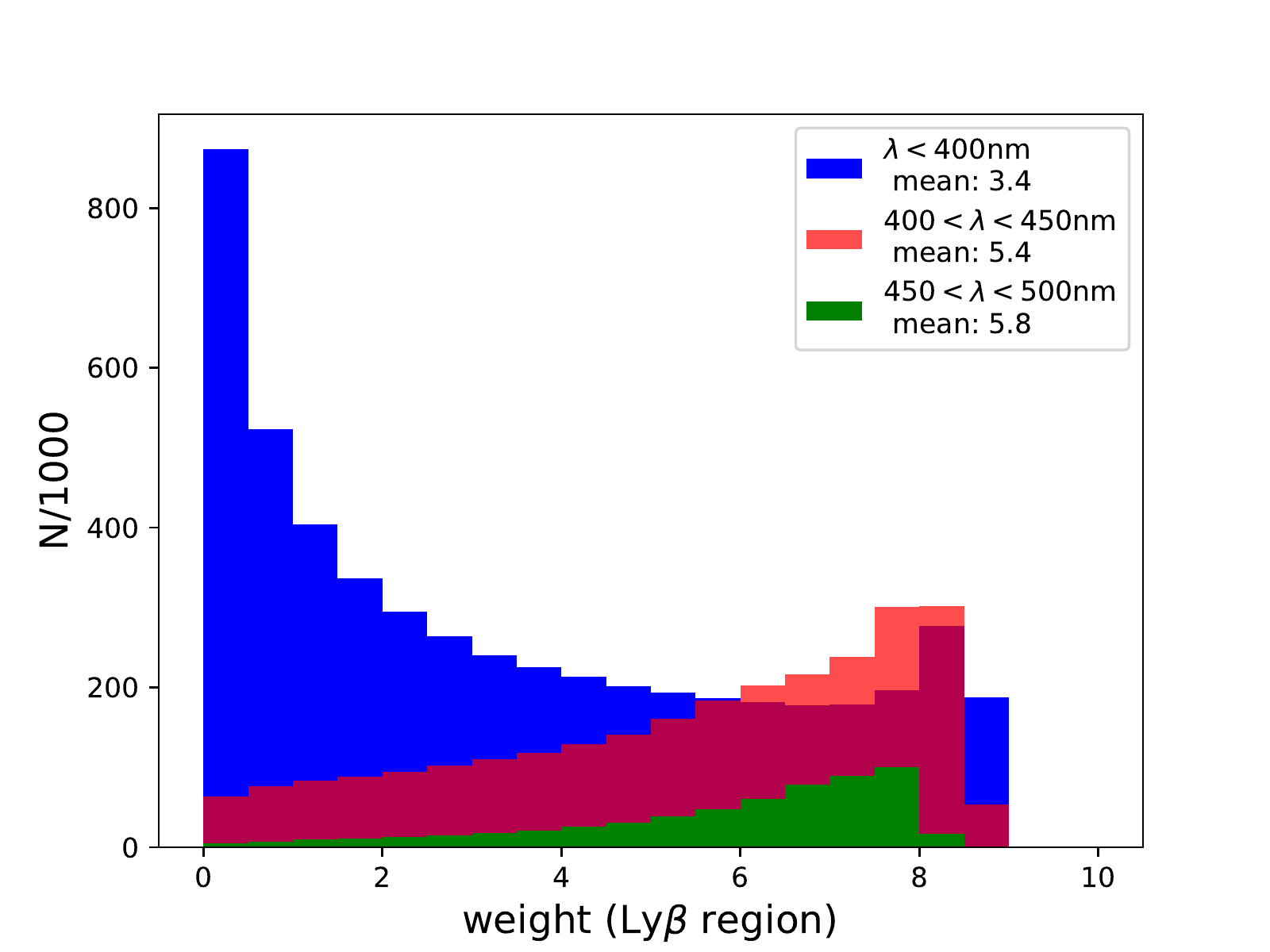}
    \caption{
      The distribution of weights
      (equation \ref{equation::definition_of_weight_auto})
      in four wavelength bands.
      The left and right panels are for the \lya~and \lyb~regions,
      respectively.
      For $\lambda<400$~nm (blue), the distribution has  a peak at zero weight
      (corresponding to noisy spectra) and a second peak at
      the maximum
      weight allowed by LSS.  With increasing wavelength and the accompanying
      decreasing noise, the weight distribution becomes more and more
      concentrated near the maximum weight.
      Compared to the \lya~forest, the maximum weight of the \lyb~forest 
      is reduced by a factor $\approx2$ because of 
      the increased absorption fluctuations and the wavelength distribution
      is shifted to lower wavelengths.
    }
    \label{figure::weights}
\end{figure*}

The functions $\overline{C}$, $\eta$, $\sigma^{2}_{\mathrm{LSS}}$
and $\epsilon$ are computed iteratively.
One starts with an initial estimate
of $\overline{C}(\lambda_{\rm RF})$ as the mean spectrum
using as weights an initial estimate of the $1/\sigma^2_q(\lambda)$.
The first estimate of the quasar parameters $a_q$ and $b_q$ are then calculated.
The resulting $\delta_q(\lambda)$ are calculated
and their variance determined in bins of $\tilde\sigma_{\mathrm{pip},q}$ and $\lambda$.
The functions $\eta(\lambda)$, $\epsilon(\lambda)$ and
$\sigma_{\rm LSS}(\lambda)$ are  then determined by fitting 
the variance of $\delta_q(\lambda)$ as a function of
$\tilde\sigma_{{\mathrm pip},q}$ and $\lambda$.
The mean spectrum, $\overline{C}(\lambda_{\rm RF})$ 
is then recalculated with the new weights.
This process is repeated until stable values are obtained after about
five iterations.
This process is executed independently
for the Ly$\alpha$ and Ly$\beta$ regions.
Figure~\ref{figure::weights} shows histograms of $\sigma^{-2}_{q}$
after correction for bias evolution
(eqn.~\ref{equation::definition_of_weight_auto}).

This analysis does not determine separately the mean transmission,
$\overline{F}(z)$ and the quasar continuum, $C_q(\lambda)$.
In Figure~\ref{figure::exemple_data_forest}, the latter is
calculated assuming the $\overline{F}(z)$ measured by
\citet{2012MNRAS.422.3019C}
and shown purely for illustration purposes.

The $\delta_q(\lambda)$ for
forests with identified DLAs are calculated masking
pixels where
a DLA reduces the transmission by more than 20\%.
The masked pixels are not used for the calculation
for correlation functions.
The absorption in the wings is corrected  using a 
Voigt profile following
the procedure of \citet{2012A&A...547L...1N}.  

The catalog of $\delta_q(\lambda)$
are publicly available with a data format described
in Appendix~\ref{section::catalog_of_fluctuations_of_transmitted_flux_fraction}.
The catalog for data used in this analysis consists of $34.3$ and 
$8.4$ million (analysis) pixels in the Ly$\alpha$ and Ly$\beta$
regions, respectively. 
The summary of the Ly$\alpha$ and Ly$\beta$ regions is given in
Table~\ref{table::definition_forests}.
The redshift distribution of the pixels, assuming Ly$\alpha$
absorption, is presented in the left panel of Figure~\ref{figure::histo_pairs_tracer}.

As demonstrated in the DR12 Ly$\alpha$ analysis
(\Bautista , \duMasdesBourboux), the quasar fitting of
equation~\ref{equation::definition_quasar_continuum} biases 
the mean and the spectral slope of each individual $\delta_q$ toward zero 
for each line-of-sight, resulting 
in a distortion of the correlation function.
This distortion can be modeled  if we make the biases exact by redefining
the $\delta_q$, per spectral region:
\begin{equation}
    \delta_{q}(\lambda_i)
    \rightarrow
    \sum_{j} \eta^q_{ij}\delta_q(\lambda_j) \; ,
    \label{equation::projection_1}
\end{equation}
where
\begin{equation}
    \eta^q_{ij}
    =
    \delta^K_{ij}
    - \frac{
    w_{j}
    }{
    \sum\limits_{k} w_{k}
    }
    -
    \frac{
    w_{j} \left( \Lambda_{i}-\overline{\Lambda_{q}} \right) \left(\Lambda_j-\overline{\Lambda_{q}} \right)
    }{
    \sum\limits_{k} w_{k} \left( \Lambda_{k}-\overline{{\Lambda_{q}}} \right)^{2}
    }
    \; ,
    \label{equation::projection_1_detail}
\end{equation}
and $\overline{\Lambda_q}$ is the mean of $\Lambda=\log\lambda$
for spectrum $q$.
In the definition of the projection (eqn.~\ref{equation::projection_1_detail}),
the weights are the ones used in the measurement of the correlation
functions:
\begin{equation}
    w_{i}
    =
    \sigma_q^{-2}(\lambda_i)
    \left( \frac{1+z_{i}}{1+2.25} \right)^{\gamma_{\mathrm{Ly}\alpha} -1} \;,
	\label{equation::definition_of_weight_auto}
\end{equation}
where the redshift evolution of the \lya~bias is taken into account
($\gamma_{\mathrm{Ly}\alpha} = 2.9$, \citealt{2006ApJS..163...80M}).
In Sect.~\ref{section::the_distortion_matrix}, we will use
the two equations~\ref{equation::projection_1} and
\ref{equation::projection_1_detail}
to establish the relation between the measured
correlations and the true correlations.

  The mapping applied by equation~\ref{equation::projection_1} changes
  the evolution of mean $\delta_q$ per wavelength bin slightly,
  allowing it to deviate 
 from zero.
We reintroduce this property for the cross-correlation
(Section~\ref{subsection::The_Lya_quasar_cross_correlation}) by
redefining the $\delta_q$ per observed wavelength bin:
\begin{equation}
    \delta_q(\lambda_{i})
    \rightarrow
    \delta_q(\lambda_{i})
    - \overline{\delta(\lambda)},
    \label{equation::projection_2}
\end{equation}
where the weighted average, $\overline{\delta(\lambda)}$, is computed using
the same weights as the ones used when computing the correlation function
(eqn.~\ref{equation::definition_of_weight_auto}).
This modification ensures that the cross-correlation approaches
zero at large scale,
whatever the redshift distribution of the quasars.

%
%
\section{Measurement of the auto- and cross-correlations}
\label{section::Measurement_of_the_auto_and_cross_correlations}

This section presents the measurements of the Ly$\alpha$ auto-correlation
functions and of the Ly$\alpha$-quasar cross-correlation
functions.
In Section~\ref{section::Quasar_samples_and_data_reduction}, we presented
the catalog of tracer quasars and the two catalogs of Ly$\alpha$ absorption
in pixels, covering the Ly$\alpha$ and Ly$\beta$ spectral regions.
Since these two spectral regions have different sizes,
levels of noise, shapes of the quasar continuum,
and covariance matrices, we determine the measurements independently. We thus use
Ly$\alpha$ absorption in the Ly$\alpha$ region
to compute the auto-correlation \lyalyalyalya~and
the cross-correlation
\lyalyaq. We use Ly$\alpha$ absorption
in the Ly$\beta$ region to compute two
additional correlation functions \lyalyalyalyb~and \lyalybq.

\subsection{From redshifts and angles to distances}
\label{subsection::From_redshifts_and_angles_to_distances}

In principle, the correlation functions can be
computed as a function of redshift and angular separation,
$\xi(\Delta z, \Delta \theta)$, as those are directly observed quantities.
However, because of the redshift dependence of the
comoving angular diameter distance $\DM(z)=(1+z)D_A(z)$ and
of the Hubble distance $\DHub(z)=c/H(z)$,
this would widen the BAO peak unless
the data were sorted into multiple redshift bins.
To avoid degradation of the BAO feature, we convert angular
separations to $\rperp$ and redshifts separations to $\rpar$ by adopting
a ``fiducial'' cosmology.
This acts as an optimal data compression designed to maximize the BAO signal 
if the fiducial cosmology is correct.  As demonstrated in
Appendix~\ref{section::physical_model_to_the_broadband_of_the_correlation_functions},
the conversion from observed to comoving coordinates does not bias the 
measurements of the BAO distance scale.

The fiducial cosmology that we adopt is the
$\Lambda$CDM cosmology
of \citet{2016A&A...594A..13P}, hereafter ``\Planck''.
If the fiducial cosmology approximates the true cosmology, then
the BAO peak will be
at a constant separation,
$r_{\mathrm{BAO}} \sim 100 \, h^{-1}\mathrm{Mpc}$, for all redshifts.
The cosmological parameters for this model are shown in the first part of
Table~\ref{table::cosmology_parameters} and the derived
parameters in the second part; they are computed using the
``Code for Anisotropies in the Microwave Background''
\citep[CAMB:][]{2000ApJ...538..473L}.
The same assumed cosmology is used to produce and analyze the mock data of
Section~\ref{section::Validation_of_the_analysis_with_mocks}.
This cosmology is the same as the one used in DR12 and DR14 studies of
Ly$\alpha$ BAO.
\begin{table}
    \caption{
        Parameters of the flat-$\Lambda$CDM cosmological model,
        from \citet{2016A&A...594A..13P}, used
        for the production and analysis of the mock data
        and the analysis of the data.
        }
    \centering
    \begin{tabular}{l c }
        Parameter & Planck (2016) cosmology               \\
                  &  $\left( \mathrm{TT+lowP} \right)$    \\

        \noalign{\smallskip}
        \hline \hline
        \noalign{\smallskip}

        $\Omega_{m} h^2$          &     $0.14252$    \\
        $=\Omega_{c} h^2$         &     $0.1197$    \\
        $\;+\Omega_{b} h^2$       &     $0.02222$    \\
        $\;+\Omega_{\nu} h^2$     &     $0.0006$    \\
        $h$                       &     $0.6731$    \\
        $N_{\nu}$                 &     $3$        \\
        $\sigma_{8}$              &     $0.8299$    \\
        $n_{s}$                   &     $0.9655$    \\

        \hline \noalign{\smallskip}
        
        $\Omega_{m}$            &    $0.31457$    \\
        $\Omega_{r}$            &    $7.975\,10^{-5}$    \\
        $r_d~[\mathrm{Mpc}]$    &    $147.33$    \\
        $r_d~[\hMpc]$           &    $99.17$    \\
        $D_{H}(z=\zeff)/r_{d}$  &    $8.6011$    \\
        $D_{M}(z=\zeff)/r_{d}$  &    $39.2035$    \\
        $f(z=\zeff)$            &    $0.9704$    \\

    \end{tabular}
\tablecomments{
    The first part of the table gives the cosmological parameters,
        the second part gives derived quantities used in this paper.
        They are computed using CAMB \citep{2000ApJ...538..473L}.
        }
    \label{table::cosmology_parameters}
\end{table}

The separation between two tracers is determined along and across the line-of-sight,
$\rvec = (r_{\parallel},r_{\perp})$.
For tracers $i$ and $j$, of redshift $z_{i}$ and $z_{j}$ and offset by an observed angle
$\Delta \theta$, the separation is computed as follows
\begin{equation}
    r_{\parallel} = \left[ D_{c}(z_{i})-D_{c}(z_{j}) \right]
    \cos
    \left( \frac{\Delta \theta}{2} \right),
    \label{equation::definition_distances_rp}
\end{equation}
and
\begin{equation}
    r_{\perp} = \left[ D_{M}(z_{i})+D_{M}(z_{j}) \right]
    \sin
    \left( \frac{\Delta \theta}{2} \right).
    \label{equation::definition_distances_rt}
\end{equation}
where $D_c(z)=\int_0^zdz/H(z)$ is the comoving distance.
We will also refer to $(r,\mu)$ in this study; they are defined as
$r^{2} = r_{\parallel}^{2} + r_{\perp}^{2}$ and as
$\mu = r_{\parallel}/r$.
The quantity $\mu$ is the cosine of the angle formed by
the median line-of-sight of both tracers and the vector $\rvec$.

For quasars, the redshift is defined to be \zlyawg~
(see Section~\ref{subsection::quasar_redshifts}). We test
the effects of other estimators of redshifts on the measurement of the BAO
in Appendix~\ref{section::Systematic_tests_on_BAO}.
For spectral pixels, the main absorber, and the one used to measure BAO,
is Ly$\alpha$. We thus define the redshift from its rest-frame wavelength.
For each spectral pixel $i$, of observed wavelength $\lambda_{i}$, the redshift
is then given by $z_{i} = \lambda_{i}/\lambda_{\mathrm{Ly}\alpha}-1
= \lambda_{i}/121.567-1$. The consequences of the presence of other absorption,
like \mbox{SiII(126)} or the \mbox{CIV} doublet, are discussed in
Section~\ref{section::Model_of_the_correlations}.

As given in Table~\ref{table::cosmology_parameters}, the BAO peak is
expected at the separation $r_d \sim 100 \, h^{-1}\mathrm{Mpc}$; furthermore,
we know from linear theory
  that the peak has a width (F.W.H.M) of $\approx20 \, h^{-1}\mathrm{Mpc}$.
For both these reasons, we compute the different correlations up to
$\pm 200 \, h^{-1}\mathrm{Mpc}$ along and across the line-of-sight,
i.e., twice the expected BAO scale, with a bin size of $4\, h^{-1}\mathrm{Mpc}$.
At the effective redshift of this study,
$z_{\mathrm{eff}.}=\zeffshort{}$, the BAO scale is
$\Delta \theta_{\mathrm{BAO}} \sim 1.5 \, \mathrm{deg}$ across the line-of-sight,
and 
$\Delta z_{\mathrm{BAO}} \sim 0.12$
(corresponding to $\approx50$ analysis pixels) along the line-of-sight.

\subsection{The \texorpdfstring{Ly$\alpha$}{Lya} auto-correlation}
\label{subsection::The_Lya_auto_correlation}
\label{subsection::auto:the_correlation}
\label{section::the_covariance_matrix_auto}

For the estimator of the 3D auto-correlation of Ly$\alpha$ absorption in
spectral pixels, we use
\begin{equation}
	\xi_{A} = \frac{
    \sum\limits_{(i,j) \in A} w_{i}w_{j} \, \delta_{i}\delta_{j}
    }{
	\sum\limits_{(i,j) \in A} w_{i}w_{j}
    }.
	\label{equation::xi_estimator_auto}
\end{equation}
This is the standard ``covariance'' estimator when the mean
has been subtracted: $\left<\delta_{i}\right> = 0$, by definition of the
projection of equation~\ref{equation::projection_1}.

In this equation, $i$ and $j$ refer to two spectral pixels of
the flux-transmission field,
 $\delta_{i}$ and $\delta_{j}$, from
eqn.~\ref{equation::definition_delta}.
The weights $w_{i}$ and $w_{j}$
are defined
by eqn. \ref{equation::definition_of_weight_auto}
where the factor $(1+z)^\gamma$  $\gamma_{\mathrm{Ly}\alpha} = 2.9$
is chosen to favor high-redshift pixels where the amplitude of the correlation
function is
greatest \citep{2006ApJS..163...80M}.
  We have tested that small changes in the assumed redshift evolution
of the weights do not translate into noticeable differences in the BAO results
presented in the next sections.

As explained in
Section~\ref{subsection::From_redshifts_and_angles_to_distances},
the computation of the correlation is done for all possible pairs of pixels
$(i,j)$, of separation $(r_{\parallel},r_{\perp})$, within
$[0,200]\,h^{-1}\mathrm{Mpc}$ in both directions. Each bin $A$ is
$4\,h^{-1}\mathrm{Mpc}$ wide in both directions.
As a result, each correlation function has $N_{\mathrm{bin}} = 50 \times 50 = 2500$ bins.
The sum runs over all possible pairs of pixels from different
lines-of-sight.
We exclude pairs
of pixels from the same line-of-sight
because of 
correlated continuum errors that could bias our measurement
of the correlation function.

In previous measurements of the Ly$\alpha$ auto-correlation, e.g. 
\Bautista~and \deSainteAgathe, pairs of pixels involving the same
spectrograph and $r_{\parallel} < 4 \,h^{-1}\mathrm{Mpc}$ were avoided
to minimize spurious correlations due to the sky-subtraction procedure.
In this analysis, we keep these pairs, allowing us to model their
direct effect on the lowest $\rpar$ bins and their indirect
effect on other bins through distortion due to continuum fitting.
This procedure is detailed in Sect. \ref{contribution_of_sky_residuals}.

The resulting correlations have $6.8 \times 10^{11}$ pairs of pixels
for the \lyalyalyalya\ auto-correlation and
$3.8 \times 10^{11}$ 
for the \lyalyalyalyb\ auto-correlation.
The lower number of pixels in the \lyb~region combined with
the increased absorption fluctuations and Poisson noise translates
to a variance in the \lyalyalyalyb~correlation function that is
approximately three times larger than that in the \lyalyalyalya~function.

The right panel of Figure~\ref{figure::histo_pairs_tracer} shows the
normalized redshift distribution of the pairs within the BAO region.
The Ly$\alpha$(Ly$\alpha$) $\times$ Ly$\alpha$(Ly$\alpha$) distribution
is presented in orange, Ly$\alpha$(Ly$\alpha$) $\times$ Ly$\alpha$(Ly$\beta$) in green,
and the sum of the two in blue.
Figure~\ref{figure::footprint} presents the comparison of the weighted
number of pairs between this analysis and the DR12 auto-correlation (\Bautista)
reproduced using the different improvements of this analysis.
The ratio is shown over the BOSS+eBOSS survey footprint, sampled by
HEALPix pixels \citep{2005ApJ...622..759G}.

\begin{figure}
    \centering
    \includegraphics[width=0.98\columnwidth]{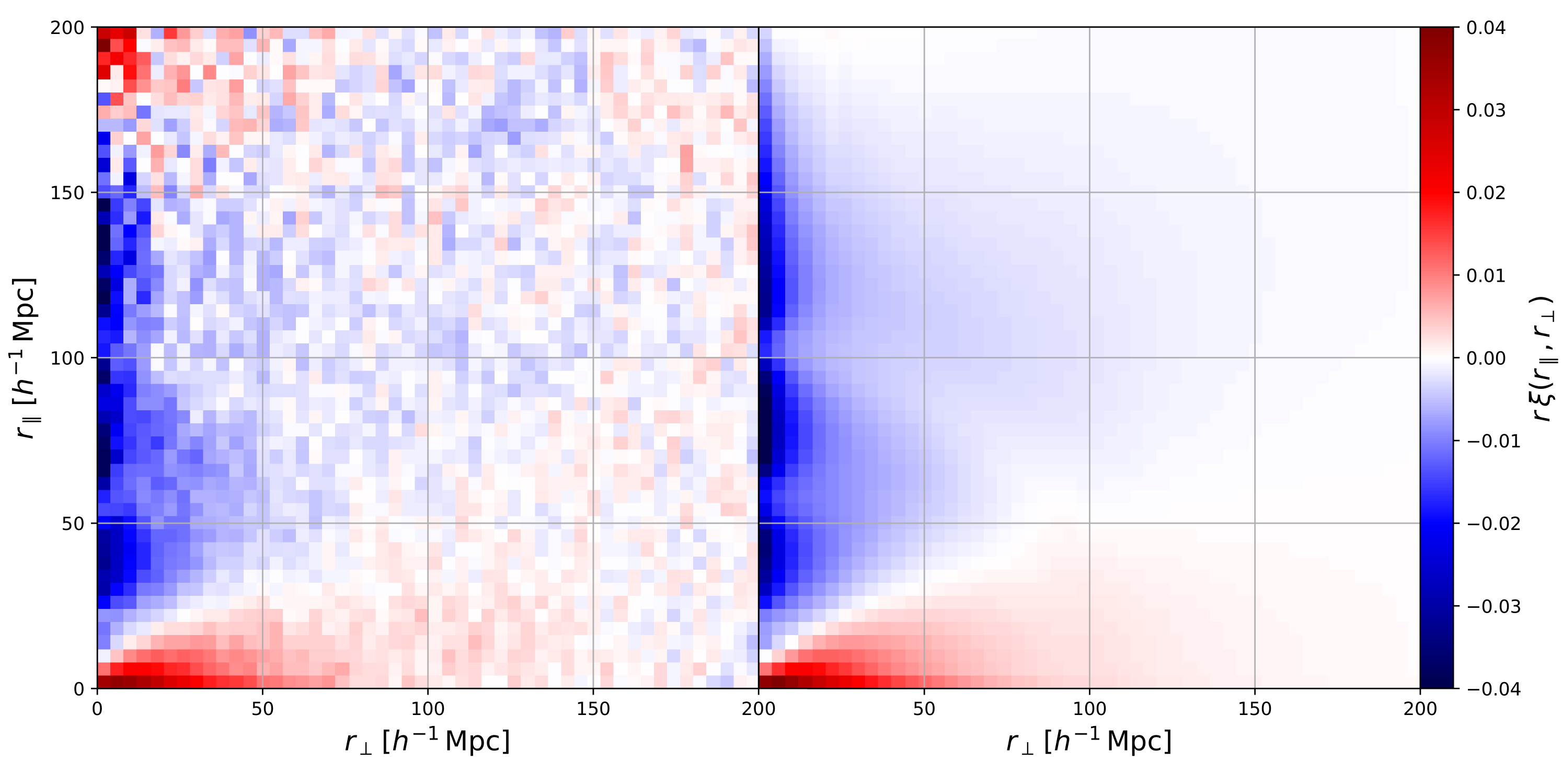}
    \caption{
    Measured (left) and best fit model (right) Ly$\alpha$
    auto-correlation function for two pixels in the Ly$\alpha$ region:
    Ly$\alpha$(Ly$\alpha$) $\times$ Ly$\alpha$(Ly$\alpha$).
    The correlation is multiplied by the separation $|r|$ and the color bar
    is saturated and symmetric around zero for visualization purpose.
    The BAO can be observed as a quarter of a ring at
    $r \sim 100 \, h^{-1}\mathrm{Mpc}$.
    }
    \label{figure::auto_2d}
\end{figure}

Figure~\ref{figure::auto_2d} presents the 
Ly$\alpha$(Ly$\alpha$) $\times$ Ly$\alpha$(Ly$\alpha$) auto-correlation
for the data on the left and for the best-fit model on the right
(Section~\ref{section::Fit_to_the_data}).
The correlation is multiplied by the separation $|r|$ for visual purposes,
and the color bar is saturated and symmetric around zero.
Even though each individual bin of the correlation is noisy when presented in
2D, the BAO scale is seen in the data at large $\mu$
($r\approx\rpar\approx100\,\hMpc$) by the transition from blue,
negative values, to white, zeros.

\begin{figure*}
    \centering
    \includegraphics[width=.95\textwidth]{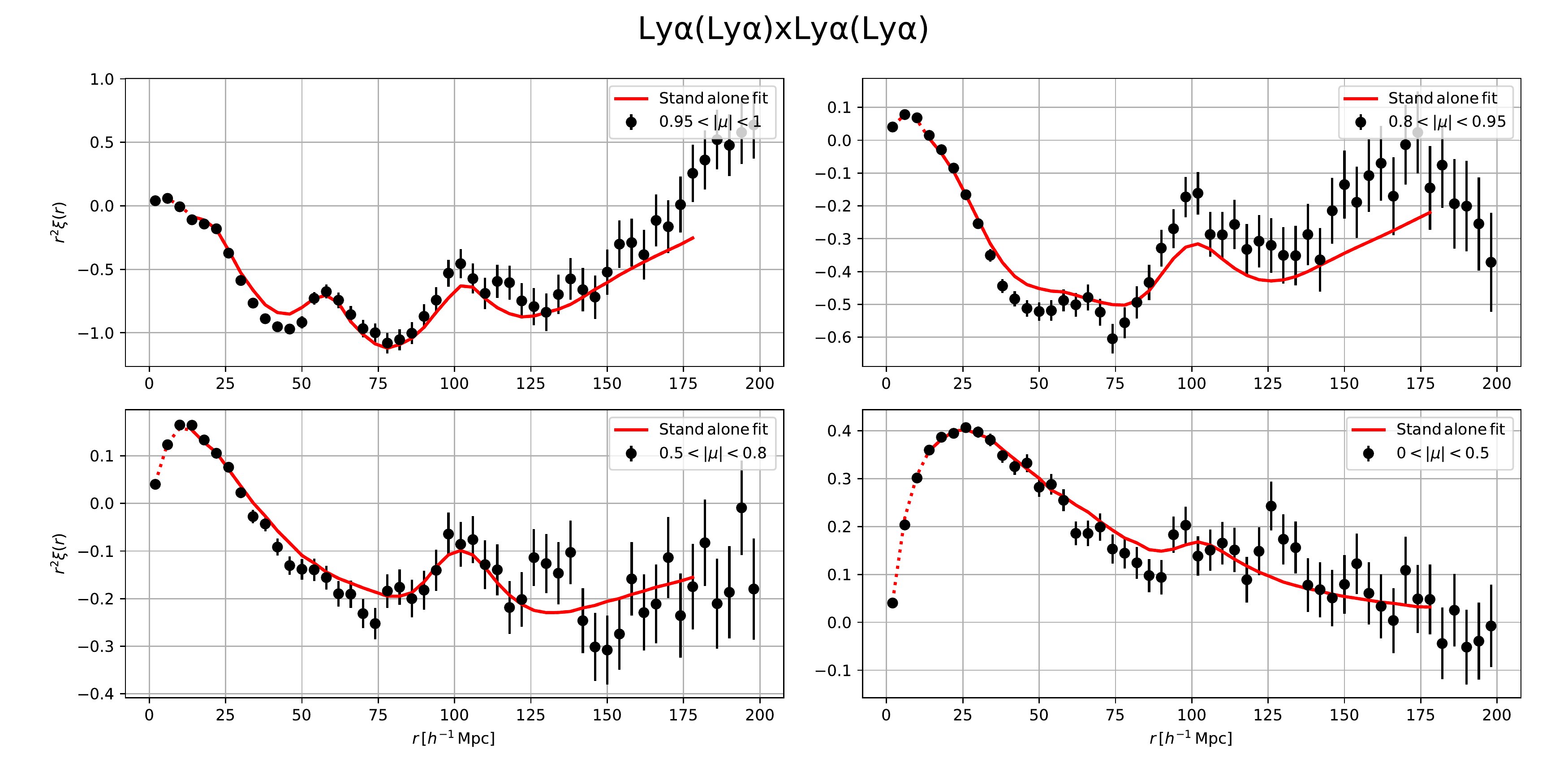} \\
    \includegraphics[width=.95\textwidth]{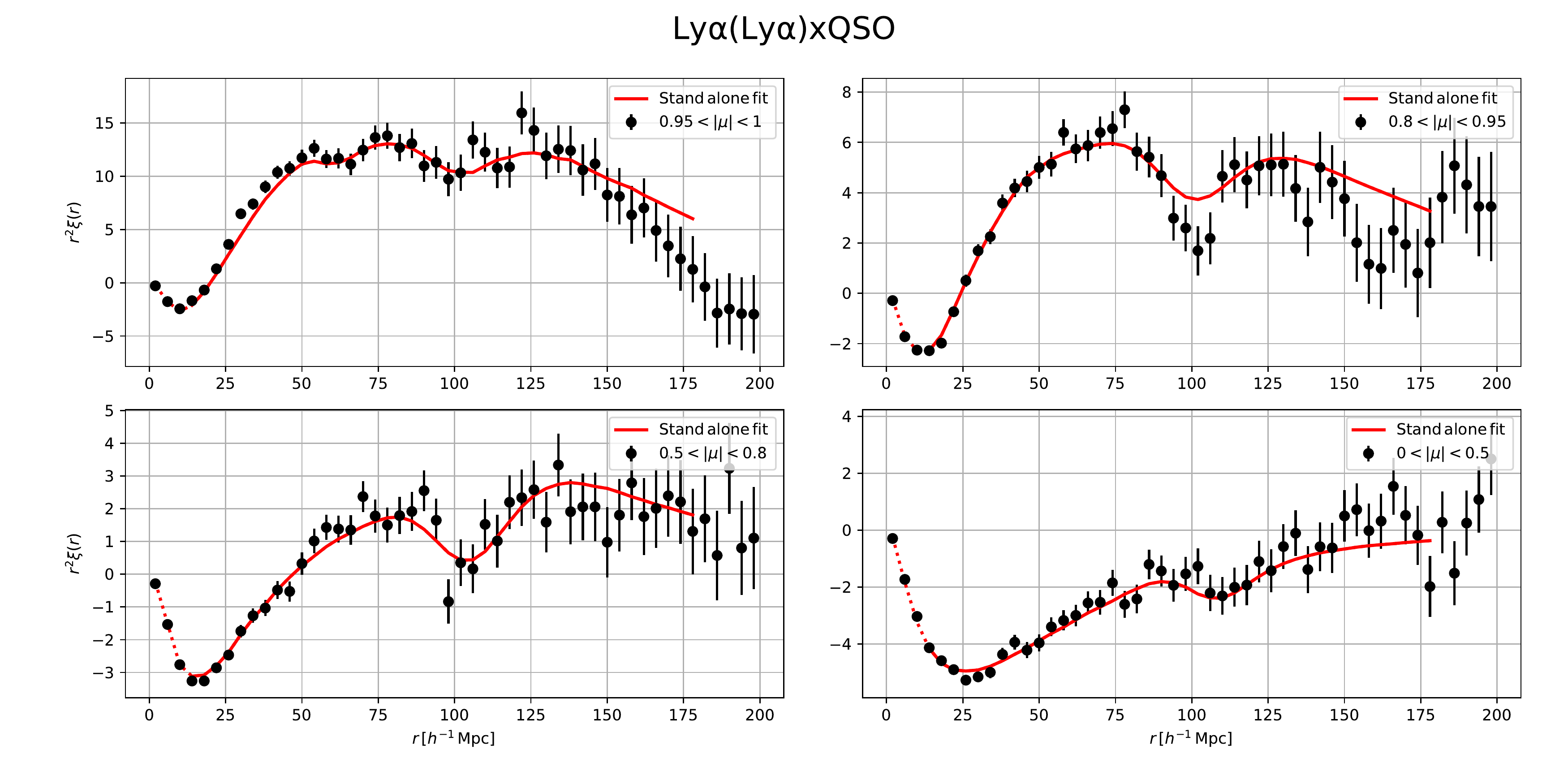}
    \caption{
    Ly$\alpha$ auto-correlation function (top four panels) and Ly$\alpha$-quasar
    cross-correlation (bottom four panels), for pixels in the Ly$\alpha$ region:
    \lyalyalyalya
    and 
    \lyalyaq.
    The correlations are multiplied by the separation $r^{2}$ to better
    see the BAO scale.
    The black points give the measured correlation of figure~\ref{figure::auto_2d},
    and figure~\ref{figure::cross_2d},
    and the red curves give the best fit models, in four wedges of
    $|\mu| = |r_{\parallel}/r|$.
    The dashed red lines give the best fit models,
    interpolated to lower and higher separation than the fitted range:
    $r \in [10,180] \, \hMpc{}$.
    Along the line-of-sight, $|\mu| \in [0.95,1]$, the contribution from
    metals: SiII(119), SiII(119.3), and SiIII(120.7), can be observed as
    extra knees or peaks at respectively $r \sim 20$ and $60 \, \hMpc{}$.
    }
    \label{figure::auto_4_wedges__cross_4_wedges}
\end{figure*}

The measured auto-correlation,
Ly$\alpha$(Ly$\alpha$) $\times$ Ly$\alpha$(Ly$\alpha$),
is also presented in the top four panels
of Figure~\ref{figure::auto_4_wedges__cross_4_wedges}. These panels show
the 2D correlation of Figure~\ref{figure::auto_2d} reduced to a weighted
1D correlation
for four different wedges of $|\mu| = |r_{\parallel}/r|$.
In the same figure, the best fit model is shown in red and is discussed
in Section~\ref{section::Fit_to_the_data}.
The BAO scale peak at $r \sim 100 \, \hMpc{}$ is visible, especially
for $\mu>0.8$.
Four similar panels in Appendix~\ref{section::more_plots} present 
the  auto-correlation
Ly$\alpha$(Ly$\alpha$) $\times$ Ly$\alpha$(Ly$\beta$)
(Figure~\ref{figure::autoLyaLyainLyb_4_wedges__crossLyainLyb_4_wedges}).

The auto-correlation,
\lyalyalyalya,
is also measured in two redshift bins.
This allows us to look at the evolution of the different parameters
of the best fit models, e.g. the bias of Ly$\alpha$, along with allowing systematic tests.
 In their DR14 study, \deSainteAgathe~showed that for the
 Ly$\alpha$ auto-correlation, the optimal way to assign pixel pairs
 to the high- or low-redshift measurement
 is to cut on the mean of the maximum redshift of the two forests,
 rather than simply on the redshift of the two pixels.
This minimizes the cross-covariance 
between the  low-z and high-z
auto-correlations
(Section~\ref{subsection::the_cross_covariance_between_the_correlations}).
As in \deSainteAgathe, we chose $z_{\mathrm{cut}}=2.5$, in order to approximately
get the same weighted number of pairs in both redshift splits.
The resulting auto-correlation function is presented in the four top
panels of Figure~\ref{figure::auto_4_wedges_zBins__cross_4_wedges_zBins},
in Appendix~\ref{section::more_plots}.

The covariance matrix of each auto-correlation function
has $N_{\mathrm{bin}}^{2} = 2500\times2500 = \num{6250000}$ elements.
For two bins $A$ and $B$ of the correlation function $\xi$, 
the covariance is defined by:
\begin{equation}
	C_{A B}
	=
	\langle \xi_{A} \xi_{B} \rangle
	-
	\langle \xi_{A} \rangle
	\langle \xi_{B} \rangle.
	\label{equation::covar_matrix_estimator_auto}
\end{equation}
In this study, the covariance matrix is estimated by dividing the sky into sub-samples
defined by HEALPix pixels and computing their weighted covariance.
Using $\texttt{nside}=16$ over the eBOSS footprint, we get around $880$ sub-samples,
each covering $3.7\times3.7 = 13.4 \, \mathrm{deg^{2}}$ on the sky.
This solid angle is equivalent to
a $250 \times 250 \, (h^{-1}\,\mathrm{Mpc})^{2}$
patch at $z_{\mathrm{eff}} = \zeffshort{}$.
  \lya~pixel pairs are assigned to HEALPix pixels according to the \lya~pixel
  with the smallest right-ascension.
The covariance is then given by the following, neglecting the small correlations
between sub-samples:
\begin{equation}
	C_{AB} = \frac{1}{W_{A} W_{B}} \sum\limits_{s} W_{A}^{s} W_{B}^{s} 
	\left[ \xi^{s}_{A} \xi^{s}_{B}
	- \xi_{A} \xi_{B} \right].
	\label{equation::covar_xi_estimator_subsampling_auto}
\end{equation}
Here, $s$ is a sub-sample with summed  weight $W_{A}^{s}$
and  measured correlation $\xi^{s}$, and $W_A=\sum_sW_A^s$.

  The covariance is dominated by the diagonal elements that are
  of order $C_{AA}\approx\langle\delta^2\rangle^2/N_A$ where
  $\langle\delta^2\rangle$ is the pixel variance and
  $N_A$ is the number of pixel pairs in the bin $A$.
  Deviations from this simple expression are due to intra-forest
  correlations that make the effective number of independent pairs
  less than $N_A$.
We find
\begin{equation}
    Var_{A} \approx \frac{
    \langle\delta^{2}\rangle^{2}
    }{
    fN^{\mathrm{pair}}_{A}
    } \;,
\end{equation}
where $\langle\delta^{2}\rangle\approx0.13\;(0.24)$ and  $f\approx0.4\;(0.8)$
for the \lya~and \lyb~regions.
For the \lyalyalyalya~correlation this is equivalent to
$Var_{A}\approx 1.5\times10^{-10}(100\,\hMpc/\rperp)$,
where the $\rperp$ dependence reflects the approximate
proportionality between $N_A$ and $\rperp$.

  Off-diagonal elements of the covariance matrix are due to intra-forest
  correlations which break the independence of  pixel pairs in different
  bins, $A\neq B$.\footnote{
    The role of intra-forest correlations
    in the covariance matrix is made explicit in the ``Wick'' calculation
    of the covariance in Appendix \ref{section::covariance_matrix_tests}.
  }
Intra-forest correlations reflect both physical correlations
  of absorption and the effect of continuum fitting.
  Imperfections in the continuum fitting increase both
  $\langle\delta^2\rangle$ and the intra-forest
  correlations, and are therefore reflected in
  the covariance matrix, as long as the imperfections themselves
  are not correlated between different forests.

  The size of the off-diagonal elements is made clearer
  through
the normalized covariance matrix, i.e. the correlation
matrix, with elements in $[-1,1]$:
\begin{equation}
	Corr_{A B}
	=
	\frac{C_{A B}}{
    \sqrt{Var_{A}Var_{B}}
    },
	\label{equation::correlation_matrix_estimator_auto}
\end{equation}
where $Var_A=C_{AA}$ is the variance.
The largest values  are $Corr_{AB}\approx0.4$ for $\rperp^A=\rperp^B$
and $|\rpar^A-\rpar^|=4\hMpc$.
Elements with $\rperp^A\neq\rperp^B$ are very small, $<0.03$.

\begin{figure*}
    \centering
    \includegraphics[width=.48\textwidth]{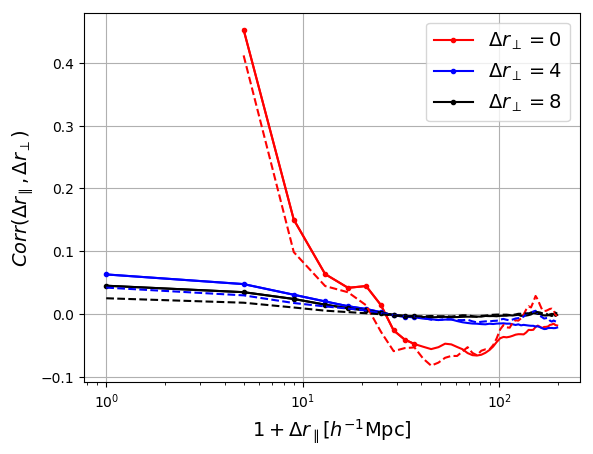}
    \includegraphics[width=.48\textwidth]{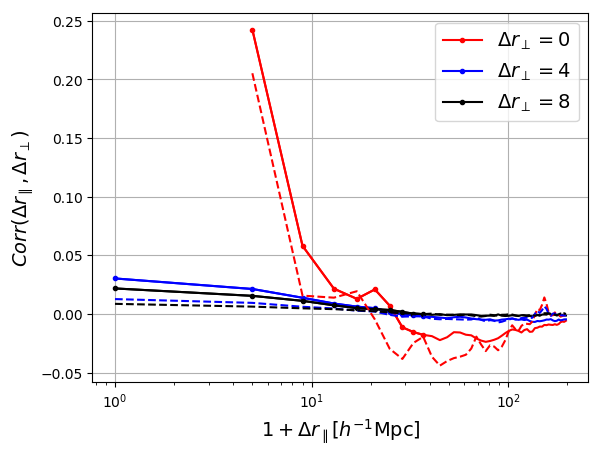}
    \caption{
    The normalized covariance matrix, i.e. the correlation matrix
    (eqn.~\ref{equation::correlation_matrix_estimator_auto}),
    as a function of $(\Delta\rperp,\Delta\rpar)$ for the auto-correlation (left)
    and cross-correlation (right) as estimated by subsampling
(eqn. \ref{equation::covar_xi_estimator_subsampling_auto}).
    The solid lines are for the \lyalyalyalya~and \lyalyaq~functions and the
    dashed lines for the \lyalyalyalyb~and \lyalybq~functions.
    The correlations have been smoothed as explained in the text.
    }
    \label{figure::correlation_matrix}
\end{figure*}

The estimates of the off-diagonal elements of the correlation matrix from
equation~\ref{equation::correlation_matrix_estimator_auto}
are noisy and we smooth them by modeling them
as a function of the difference of separation along and across the line-of-sight:
$Corr_{AB} = Corr(r^{A}_{\parallel},r^{A}_{\perp},r^{B}_{\parallel},r^{B}_{\perp})
= Corr(\Delta r_{\parallel}, \Delta r_{\perp})$,
with $\Delta r_{\parallel} = |r^{A}_{\parallel}-r^{B}_{\parallel}|$
and
$\Delta r_{\perp} = |r^{A}_{\perp}-r^{B}_{\perp}|$.
They are presented in the left panel of Figure~\ref{figure::correlation_matrix},
where they are shown to  decrease rapidly as a function of
$\Delta r_{\parallel}$ and $\Delta r_{\perp}$.

We have also calculated the covariance matrix using other techniques
presented in Appendix~\ref{section::covariance_matrix_tests}.
We find no significant change from the estimates of the
covariance matrix that is used in the main analysis.

\subsection{The \texorpdfstring{Ly$\alpha$}{Lya}-quasar cross-correlation}
\label{subsection::The_Lya_quasar_cross_correlation}

For the \lya-quasar cross-correlation
we use the same estimator as the one from
previous studies (\citealt{2012JCAP...11..059F,2013JCAP...05..018F}, 
\duMasdesBourboux, \Blomqvist ).
It is defined to be the weighted mean of the
flux-transmission field
at a given separation from a quasar:
\begin{equation}
	\xi_{A} = \frac{
    \sum\limits_{(i,j) \in A} w_{i}w_{j} \, \delta_{i}
    }{
	\sum\limits_{(i,j) \in A} w_{i}w_{j}
    }.
	\label{equation::xi_estimator_cross}
\end{equation}
In this equation, $i$ indexes a flux pixel and $j$ a quasar.
The weights $w_{i}$ are as defined in
equation~\ref{equation::definition_of_weight_auto} for the \lya\ absorption fluctuations.
In their study, \citet{2019ApJ...878...47D} fitted the redshift evolution of the
quasar bias and found a best fit power-law of index
$\gamma_{\mathrm{quasar}} = 1.44\pm0.08$.
The quasar weights are then defined to be:
\begin{equation}
    w_{j} = \left( \frac{1+z_{j}}{1+2.25} \right)^{\gamma_{\mathrm{quasar}} -1}.
	\label{equation::definition_of_weight_cross}
\end{equation}

In a similar way as for the auto-correlation, the cross-correlation
is computed for all pixel - quasar pairs, though omitting
pairings of pixels with their own background
quasar whose mean correlation vanishes
due to the continuum fitting procedure. 
The correlation is computed for all pairs within
$r_{\perp} \in [0,200]\,\hMpc{}$. Unlike the auto-correlation, the
cross-correlation is not symmetric by permutation of the two tracers. We
thus have the opportunity to define positive values of the separation along the line-of-sight,
$r_{\parallel}$, when the tracer quasar is in front of the Ly$\alpha$
pixel tracer, i.e. $z_{\mathrm{Ly\alpha}}>z_{\mathrm{quasar}}$.
The line-of-sight separation then ranges over $r_{\parallel} \in [-200,200]\,\hMpc{}$.
With a width of $4\,\hMpc{}$, the correlation is computed on
$N_{\mathrm{bin}} = 100\times50 = 5000$ bins.

The weighted distribution of the pair redshifts in the BAO region
is similar for the cross-correlations to what is presented for the
auto-correlation in the right panel of Figure~\ref{figure::histo_pairs_tracer}.
The comparison on the sky of the weighted number of pairs between DR12 (\duMasdesBourboux)
and this study is also similar to what is shown for the auto-correlation
in Figure~\ref{figure::footprint}.

\begin{figure}
    \centering
    \includegraphics[width=0.98\columnwidth]{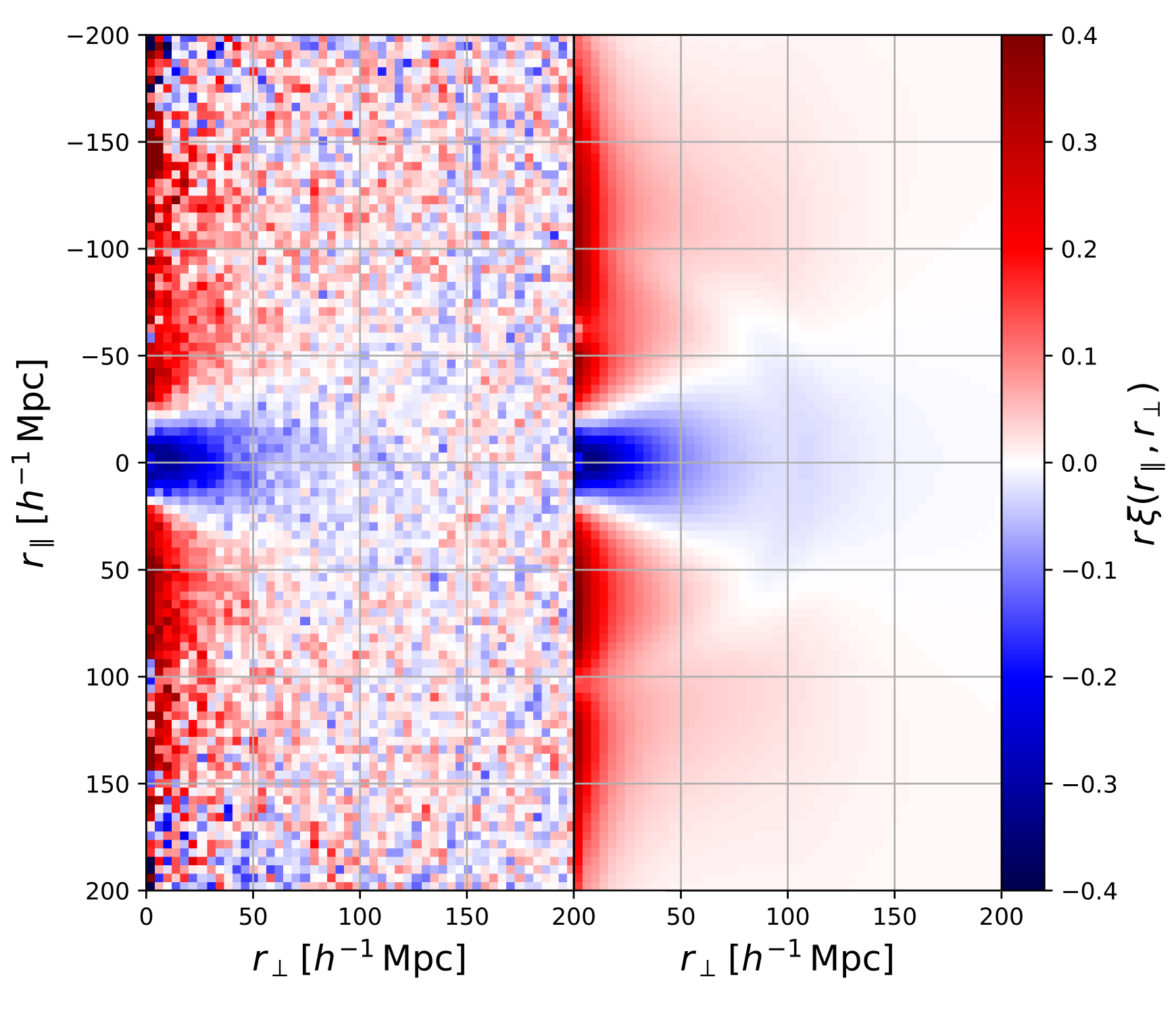}
    \caption{
    Measured (left) and best fit model (right) Ly$\alpha$-quasar
    cross-correlation function where the pixel is in the Ly$\alpha$ region:
    Ly$\alpha$(Ly$\alpha$) $\times$ quasar.
    The correlation is multiplied by the separation $|r|$ and the color bar
    is saturated and symmetric around zero for visualization purpose.
    The BAO can be observed as half a ring at
    $r \sim 100 \, h^{-1}\mathrm{Mpc}$.
    }
    \label{figure::cross_2d}
\end{figure}

Figure~\ref{figure::cross_2d} presents in 2D the measured cross-correlation,
Ly$\alpha$(Ly$\alpha$) $\times$ quasar, and its best-fit model
(Section~\ref{section::Fit_to_the_data}). In such a display, the BAO
scale would be seen as a half ring of radius $r \sim 100\,\hMpc{}$.
However, the signal is difficult to see due to the noise in the data.
We show in the bottom four panels of Figure~\ref{figure::auto_4_wedges__cross_4_wedges},
the same cross-correlation but averaged over four different wedges of $|\mu|$.
There the BAO scale can be clearly observed as a dip at
$r \sim 100\,\hMpc{}$. 
We note that the correlation is reversed,
i.e. negative, with respect to the matter
correlation function, because the bias of Ly$\alpha$ absorption is negative and the
bias of quasars is positive, giving a negative product of biases.
The cross-correlation Ly$\alpha$(Ly$\beta$) $\times$ quasar is presented
in Figure~\ref{figure::autoLyaLyainLyb_4_wedges__crossLyainLyb_4_wedges}
of Appendix~\ref{section::more_plots}.

As in \Blomqvist, we alternatively compute the cross-correlation,
Ly$\alpha$(Ly$\alpha$) $\times$ quasar, in two redshift bins.
For this measurement, it is important that $\langle\delta\rangle = 0$
per bin of observed wavelength.
To ensure this, we recompute the $\delta$
of Section~\ref{subsection::Measurement_of_the_fluctuations_of_the_transmission_field},
independently for the two redshift bins.
We split lines-of-sight by their background quasar redshift, $z_{q}$,
and select the splitting redshift, $z_{\mathrm{cut}} = 2.57$, to
have approximately the same weighted number of pairs in each binned
cross-correlation.
This definition of redshift bins allows us to minimize the cross-covariance
between the two samples
(Section~\ref{subsection::the_cross_covariance_between_the_correlations}).
Figure~\ref{figure::auto_4_wedges_zBins__cross_4_wedges_zBins} of
Appendix~\ref{section::more_plots} presents, in its four bottom panels,
the two redshift bins of the cross-correlation, in four wedges of $\mu$.

The covariance matrix is computed using the same estimators
as in Section~\ref{section::the_covariance_matrix_auto} for
the auto-correlation. The covariance matrix is dominated by its
diagonal elements, i.e. the variance. This latter is approximately
inversely proportional to the number of pairs: 
\begin{equation}
    Var_{A}
    \approx
    \frac{ \langle\delta^{2}\rangle}{0.7 N^{\mathrm{pair}}_{A}} \;,
    \label{equation::var_cross}
\end{equation}
where the factor $0.7$ gives the effective loss of number of pairs,
due to correlations between neighboring pixels.
  For the DR16 data set, the resulting variance 
  for the \lyalyaq~correlation
is $Var_{A}\approx 8.6\times10^{-8}(100\,\hMpc/\rperp)$.
There are a total of $\approx 1.2\times10^{9}$ quasar-pixel pairs in
the BAO region, $80<r<120\,\hMpc$.

  The off-diagonal elements of the covariance matrix
  are presented in the right panel of Figure~\ref{figure::correlation_matrix}.
As with those for the auto-correlation, they decrease rapidly as a function of
$\Delta r_{\parallel}$ and $\Delta r_{\perp}$.

\subsection{The cross-covariance between the correlations}
\label{subsection::the_cross_covariance_between_the_correlations}

We modify the estimator of the covariance matrix via sub-sampling
defined by eqn.~\ref{equation::covar_xi_estimator_subsampling_auto} to compute
the cross-covariance between the different measured correlation functions.
\begin{equation}
	C^{12}_{AB} = \frac{1}{W_{A} W_{B}} \sum\limits_{s} W_{A}^{s} W_{B}^{s} 
	\left[ \xi^{1,s}_{A} \xi^{2,s}_{B}
	- \xi^{1}_{A} \xi^{2}_{B} \right].
	\label{equation::covar_xi_estimator_subsampling_cross_covariance}
\end{equation}
In this equation, $\xi^{1}$ and $\xi^{2}$ are two different measured
correlation functions, e.g. $\xi^{1}$ could be the auto-correlation
Ly$\alpha$(Ly$\alpha$) $\times$ Ly$\alpha$(Ly$\alpha$)
and $\xi^{2}$ the cross-correlation
Ly$\alpha$(Ly$\alpha$) $\times$ quasar.

The different correlation functions are found to be marginally correlated.
For example, the correlations between the \lyalyalyalya\ and \lyalyaq\
functions are less than $1\%$, as shown in
Figure~\ref{figure::compare_covariance_data_mocks}.

\subsection{The distortion-matrices}
\label{section::the_distortion_matrix}

The continuum fitting (eqn.~\ref{equation::definition_quasar_continuum})
  and the projection (eqn.~\ref{equation::projection_1})
mix the $\delta_q$ within a given forest and thereby modify significantly the
correlation function.  The transformation (eqn.~\ref{equation::projection_2})
also has a small effect on the correlations.
Under the assumption that the $\delta$ transformations are linear,
the true and distorted correlation functions are related by 
a ``distortion matrix'', $D_{AB}$:
\begin{equation}
\hat\xi_{\rm distorted}(A)=\sum_B D_{AB} \xi_{\rm true}(B)\; .
\end{equation}

To the extent that the projection given by
eqns.~(\ref{equation::projection_1}) and
  (\ref{equation::projection_1_detail}) captures the full distortion, 
the distortion matrices for the auto- and cross-correlations
can be read off from the coefficients in  (\ref{equation::projection_1_detail}):
\begin{equation}
D^{auto}_{AB} =  W_A^{-1} \sum_{ij\in A} w_i w_j\; \Big(\sum_{i'j' \in B} \eta_{ii'} \eta_{jj'}\Big)\;,
\end{equation}
\begin{equation}
  D^{cross}_{AA^\prime} = W_A^{-1}
  \sum\limits_{(i,j)\in A}w_{i}w_j
  \Big( \sum\limits_{(i^\prime,j)\in A^{\prime}}\eta_{ii^\prime} \Big) \; .
\label{equation::distortionmatrix}
\end{equation}

In the fits of the data described in Section~\ref{section::Fit_to_the_data},
the physical model of the correlation function
(Section~\ref{section::Model_of_the_correlations}) is multiplied by the
distortion matrix before comparison with the data.
This procedure is validated with the analysis of the mock data
sets (Section~\ref{section::Validation_of_the_analysis_with_mocks}),
where it is shown that the BAO parameters are accurately recovered
in the presence of distortion.

%
%

\section{Model of the correlations}
\label{section::Model_of_the_correlations}

This section presents the theoretical model of the auto- and
cross-correlations. Even though this analysis focuses on measuring
the BAO scale along and across the line-of-sight, much effort has also
been invested into better understanding the global shape of the
correlation functions.
The improved modeling of the global shape allows us to better
understand and test possible sources of systematic errors and
to better measure BAO independently of the overall correlation function.
Apart from the contribution of the sky calibration to the auto-correlation
(Section~\ref{contribution_of_sky_residuals}),
  the model
is the same as that used in the  DR12 and DR14 analyses
(\Bautista; \duMasdesBourboux; \deSainteAgathe; \Blomqvist).

  A cosmological analysis of \lya-forest correlations is possible
  only because the dominant \lya-absorption has two essential
  characteristics: it traces the underlying matter fluctuations and,
  for a fixed set of cosmological parameters, there is
  a unique mapping between wavelength and distance.
  Absorption by metals also traces matter fluctuations but has a
  different wavelength-distance relation.
  Correlations due to instrumental and analysis effects do
  not relate at all to cosmology.
  Fortunately, the BAO feature due to the \lya~correlations
  is not degenerate with any of these secondary effects.

  For the dominant \lya-\lya~or \lya-quasar correlations,
  the BAO peak in the space of angular and redshift separation
  appears at $\Delta\theta\sim r_d/D_H(z)$ and $\Delta z\sim r_d/D_M(z)$,
  where $D_H$ and $D_M$ are the Hubble and comoving angular diameter
  distances calculated assuming \lya~absorption.
  We want to measure 
  $r_d/D_H(z)$ and $r_d/D_M(z)$ in a way that does not depend significantly
  on the smooth part of the correlation function underneath
  the BAO peak.
  To do this,
  we follow the procedure described in Section~\ref{section::measuring_the_full_correlation_function}
  of separating the correlation function into two components:
\begin{equation}
    \xi(r_{\parallel},r_{\perp},\alpha_{\parallel},\alpha_{\perp}) =
    \xi_{\mathrm{sm}}(r_{\parallel},r_{\perp})
    +
    \xi_{\mathrm{peak}}(\alpha_{\parallel}r_{\parallel},\alpha_{\perp}r_{\perp}),
    \label{equation::BAO_smooth_peak_separation}
\end{equation}
where $\xi_{\mathrm{sm}}$, is the smooth correlation function, without
the BAO peak, and $\xi_{\mathrm{peak}}$ is the peak-only correlation.
The BAO parameters in this equation are:
\begin{equation}
    \alpha_{\parallel} = \frac{ \left[D_{H}(z_{\mathrm{eff}})/r_{d}\right] }
    {\left[D_{H}(z_{\mathrm{eff}})/r_{d}\right]_{\mathrm{fid}}}
    \hspace*{3mm}{\mathrm{and}}\hspace*{5mm}
    \alpha_{\perp} = \frac { \left[D_{M}(z_{\mathrm{eff}})/r_{d}\right] }
    {\left[D_{M}(z_{\mathrm{eff}})/r_{d}\right]_{\mathrm{fid}}}.
    \label{equation::BAO_definition}
\end{equation}

In the standard fits, we assume that
the correlation function for $\aperp=\apar=1$ is that of the
fiducial cosmology of Table~\ref{table::cosmology_parameters}
as calculated by 
CAMB\footnote{\url{https://github.com/cmbant/CAMB}}.
The smooth and the peak part of the correlation are obtained
at the level of the matter power spectrum, as described below
(Section~\ref{section::measuring_the_full_correlation_function}).

Though the BAO parameters
$(\alpha_{\parallel}, \alpha_{\perp})$ depend of the assumed cosmology,
the measured
$\left( D_{H}(z_{\mathrm{eff}})/r_{d}, D_{M}(z_{\mathrm{eff}})/r_{d} \right)$ do not.
This was studied in detail by \citet{2020MNRAS.494.2076C} in the context
of galaxy correlations.
We further illustrate this property in
Appendix~\ref{section::physical_model_to_the_broadband_of_the_correlation_functions}
where we use a different fiducial cosmology.

  One feature of the peak-smooth splitting 
  (eqn. \ref{equation::BAO_smooth_peak_separation}) is that unless
  the fit yields $\aperp=\apar=1$, the best fit correlation function
  does not obviously correspond to any physical cosmological model.
  In Appendix~\ref{section::physical_model_to_the_broadband_of_the_correlation_functions}
  we choose a physical model with values of $(h,\omegak h^2)$ that
  yield $\aperp=\apar=1$, thereby yielding a best fit that is
  a physical correlation function.
  As demonstrated in
  Appendix~\ref{section::physical_model_to_the_broadband_of_the_correlation_functions},
  changing the model of the smooth continuum through a change in cosmological parameters
  produces a difference in the derived values of
  $\left( D_{H}(z_{\mathrm{eff}})/r_{d}, D_{M}(z_{\mathrm{eff}})/r_{d} \right)$
  of less than one part in 300,
  negligible compared to the statistical precision.

  In addition to the primary correlations from which we measure the
  BAO peak, the full correlation function receives sub-dominant
  contributions from other effects that we describe here.
The theoretical model of the correlation function, $\xi^{t}$, of
equation~\ref{equation::model_correlation_separation_space_auto}, is composed of
different correlations.
For the auto-correlation of Ly$\alpha$, it is given by:
\begin{equation}
    \xi^{t} =
    \xi^{\mathrm{Ly}\alpha \times \mathrm{Ly}\alpha}
    + \sum\limits_{m} \xi^{\mathrm{Ly}\alpha \times m}
    + \sum\limits_{m_{1}, m_{2}} \xi^{\mathrm{m_{1}} \times \mathrm{m_{2}}}
    + \xi^{\mathrm{sky}},
    \label{equation::model_correlation_separation_space_auto}
\end{equation}
where $\xi^{\mathrm{Ly}\alpha \times \mathrm{Ly}\alpha}$
(Section~\ref{section::bias_parameters}) is the model of
the auto-correlation of Ly$\alpha$ absorption, and 
$\xi^{\mathrm{Ly}\alpha \times m}$ and $\xi^{\mathrm{m_{1}} \times \mathrm{m_{2}}}$
(Section~\ref{section::metals})
give the contribution of other
absorbers in the Ly$\alpha$ and Ly$\beta$ regions.
The correlation $\xi^{\mathrm{sky}}$
(sec.~\ref{contribution_of_sky_residuals}) models the correlations
due to the sky-subtraction procedure of the eBOSS pipeline. 

In a similar way, for the measured Ly$\alpha$-quasar cross-correlation
the theoretical model is given by:
\begin{equation}
    \xi^{t} =
    \xi^{\mathrm{Ly}\alpha \times \mathrm{QSO}}
    + \sum\limits_{m} \xi^{\mathrm{QSO} \times m}
    + \xi^{\mathrm{TP}}.
    \label{equation::model_correlation_separation_space_cross}
\end{equation}
In this equation, the first term gives the cross-correlation between
Ly$\alpha$ and quasars, the second gives the
contribution of the cross-correlation between quasars and other absorbers in the
Ly$\alpha$ and Ly$\beta$ spectral regions (Section~\ref{section::metals}).
Finally, the third term models the ``transverse proximity'' (TP)
effect of quasar
radiation on the surrounding gas
(Sec.~\ref{proximity_effect_of_quasars_on_lya_absorption}).

\subsection{The power spectra}
\label{section::measuring_the_full_correlation_function}

  The components of the correlation function that reflect the
  underlying matter correlations are 
given by the Fourier transform of the
tracer biased power-spectrum:
\begin{equation}
    \hat{P}(\kvec) =
    b_{i}b_{j}
    \left( 1+\beta_{i}\mu_{k}^{2}\right)
    \left( 1+\beta_{j}\mu_{k}^{2}\right)
    P_{\rm QL}(\kvec)F_{\rm NL}(\kvec)G(\kvec),
    \label{equation::simple_model_3d_correlation_function}
\end{equation}
where the vector $\kvec = (k_{\parallel},k_{\perp}) = (k, \mu_{k})$  
of modulus $k$, has components along and across the line-of-sight,
$(k_{\parallel},k_{\perp})$, with
$\mu_{k} = k_{\parallel}/k$.
The bias and redshift-space distortion
parameters, $(b,\beta)$, are for the tracer $i$ or $j$.
$P_{\rm QL}$ is the quasi-linear power spectrum defined below,
$F_{\rm NL}$ corrects for non-linear effects at large $\kvec$, and
$G(\kvec)$ is a damping term that accounts for averaging of the
correlation function in individual $(\rpar,\rperp)$ bins.
For the auto-correlation, the main tracer is the Ly$\alpha$ absorption,
$i=j$, for the cross-correlation two main tracers play a role:
Ly$\alpha$ absorption and quasars, $i \neq j$.

For the cross-correlation power spectrum we do not include
relativistic effects that lead to an asymmetry in $\rpar$
(quasar more distant or less distant than the forest).
These effects have been calculated to be sufficiently
small to be negligible for this study
\citep{2020JCAP...04..006L}.
They were included in the study of \Blomqvist~ and were shown to
be partially degenerate with the parameter ($\Delta r_{\parallel,QSO}$)
describing systematic
errors in quasar redshift.

The quasi linear power spectrum provides for the aforementioned decoupling of the peak component:
\begin{equation}
  P_{\rm QL}(\kvec,z) = P_{\rm sm}(k,z) +
  \exp\left[-\frac{k_{\parallel}^2\Sigma_{\parallel}^2+k_{\perp}^2\Sigma_{\perp}^2}{2}\right]
  P_{\rm peak}(k,z)\ .
\end{equation}
The smooth component, $P_{\rm sm}$, is derived from the
linear power spectrum, $P_{\rm L}(k,z)$, via the side-band
technique \mbox{\citep{2013JCAP...03..024K}}
which is
implemented in \picca, with the help of
\texttt{nbodykit}\footnote{\url{https://github.com/bccp/nbodykit}}
\citep{2018AJ....156..160H}.
The CAMB $P_{\mathrm{L}}(z_{\rm eff})$
is Fourier transformed into the correlation function
where two side-bands are defined on both side of the BAO.
A smooth function is then fitted to connect the two side-bands,
allowing us to
produce a correlation function, without the BAO peak.
This latter correlation is Fourier transform back to a smooth power-spectrum,
$P_{\mathrm{sm}}$, that does not have the BAO wiggle features.
The peak power-spectrum is thus defined as
$P_{\mathrm{peak}} = P_{\mathrm{L}}-P_{\mathrm{sm}}$.

The correction for non-linear broadening of the
BAO peak \citep{2007ApJ...664..660E}
is parameterized by
$(\Sigma_{\parallel},\Sigma_{\perp})$,
with $\Sigma_\perp=3.26~\hMpc$ and
\begin{equation}
\frac{\Sigma_{\parallel}}{\Sigma_{\perp}}=1+f\ ,
\end{equation}
where $f=d(\ln g)/d(\ln a)\approx\Omega_{\rm m}^{0.55}(z)$ is the
linear growth rate of structure, resulting in $\Sigma_\parallel=6.42~\hMpc$.

  The function $F_{\rm NL}(\kvec)$ 
in eqn. \ref{equation::simple_model_3d_correlation_function}
accounts for non-linear effects at small scales.
For the auto-correlation,
the most important effects are 
 thermal broadening,
peculiar velocities and non-linear structure growth.
We use
eqn. (3.6) of \citet{2015JCAP...12..017A}
with parameter values from their Table~7
interpolated to our effective redshift $z=2.334$.
For the cross-correlation, the most important effect 
is that of quasar non-linear velocities and,
following \citep{2009MNRAS.393..297P}, we adopt
\begin{equation}
F^{\rm cross}_{\rm NL}(k_{\parallel}) = \frac{1}{1+(k_{\parallel}\sigmav)^2}\ ,
\label{equation::FNL_cross}
\end{equation}
where $\sigmav$ is a free parameter.
This function also takes into account statistical quasar
redshift errors.

The final term in eqn. \ref{equation::simple_model_3d_correlation_function}
accounts for the effect of the binning of the correlation function
on the separation grid.
We assume the distribution to be homogeneous on each bin\footnote{In fact,
  in the perpendicular direction the distribution is approximately
  proportional to $r_\perp$; however, assuming homogeneity produces
  a sufficiently accurate correlation function (B17).} and compute
the function $G(\kvec)$ as 
the product of the Fourier transforms of the
rectangle functions that model a uniform square bin: 
\begin{equation}
  G(\kvec) = {\rm sinc} \left(\frac{k_{\parallel} R_{\parallel}}{2}\right)
                  {\rm sinc} \left(\frac{k_{\perp} R_{\perp}}{2}\right),
\end{equation}
where $R_{\parallel}$ and $R_{\perp}$ are the radial and transverse
widths of the bins, respectively.

\subsection{Ly$\alpha$ and quasar bias parameters}
\label{section::bias_parameters}

The dominant \lya~absorption can be viewed as the sum of
contributions from the diffuse IGM and from
high-column-density (HCD)\footnote{In this and other eBOSS publications we use the term 
  \textit{High Column Density systems} to describe systems with a neutral
  hydrogen column density above $10^{17.2}{\rm cm^{-2}}$,
  i.e., including both Lyman 
  Limit Systems (LLS) and Damped Lyman-$\alpha$ systems (DLAs).}
systems.
HCD absorbers
are expected to trace the 
underlying density field and
their effect on the flux-transmission field depends on whether 
they are identified and given the special treatment
described in Section~\ref{section::Quasar_samples_and_data_reduction}.
If they are correctly identified with the total absorption region masked
and the wings correctly modeled, they can be expected to have no
significant effect on the field.
Conversely, if they are not identified, the measured
correlation function will be modified because
their absorption is spread  along the radial direction.
This broadening effect
introduces a $\kpar$ dependence of the effective bias
\citep{2012JCAP...07..028F}:
\begin{equation}
\left\{
\begin{array}{ll}
\blya^\prime = \blya  + \bhcd \Fhcd (\kpar) \\[2mm]
\blya^\prime\betalya^\prime =
\blya \betalya  + \bhcd \betahcd\Fhcd (\kpar)~~\raisebox{8pt}{,}
\end{array}
\right.
\end{equation}
where $(\blya,\betalya)$ and
$(\bhcd,\betahcd)$ are the bias parameters associated with
the IGM and HCD systems and $\Fhcd$ is a function that
depends on the number and column-density distribution of HCDs.
\citet{Rogers18} numerically calculated $\Fhcd$ using hydrodynamical
simulations and we find that a simple exponential form, 
$\Fhcd =\exp(-\Lhcd \kpar)$, well approximates their results.
Here $\Lhcd$ is a typical length scale for unmasked HCDs.
For eBOSS spectral resolution,
DLA identification is possible for DLA widths
(wavelength interval for absorption greater than 20\%)
greater than $\sim2.0$~nm, corresponding to $\sim14~\hMpc$~in our sample.
Degeneracies between $\Lhcd$ and other anisotropic parameters make the
fitting somewhat unstable.
We therefore impose $\Lhcd=10~\hMpc$ while fitting for the bias parameters
$\bhcd $ and $\betahcd$.
We have verified that setting $\Lhcd$ in the range
$7 <\Lhcd< 13~\hMpc$~does not significantly change the
inferred BAO peak position.
\citet{2020arXiv200402761C} also showed that fixing $\Lhcd$ has a minimal
impact on BAO results when doing a Bayesian analysis of \lya\ BAO.

In our measured correlations functions, different
$(\rpar,\rperp)$ bins
have  mean redshifts that vary over the range $2.32<\bar{z}<2.39$.
Therefore, in order to fit a unique
  function to the data we need to assume a redshift
  dependence of the bias parameters.
Following \citet{2006ApJS..163...80M}, we assume that the product of
$\blya $ and the growth factor of structures varies with redshift as
$(1+z)^{\gamma_{\alpha}-1}$, with $\gamma_{\alpha} = 2.9$,
while  we make use of the approximation that
$\betalya $ does not depend on redshift.
Because the fit of the cross-correlation is only sensitive to the
product of the quasar and Ly$\alpha$ biases, we choose
to adopt a value for the quasar bias.
Following the analysis of \citet{2019ApJ...878...47D} based on a compilation
of quasar bias measurements
\citep{2005MNRAS.356..415C,2013ApJ...778...98S,2016JCAP...11..060L,2017JCAP...07..017L}
we set $b_{\rm q}\equiv b_{\rm q}(z_{\rm eff}) = 3.77$ with a
redshift dependence given by
\begin{equation}
  b_{\rm q}(z) = 3.60 \left(\frac{1+z}{3.334} \right)^{1.44} \; .
  \label{equation::quasarbias}
\end{equation}
The quasar redshift-space distortion, assumed to be redshift independent
in the matter-dominated epochs explored here, is
\begin{equation}
\beta_{\rm q} = \frac{f}{b_{\rm q}}\ .
  \label{equation::quasarbeta}
\end{equation}
Setting $f=0.9704$ for our fiducial cosmology yields $\beta_{\rm q}=0.269$,
which we take to be redshift-independent.

\subsection{Absorption by metals}
\label{section::metals}

We model the power spectrum for correlations involving metals
with the same form we use for \lya-\lya~and \lya-quasar correlations
(eqn \ref{equation::simple_model_3d_correlation_function})
except that HCD effects are neglected.
Each metal species, $n$, has bias parameters $(b_n,\beta_n)$.
The Fourier transform of the power spectrum $P_{mn}(\kvec,z)$
for the absorber pair $(m,n)$ is then
the model correlation function of the  pair
$(m,n)$: $\xi_{\rm mod}^{m-n}(\tilde{r}_\parallel,\tilde{r}_\perp)$,
where $(\tilde{r}_\parallel,\tilde{r}_\perp)$ are the separations calculated using
the correct restframe wavelengths, $(\lambda_m,\lambda_n)$.
  For the cross-correlation,
  the same form is used except that
  the bias parameters for the metal $n$ are replaced with
  the quasar bias parameters and the separations are calculated
  using the quasar redshift.

Since we assign redshifts to pixels assuming Ly$\alpha$ absorption,
the rest-frame wavelength we ascribe to a metal transition
is not equal to the true rest-frame wavelength.
For $m\neq n$ and $n=\,{\rm Ly}\alpha$,
this misidentification results in a shift of the model correlation
function with vanishing separation of absorbers corresponding to
reconstructed separations
$\rperp=0$ and
$\rpar\approx(1+z)D_H(z)(\lambda_m-\lambda_{\rm Ly\alpha})/\lambda_{\rm Ly\alpha}$.
The values of $\rpar$ for the
major metal transition are given in Table~\ref{table::metals_in_mocks}.
For $m=n$, the reconstructed separations are scaled from the true separations
by a factor $\DHub(z)/\DHub(z_\alpha)$ for $\rpar$ and  
by a factor $\DM(z)/\DM(z_\alpha)$ for $\rperp$, where $z$ and $z_\alpha$ are
the true and reconstructed redshifts.

For each pair $(m,n)$ of contaminants,
we compute the shifted-model correlation function
with respect to the unshifted-model correlation function,
$\xi_{\rm mod}^{m-n}$,
by introducing a metal matrix  $M_{AB}$ \citep{2018JCAP...05..029B}, such that:
\begin{equation}
  \xi_{\rm mod}^{m-n}(A) \rightarrow
  \sum_B M_{AB}
  \xi_{\rm mod}^{m-n}(\tilde{r}_\parallel(B),\tilde{r}_\perp(B)),
\end{equation}
where:
\begin{equation}
M_{AB}=\frac{1}{W_A}\sum_{(m,n)\in A,(m,n)\in B}w_mw_n \;,
\end{equation}
where $W_A=\sum_{(m,n)\in A}w_mw_n$, 
$(m,n)\in A$ refers to pixel separations computed assuming $z_\alpha$,
and $(m,n)\in B$ to pixel separations computed using
the redshifts of the $m$ and $n$ absorbers, $z_m$ and $z_n$.
We take into account the redshift dependence of the weights
in the computation of $w_m$ and $w_n$.

In the fits of the data, we include terms corresponding
to metal-\lya~correlations for the four silicon transitions
listed in Table~\ref{table::metals_in_mocks}.
Since these correlations 
are only visible in $(\rpar,\rperp)$ bins corresponding to
small physical separations,
$b_m$ and $\beta_m$
cannot be determined separately.
We therefore fix $\beta_m = 0.50$
as done in \Bautista\ based on
measurements of the cross-correlation between DLAs and the
\lya\ forest \citep{2012JCAP...11..059F}.
Correlations between CIV and \lya~ give a contribution outside
the $\rpar$ range studied here, but we include effects
of the CIV-CIV auto-correlation and 
fix $\beta_{\rm CIV(eff)} = 0.27$ \citep{2018JCAP...05..029B}.

\subsection{Proximity effect of quasars on \texorpdfstring{Ly$\alpha$}{Lya} absorption}
\label{proximity_effect_of_quasars_on_lya_absorption}

The term in (\ref{equation::model_correlation_separation_space_cross})
representing
the transverse proximity effect takes the form
\citep{2013JCAP...05..018F}:
\begin{equation}
  \xi^{\rm TP} =
  \xi^{\rm TP}_{0}
  \left(
  \frac{1\,\hMpc}{r}
  \right)^{2}
  \exp(-r/\lambda_{\rm UV})\ .
\end{equation}
This form supposes isotropic emission from the quasars.
We fix $\lambda_{\rm UV}=300~\hMpc$ \citep{2013ApJ...769..146R}
and fit for the amplitude $\xi^{\rm TP}_{0}$.

\subsection{Correlations due to sky subtraction}
\label{contribution_of_sky_residuals}

The correlation $\xi^{\mathrm{sky}}$, of
equation~\ref{equation::model_correlation_separation_space_auto},
models the
correlations induced by the sky-subtraction procedure.
The sky-subtraction for each spectrum  is done independently for each
spectrograph,
i.e. per half-plate, for the 500 fibers (450 science fibers and
$\sim 40$ sky fibers).
The Poisson fluctuations in the sky spectra that are subtracted
induce correlations in spectra obtained with the same spectrograph
at the same observed-wavelength, leading to an excess
correlation in  $\rpar=0$ bins.
This was observed in the DR12 auto-correlation measurement of \Bautista,
where they decided to reject such same-spectrograph pairs in their measurement.
This effectively removes the excess correlation at $\rpar=0$.
However, because of continuum fitting, the excess correlation at $\rpar=0$
generates a smooth distortion of the correlation function for all $\rpar$
and this was not removed by the procedure of \Bautista.
Here we do not remove the same-spectrograph pairs which allows us to
fit for its amplitude and thereby take into account the smooth component
induced by continuum fitting.
  Apart from improving the fit to the data, this procedure is necessary
  because the DR16 analysis combines all measurements of each quasar so spectra
  no longer correspond to a unique spectrograph.

\begin{figure}
    \centering
    \includegraphics[width=\columnwidth]{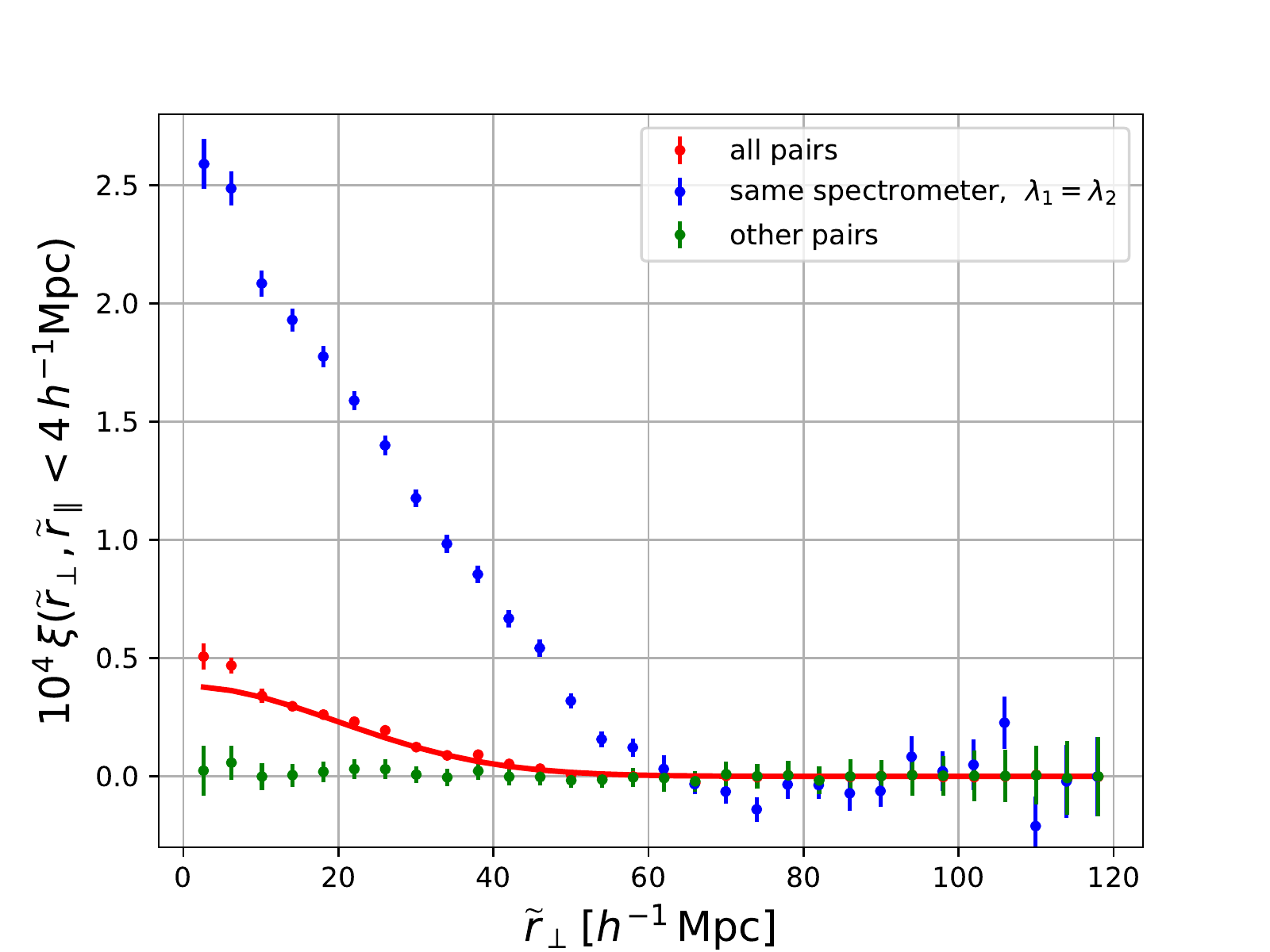}
    \caption{
      Correlations due to the sky-subtraction procedure
      in the MgII(11) spectral region:
      $\lrf\in[260,276]\,\mathrm{nm}$.
      The correlations are shown for $\tilde{r}_\parallel=0$ as a function
      of $\tilde{r}_\perp$ where $(\tilde{r}_\parallel,\tilde{r}_\perp)$  are the
      coordinates calculated assuming MgII~absorption.
      The green points show  the correlations for pairs of pixels that are not
      taken with the same spectrograph (half-plate).  Due to the small
      amount of MgII(11) absorption, the correlations are consistent
      with zero, as expected.
      The blue points show the correlations for pairs of points taken
      with the same spectrograph ($\lambda_1=\lambda_2$ pairs only).
      Here, the sky-subtraction procedure
      generates significant correlations that diminish
      with increasing angular separation (increasing $\tilde{r}_\perp$).
      The red points show the correlations for all pairs and the
      red curve the fit using the form of eqn. \ref{equation::model_sky_auto}.
    }
    \label{figure::correct_for_same_half_pairs_MgII}
\end{figure}

The sky-fiber induced correlations are easily seen 
at rest-frame wavelength longer than
the Ly$\alpha$ quasar emission line where correlations due
to absorption are small.
An example is shown in Figure~\ref{figure::correct_for_same_half_pairs_MgII}
showing correlations in the 
``MgII(11)'' spectral region, $\lrf\in[260,276]\,\mathrm{nm}$,
\citep{2019ApJ...878...47D}, where the pixels are immediately blueward
of the MgII quasar emission line,
$\lrf = 279.6 \, \mathrm{nm}$.
The correlation is consistent with zero for pairs of pixels taken
with different spectrographs but is significant for pairs from
the same spectrograph.
This spurious correlation decreases rapidly with increasing
angular separation.
This decrease is due to the eBOSS
pipeline that adds broad-band functions to the sky calibration, modeling its
variation across the size of the plate and thus decreasing the
correlations with increasing angular separation.

Since as the angular separation increases, the number of same-spectrograph
pairs decreases, and the number of different-spectrograph pairs increases,
the weighted sum of the two, in red curve in the figure, can be modeled as a
Gaussian function of $\rperp$:
\begin{equation}
    \xi^{\mathrm{sky}}(r_{\parallel},r_{\perp}) =
    \begin{cases}
        \frac{ A_{\mathrm{sky}} }{ \sigma_{\mathrm{sky}} \sqrt{2 \pi} }
        \exp{\left( -\frac{1}{2} \left( \frac{r_{\perp}}{\sigma_{\mathrm{sky}}} \right)^{2} \right)}
        & \mathrm{, if\,} r_{\parallel} = 0\\
        0
        & \mathrm{, if\,} r_{\parallel} \neq 0
    \end{cases}
    \label{equation::model_sky_auto}.
\end{equation}
The two free parameters $(A,\sigma)_{\mathrm{sky}}$ give the scale and
the width of the correlation.

We fit the measured auto-correlation, MgII(MgII(11)) $\times$ MgII(MgII(11)),
via the distorted model of equation~\ref{equation::relation_measure_model}
and the model of the sky correlation
of equation~\ref{equation::model_sky_auto}. The fit is performed in the
same condition as for the Ly$\alpha$ auto-correlation,
hence $r \in [10,180]\,\hMpc{}$ (Section~\ref{section::Fit_to_the_data}),
resulting in $1590$ bins.
We get a best fit with
$\chi^{2} / (DOF) = 1551.94 / (1590-2)$, corresponding to a probability
$p = 0.76$ when using 
only the best observation.  We find
$\chi^{2} / (DOF) = 1531.49 / (1590-2)$ and $p = 0.86$ when using 
all observations.
In both cases, we find
$(A,\sigma)_{\mathrm{sky}} \sim (1 \times 10^{-3}, 20 \, \hMpc{})$.
A null test on the auto-correlation, using all observations, proves the very
high significance of these two parameters:
$\chi^{2} = 2824.85$, and $\Delta \chi^{2} = 1293.36$, 
for a difference of two parameters.
Interestingly, a similar null-test, removing
the $\rpar=0$ bins
still yields a significant measurement of the two parameters,
$\Delta \chi^{2} = 223.50$. 
This test shows that the effect of the sky-subtractions residuals is
mainly
in the $\rpar=0$ bins
but that a non negligible
fraction is brought to other bins by the effects of continuum fitting.

For the \lyalyalyalya~correlation function,
the spurious correlation from the sky calibration in the $\rpar=0$ bins
reaches $\approx20\%$ of the physical correlation
at $\rperp\approx50\,\hMpc$.
For $\rpar\neq0$, the distortion-induced correlation is smooth so it
does not affect our BAO measurement. However, it has a non-negligible
effect on the measurement of the Ly$\alpha$ bias
parameters (Table~\ref{table::best_fit_BAO_data_different_model}),
because of its $\mu$ dependence.

\subsection{Power-law correlations}
\label{the_effective_broadband_functions}

An important test of systematic effects in the position of the BAO peak
is performed by adding polynomial ``broadband'' terms to the correlation
function.
We follow the procedure and choice of broadband forms used by
\Bautista~and adopt the form 
\begin{equation}
B(r,\mu) =
\sum_{j=0}^{\jmax} \sum_{i=\imin}^{\imax} a_{ij} \frac{L_{j}(\mu)}{r^i}
\hspace*{5mm}\; (j\,\rm{even}) ,
\label{xibroadbandeq}
\end{equation}
where the $L_{j}$ are Legendre polynomials.
Following \Bautista~we fit with
$(\imin,\imax)=(0,2)$ corresponding to a parabola in 
$r^2\xi_{\rm smooth}$ underneath the BAO peak. We set 
$\jmax=6$, giving four values of $j$  corresponding 
to approximately independent broadbands in each of the four angular ranges.

\subsection{The distorted model}

The expected measured correlation function, $\hat{\xi}$, is related to the
true
$\xi^{t}$, by the distortion matrix:
\begin{equation}
    \hat{\xi}_{A} = \sum\limits_{A^{\prime}} D_{A A^{\prime}} \xi^{t}_{A^{\prime}}.
    \label{equation::relation_measure_model}
\end{equation}
The distortion matrix, $D_{A A^{\prime}}$, is computed in
Section~\ref{section::the_distortion_matrix}
for the auto-
and the cross-correlation.
The matrix accounts for the correlations introduced by the
continuum fitting procedure.
In this equation, $A^{\prime}$ is a bin of the model and $A$ is a bin
of the measurement.\footnote{
    There are no theoretical requirements for the model
to be computed using the same binning as the measurement. However,
because of computing memory limitations we decide to keep the same binning.
In Appendix~\ref{section::Systematic_tests_on_BAO} we test the consequences
of binning the distortion matrix on a finer grid with twice as many bins
for the model as the measurement.
  }

%
%

\section{Validation of the analysis with mocks}
\label{section::Validation_of_the_analysis_with_mocks}

In this section we present a set of synthetic realizations of the eBOSS survey,
and we use them to validate our BAO analysis.
We start in \ref{subsection::raw_skewers} by presenting the methodology used
to simulate the quasar positions and the \lya\ forest fluctuations, and
continue in \ref{subsection::simulate_spectra} where we describe how we
simulate the quasar continua and instrumental artifacts.
We present the BAO fits on the simulated datasets in 
\ref{subsection::bao_in_mocks}.

\begin{figure*}[tb]
    \centering
    \includegraphics[width=0.80\textwidth]{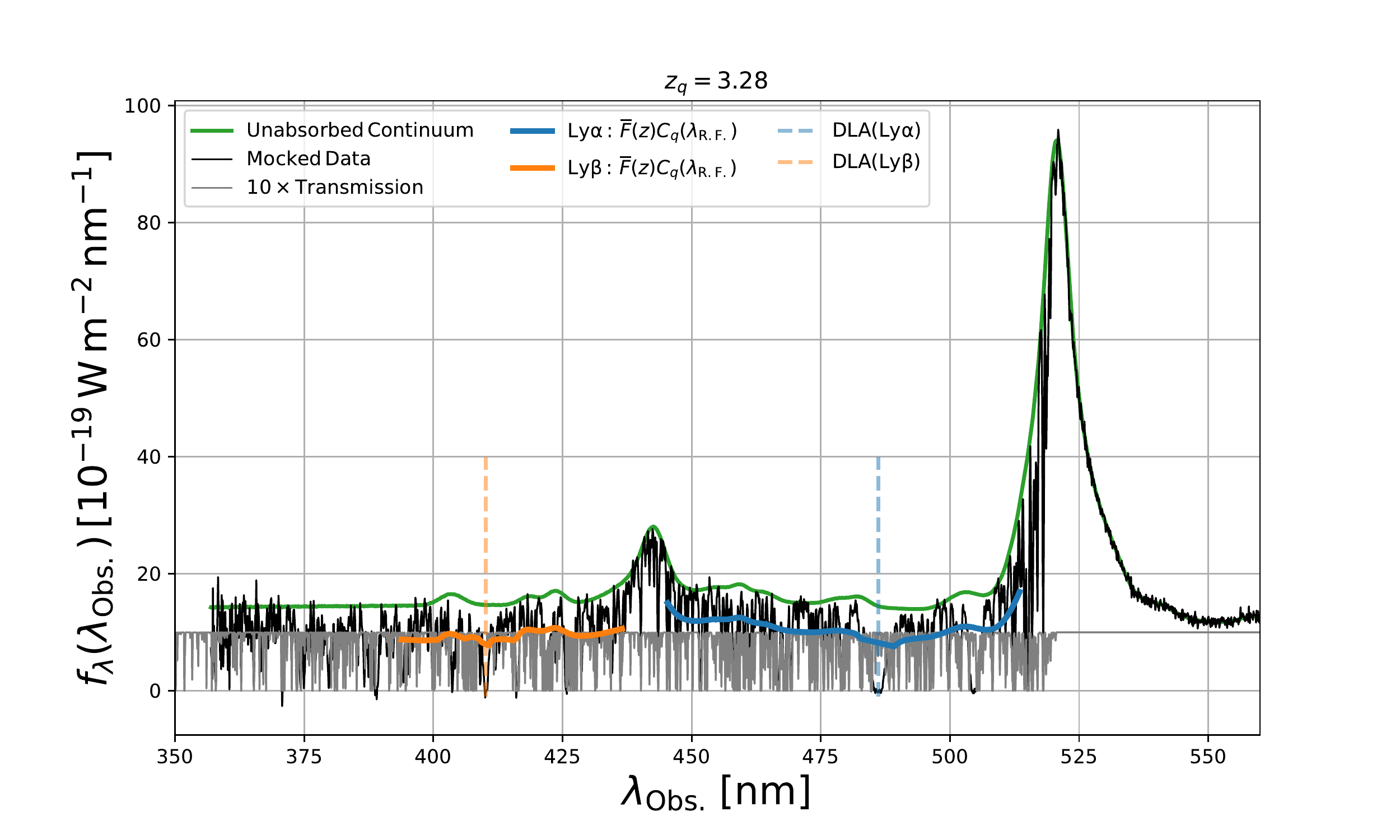}
    \caption{Example of synthetic spectra, for one of the brightest quasars
  in our sample. 
  The green line shows the unabsorbed continuum; 
  the solid blue and orange lines show the continuum multiplied by the mean
  transmitted flux fraction in the \lya\ region and in the \lyb\ region;
  the gray line
  shows the transmitted flux fraction generated with 
  \texttt{LyaCoLoRe}, multiplied by a factor of 10 for better visualization;
  the black line shows the final spectrum, including instrumental noise.
  The vertical dashed blue line shows the position of a DLA, while its
  corresponding \lyb\ absorption is marked with a vertical dashed orange line.
    }
    \label{figure::mock_spectrum}
\end{figure*}

\subsection{Gaussian random field simulations}
\label{subsection::raw_skewers}

\citet{2012JCAP...01..001F} presented a method to efficiently simulate 
\lya\ forest spectra for a given survey configuration.
This method was used in multiple analyses of the \lya\ auto-correlation
in BOSS \citep{
2011JCAP...09..001S,
2013A&A...552A..96B,
2013JCAP...04..026S,
2015A&A...574A..59D,
2017A&A...603A..12B}, and it enabled multiple tests of potential
systematics and the validation of the BAO analysis pipeline.

However, because the quasar positions behind the simulated forests
did not correspond to density peaks,
these simulated datasets did not have the correct cross-correlation
between the quasars and the \lya\ forest.
This made them unfit for BAO
studies using this cross-correlation.
For this reason, the DR12 analysis of the cross-correlation in
\duMasdesBourboux~presented a new set of simulations using a
different algorithm described in \citet{2011A&A...534A.135L} and that
included a realistic cross-correlation.
This allowed the first validation of BAO in cross-correlation using mock data,
and it showed that the auto- and the cross-correlation measurements of the
BAO scale have a very small correlation and can therefore be combined.
These simulations, however, did not have the correct survey geometry, and
assumed instead that the lines-of-sight were parallel.

We have developed two different sets of simulations that have been
generated in collaboration with the Lyman-$\alpha$ working group of the
Dark Energy Spectroscopic Instrument (DESI).
The results presented in this section use simulations described in
\citet{2020JCAP...03..068F},
developed primarily at University College London and
referred here as the {\it London mocks}.
A second set
\citep{SaclayMocks},
were developed primarily at CEA Saclay and referred here as
the {\it Saclay mocks}.
These mocks were completed after the London mocks and benefit
from refinements inspired by experience with the London mocks.
They will be used for future studies of the non-BAO
component of the correlation
function, including the effect of HCDs.

\subsubsection{London mocks}

The methodology of the London mocks is described in \citet{2020JCAP...03..068F},
and we refer the reader to that publication for further details.
We use the \texttt{CoLoRe} package 
\footnote{\url{https://github.com/damonge/CoLoRe}}
to generate a low-resolution Gaussian field on a very large box, of length
$L \approx 10\,\hGpc$, enclosing an all-sky ``light cone'' to $z=3.8$.
The same package allows us to populate the peaks of the Gaussian field with
quasars, following an input quasar bias and number density.
Finally, \texttt{CoLoRe} also provides the interpolated values of the
Gaussian field on the lines-of-sight from the center of the box towards the
quasars, as well as the interpolated radial velocity computed using the
gradient of the gravitational potential.

The boxes used in this publication had $4096^3$ cells, resulting in a
resolution of  $\approx 2.4\,\hMpc$. 
As described in \citet{2020JCAP...03..068F},
we use the \texttt{LyaCoLoRe} software
\footnote{\url{https://github.com/igmhub/LyaCoLoRe}}
to add extra small-scales fluctuations to each line-of-sight in order to
reproduce the variance in the \lya\ forest in the data.
We then apply a log-normal transformation to the resulting Gaussian field,
and use the Fluctuating Gunn-Peterson Approximation (FGPA) to compute the
optical depth in each cell.
Finally, we use the radial velocities to include redshift space distortions,
and we compute the transmitted flux fraction for each \lya\ spectrum.

\subsubsection{Saclay mocks}

The methodology of the Saclay mocks is described in
\citet{SaclayMocks},
and we refer the reader to that publication for further details.
The Saclay mocks are similar to the London mocks. 
The main difference is that velocity gradient boxes are produced in addition
to density field boxes, and redshift space distortions are implemented by
applying the log-normal transformation to the sum of the density and
velocity gradient fields before applying FGPA.
A nice feature of this implementation is that it allows for a prediction of
the correlation function of the mocks, in a way similar to
\citet{2012JCAP...01..001F}.
The code
\footnote{\url{https://github.com/igmhub/SaclayMocks}}
does not benefit from
the high level of parallelization of the CoLoRe package.
The Gaussian field boxes are then smaller and seven of them are needed to
cover the eBOSS footprint.

\subsection{Simulating eBOSS spectra}
\label{subsection::simulate_spectra}

Once we have the simulated transmitted flux fraction along each line-of-sight,
we use these to simulated synthetic quasar spectra with the relevant
astrophysical and instrumental artifacts.
The methodology here is similar to the one used in the BOSS \lya\ analyses,
described in \citet{2015JCAP...05..060B}, and it can be summarized as follows:

\begin{itemize} 
 \item Add contaminant absorption, including higher order hydrogen lines,
  metal absorbers and High Column Density (HCD)
  systems.
 \item Multiply each transmitted flux fraction by a simulated quasar continuum.
 \item Convolve the simulated spectrum with the resolution of the spectrograph,
  pixelize it, and add Gaussian noise simulating eBOSS observations.
\end{itemize}

Figure~\ref{figure::mock_spectrum} shows a simulated spectrum with different
levels of complexity.

\subsubsection{Adding contaminants}

In section \ref{subsection::raw_skewers} we have described how we simulate
our signal, the \lya\ forest transmitted flux fraction.
In order to study potential systematic biases caused by the presence of
contaminants, we have the option to
add other absorption lines from the  Lyman series,
as well as metal lines. 

We simulate the optical depth of each contaminant as a rescaled version of
the \lya\ optical depth, mapped into a different observed wavelength using
its restframe wavelength.
The scaling factors for the higher order Lyman lines are computed using the
oscillator strengths of each transition (see equation $1.1$ in 
\citealt{2013JCAP...09..016I}).
The scaling factor for \lyb~is 0.1901, and we include absorption
up to the fifth Lyman line.
The scaling factor for the metal lines have been tuned to approximately match
the level of contamination observed in the data, and are presented in 
Table~\ref{table::metals_in_mocks}.
We have only included the four Silicon lines that have a restframe wavelength
similar to that of the \lya, and that are detected in the data in both the
auto-correlation and in the cross-correlation with quasars.

\begin{table}
    \caption{Metal lines included in the synthetic spectra}
    \centering
        
    \begin{tabular}{l l l l}

    Metal line & $\lambda_m$ [nm] & Relative & $\rpar$ \\
     & & strength ($\times 10^3$) & [$\hMpc$] \\
    
    \noalign{\smallskip} \hline \noalign{\smallskip}

   SiIII & 120.7   & 1.892 & -21 \\
   SiIIa & 119.0   & 0.642 & -64 \\
   SiIIb & 119.3   & 0.908 & -56 \\
   SiIIc & 126.0   & 0.354 & +111 \\

    \end{tabular}

    \tablecomments{
   For each metal line   their
  transition wavelength and their relative strength with respect to the 
  \lya\ optical depth are given.
  The last column gives the position of maximum
  apparent correlation $\rperp=0$ and
  $\rpar\approx(1+z)D_H(z)(\lambda_m-\lambda_{\rm Ly\alpha})/\lambda$
  corresponding to
  metal and \lya~absorption at the same physical position
  but reconstructed assuming only \lya~absorption at
  $z_{\rm eff}=2.334$.
    }
    \label{table::metals_in_mocks}
    
\end{table}

Using the method described in \citet{2020JCAP...03..068F},
we generate a catalog of
HCDs, with a column density and redshift distribution in agreement with current
observations.
We included systems with column densities in the range 
$\log{N_{HI}}=[17.2,22.5]$ using the software \texttt{pyigm} \citep{pyigm},
itself calibrated using observations from \citet{2014MNRAS.438..476P}.
The clustering of HCDs in the mocks has a large scale bias consistent with
the observed clustering of DLAs 
\citep{2012JCAP...11..059F,2018MNRAS.473.3019P}.
A Voigt profile is then constructed for each absorber, and the absorption is
included in the simulated transmitted flux fraction.

\subsubsection{Simulating quasar continua}

We assign a random magnitude to each simulated quasar, following the quasar
luminosity function measured in \citet{2013ApJ...773...14R} using quasars from
the ninth data release of SDSS, truncated at a maximum magnitude of 
${\rm r}=21.3$.
We then use the publicly available software package \texttt{simqso} 
\footnote{\url{https://github.com/imcgreer/simqso}}
to generate an unabsorbed continuum for each quasar.
\texttt{simqso} is based on the simulations used in \citet{2013ApJ...768..105M},
and we refer the reader to that publication for a detailed description.
In short, each continuum is constructed by adding a set of emission lines on
top of a broken power law.

The distribution of quasar redshift errors is discussed in section 
\ref{subsection::quasar_redshifts} and in appendix 
\ref{section::quasar_redshifts}.
Redshift errors have an important impact in the cross-correlation of quasars
and the \lya\ forest, since they cause an extra smoothing of the correlations
along the line-of-sight.
In order to emulate this effect when measuring correlations in the synthetic
datasets, we add a random redshift error drawn from a Gaussian distribution
with $\sigma_z = 400\, \kms$.

\subsubsection{Simulating instrumental effects}

The next step is to simulate the instrumental artifacts, including a Gaussian
smoothing due to the finite spectral resolution of spectrograph, the
pixelisation of the spectra and the instrumental noise.

While these effects have a dramatic impact on studies of the small-scale
correlations of the \lya\ forest, only the noise level has an impact
on BAO measurements.
This justifies our choice of \texttt{specsim} \citep{specsim}, a simulator
of the DESI spectrograph, to simulate the BOSS spectrograph. 
The small differences in the resolution and pixel size do not impact BAO
measurements, and we have chosen an exposure time to reproduce the 
signal-to-noise in the eBOSS survey.

\subsection{Comparison of mocks and data}
\label{subsection:mocks_vs_data}

As described in \citet{2020JCAP...03..068F},
the synthetic \lya\ spectra were tuned
in order to reproduce large scale measurements from the BOSS DR12 results, as
well as reproducing the line-of-sight power spectrum over the relevant scales.
In Appendix~\ref{section::more_plots} we present figures comparing the
measured correlations in the mocks and in the data, including for the first
time comparisons of measurements of \lya\ absorption in the \lyb\ region.

\begin{figure*}[tb]
    \centering
    \includegraphics[width=0.48\textwidth]{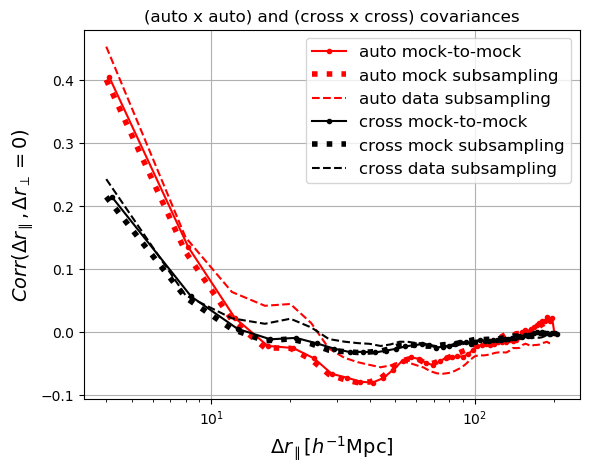}
    \includegraphics[width=0.48\textwidth]{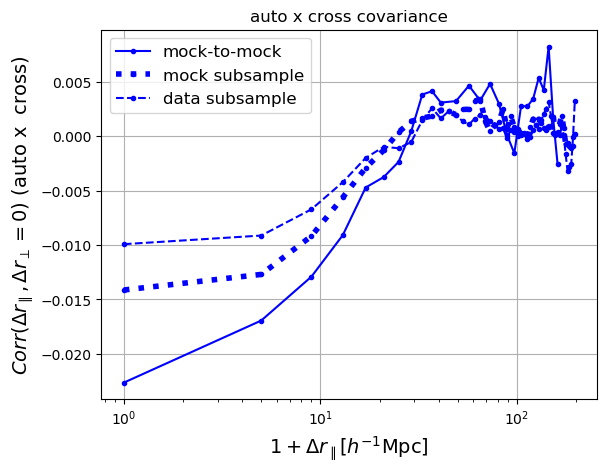}
    \caption{
      Normalized mock covariances (correlations)
      measured by mock-to-mock variations and by sub-sampling and
        data covariances measured by sub-sampling.
        The left panel shows the $\Delta\rperp=0$ correlations vs. $\Delta\rpar$
        for the \lyalyalyalya~and \lyalyaq~correlation functions.
        The right panel shows the correlations between
        the \lyalyalyalya~and \lyalyaq~functions.
        The largest correlations are those for
        $(\Delta\rperp=0,\Delta\rpar=4~\hMpc)$ with $Corr\approx0.4$
        for the auto and $\approx0.2$ for the cross. All correlations
        between the auto and cross correlations are at the percent level.
        For the mocks, the sub-sampling and mock-to-mock correlations agree at
        the percent level.
        The difference between mock and data correlations at $\Delta\rpar\approx30\,\hMpc$
        are due to different amounts of metals in the mocks and data, leading to
        differences in the intra-forest correlation function.
    }
    \label{figure::compare_covariance_data_mocks}
\end{figure*}

  An important role of the mocks is to verify the sub-sampling
  procedure for calculating the covariance matrix of the correlation
  functions. This can be done by comparing the sub-sampling
  calculation with the mock-to-mock variations of the correlation
  functions.
  We see no difference between the two at more than the 1\% level.
  For example,
in the left panel of Figure~\ref{figure::compare_covariance_data_mocks} we
illustrate off-diagonal elements of the correlation (normalized covariance)
matrix, that has been smoothed as described in 
Sect.~\ref{section::the_covariance_matrix_auto}
for the \lyalyalyalya~and \lyalyaq~correlation functions.
The correlations measured by sub-sampling and by mock-to-mock
variations are in excellent agreement, verifying our procedure.

Fig.  \ref{figure::compare_covariance_data_mocks} shows that
the covariances for mocks and data have a very similar structure,
the differences being due in part to
the smaller metal component in the mocks than in the data leading to
less correlation between nearby wavelengths.
The mock variances are a factor two smaller for the mocks
for the auto-correlation and a factor 1.5 for the
cross-correlations.  This is due to fewer  low-flux spectra
in the mocks compared to the data.

\subsection{Analysis of BAO in the mocks}
\label{subsection::bao_in_mocks}

\begin{table*}
        \caption{
          Fit results for mock data sets.}
        \centering
        
        \begin{tabular}{l l l l l l }

        Mock set 
        &       $\overline{\alpha_{\parallel}}\;\;  \overline{\sigma}$ 
        &       $\overline{\aperp} \;\; \overline{\sigma}$
        &       $\overline{b_{\rm \eta\,Ly\alpha}} \;\; \overline{\sigma}$ 
        &       $\overline{\beta_{\rm Ly\alpha}} \;\; \overline{\sigma}$
        &       $\overline{\chi^{2}_{\rm min}}/DOF, \, \overline{proba}$
        \\

\noalign{\smallskip} \hline \hline \noalign{\smallskip}

\lyalya$\times$ \lyalya  \\ 
raw          &  1.012  \;  0.021  &  0.985  \;  0.028  &  -0.2  \;  0.001  &   \;  1.568 0.021  &  1629.67 /( 1590 - 4 )  \;  0.27  \\ 
+cont+noise  &  1.003  \;  0.027  &  0.995  \;  0.04  &  -0.201  \;  0.002  &   \;  1.486 0.028  &  1603.57 /( 1590 - 4 )  \;  0.41  \\ 
   +metals   &  1.012  \;  0.029  &  0.987  \;  0.05  &  -0.202  \;  0.002  &   \;  1.485 0.03  &  1600.32 /( 1590 - 8 )  \;  0.43  \\ 
+HCD+$\sigmav$  &  1.004  \;  0.029  &  1.001  \;  0.041  &  -0.205  \;  0.003  &   \;  1.48 0.051  &  1597.99 /( 1590 - 10 )  \;  0.42  \\ 

\noalign{\smallskip} \hline \hline \noalign{\smallskip}

\lyalya$\times$ quasar  \\ 
raw          &  1.008  \;  0.025  &  0.999  \;  0.024  &  -0.189  \;  0.003  &   \;  1.568 0.041  &  3227.31 /( 3180 - 6 )  \;  0.29  \\ 
+cont+noise  &  1.008  \;  0.029  &  0.992  \;  0.033  &  -0.192  \;  0.004  &   \;  1.491 0.061  &  3203.5 /( 3180 - 6 )  \;  0.39  \\ 
   +metals   &  1.006  \;  0.029  &  0.994  \;  0.033  &  -0.193  \;  0.004  &   \;  1.51 0.063  &  3194.72 /( 3180 - 10 )  \;  0.4  \\ 
+HCD+$\sigmav$  &  1.003  \;  0.033  &  0.998  \;  0.033  &  -0.199  \;  0.007  &   \;  1.48 0.081  &  3212.8 /( 3180 - 10 )  \;  0.35  \\ 

\noalign{\smallskip} \hline \hline \noalign{\smallskip}

\lyalya$\times$ \lyalyb  \\ 
raw          &  1.005  \;  0.025  &  0.996  \;  0.034  &  -0.2  \;  0.002  &   \;  1.588 0.026  &  1609.71 /( 1590 - 4 )  \;  0.41  \\ 
+cont+noise  &  1.014  \;  0.049  &  0.983  \;  0.069  &  -0.202  \;  0.003  &   \;  1.509 0.05  &  1592.39 /( 1590 - 4 )  \;  0.47  \\ 
   +metals   &  1.02  \;  0.049  &  0.994  \;  0.065  &  -0.203  \;  0.004  &   \;  1.528 0.054  &  1589.8 /( 1590 - 8 )  \;  0.46  \\ 
+HCD+$\sigmav$  &  1.009  \;  0.054  &  1.019  \;  0.087  &  -0.206  \;  0.004  &   \;  1.502 0.085  &  1609.88 /( 1590 - 10 )  \;  0.35  \\ 

\noalign{\smallskip} \hline \hline \noalign{\smallskip}

\lyalyb$\times$ quasar  \\ 
raw          &  1.028  \;  0.042  &  1.009  \;  0.044  &  -0.189  \;  0.005  &   \;  1.595 0.073  &  3189.37 /( 3180 - 6 )  \;  0.45  \\ 
+cont+noise  &  1.008  \;  0.07  &  1.015  \;  0.082  &  -0.193  \;  0.01  &   \;  1.527 0.146  &  3183.49 /( 3180 - 6 )  \;  0.47  \\ 
   +metals   &  0.994  \;  0.071  &  1.002  \;  0.093  &  -0.19  \;  0.01  &   \;  1.495 0.149  &  3216.12 /( 3180 - 10 )  \;  0.36  \\ 
+HCD+$\sigmav$  &  1.011  \;  0.08  &  1.013  \;  0.099  &  -0.192  \;  0.015  &   \;  1.447 0.186  &  3195.57 /( 3180 - 10 )  \;  0.4  \\ 

\noalign{\smallskip} \hline \hline \noalign{\smallskip}

all combined   \\ 
raw          &  1.009  \;  0.012  &  0.995  \;  0.014  &  -0.203  \;  0.001  &   \;  1.628 0.015  &  9948.08 /( 9540 - 7 )  \;  0.02  \\ 
+cont+noise  &  1.005  \;  0.017  &  0.992  \;  0.022  &  -0.206  \;  0.002  &   \;  1.553 0.023  &  9788.16 /( 9540 - 7 )  \;  0.1  \\ 
   +metals   &  1.01  \;  0.018  &  0.989  \;  0.023  &  -0.206  \;  0.002  &   \;  1.558 0.025  &  9822.24 /( 9540 - 11 )  \;  0.08  \\ 
+HCD+$\sigmav$  &  1.005  \;  0.019  &  0.998  \;  0.023  &  -0.205  \;  0.002  &   \;  1.464 0.036  &  9625.56 /( 9540 - 13 )  \;  0.31  \\

        \end{tabular}
        
\tablecomments{
        Shown are the mean values and mean standard deviations
($\Delta\chi^2=1$)
          for $\apar$, $\aperp$, $b_{\rm \eta\,Ly\alpha}$, and $\betalya$.
  Results are shown for ``raw'' mocks (no quasar continuum or noise added to the transmission field) and for three progressively
  more realistic mocks (adding continua, noise, metals, HCDs and quasar velocity dispersion).
  Means are based on 100 mocks for (+cont+noise) and (HCD+$\sigmav$)
  and 10 otherwise.
}
    \label{table::best_fit_BAO_mocks}

\end{table*}

In Table~\ref{table::best_fit_BAO_mocks} we present the results of our
BAO analysis on multiple synthetic realizations of our dataset.
In order to study the effect of the different systematics included in the mock
spectra, we run our analysis on different versions of each synthetic dataset:
\begin{itemize}
 \item \lya: analysis run directly on the simulated transmitted flux fractions,
  including only \lya~absorption. 
  This analysis does not need to do a continuum fitting, and is therefore
  unaffected by the continuum distortions discussed in section 
  \ref{section::the_distortion_matrix}.
 \item + continuum + noise: analysis run on simulated spectra, with quasar
  continua and instrumental noise.
 \item + metals: analysis run on the same simulated spectra than above,
  but including also absorption from metal lines.
 \item + HCDs + $\sigmav$: analysis run on the same simulated spectra
  than above, but including also absorption from HCDs and including random
  redshift errors for the quasars.
  When analyzing the real data, we mask HCDs identified with an algorithm
  that is able to find the largest systems, with $\log N_{HI}>20.3$. 
  Similarly, when computing the delta field in the mock spectra, HCDs with 
  $\log N_{HI}>20.3$ are corrected using the input $N_{HI}$ and pixels with
  absorption larger than 20\% are masked.\footnote{For one mock set,
    we have applied our DLA detector and verified that masking according
    to detected DLA parameters or to the input DLA parameters leads to nearly
  identical results.}
\end{itemize}

We ran all four analyses (\lya\ auto-correlations and quasar cross-correlations
using the \lya\ and the \lyb\ regions) on ten realizations for each of the
synthetic datasets described above. 
We analyze 90 extra realizations for two of the settings above, 
( +continuum+noise) and ( +HCDs +$\sigmav$),
allowing us to test
possible systematic biases on the BAO measurements at a level 10 times smaller
than our statistical uncertainty.

The main conclusion of this analysis is that we are able to recover the
right BAO scale even in the presence of contaminants, and that our model
is able to describe the correlations measured on the synthetic datasets.

\subsubsection{Correlations in a BAO measurement}

Anisotropic BAO studies provide a measurement of the line-of-sight BAO scale,
$\alpha_\parallel$, and of the transverse scale, $\alpha_\perp$.
In BAO analysis using galaxy clustering these parameters are 
anti-correlated at the 40\% level, i.e., they have a correlation coefficient
around $\rho \left( \alpha_\parallel, \alpha_\perp \right) = -0.4$.

\begin{table}
     \caption{Correlations between the BAO parameters measured in the different
  correlation functions as measured in mocks.
    }
   \centering
        
    \begin{tabular}{l l }
    Correlation &$\rho \left( \alpha_{\parallel}, \alpha_{\perp} \right)$  \\
    functions &    \\
    
    \noalign{\smallskip} \hline \noalign{\smallskip}

    auto $\times$ auto         &  $-0.572 \pm 0.093$ \\
    cross $\times$ cross                  &  $-0.470 \pm 0.093$ \\
    auto Ly$\beta$ $\times$ auto Ly$\beta$         &  $-0.615 \pm 0.064$ \\
    cross Ly$\beta$ $\times$ cross Ly$\beta$         &  $-0.347 \pm 0.052$ \\

    \end{tabular}


    \label{table::correlation_parameter_BAO}

\end{table}

In Table~\ref{table::correlation_parameter_BAO} we use the BAO analyses on
the 100 synthetic datasets to measure the correlation coefficient
$\rho \left( \alpha_\parallel, \alpha_\perp \right)$ in the four BAO
measurements discussed in this publication.
One can see that the results are also anti-correlated, with values ranging
from $-0.347$ to $-0.615$.
  The correlations involving $(\apar,\aperp)$ from distinct pairs of
  correlation functions are
consistent with zero.

\subsubsection{Covariance between different large-scale correlations}

The BAO results from the \lya\ auto-correlation and from its cross-correlation
with quasars were first combined in \citet{2014JCAP...05..027F}.
The authors used a Fisher matrix approach to argue that cosmic variance was
sub-dominant in both results, and therefore the results could be considered
independent.

This assumption was confirmed in \citet{2017A&A...608A.130D}, where the
covariance between the measurements was measured to be very small on synthetic
datasets.
In the right panel of Figure~\ref{figure::compare_covariance_data_mocks}
we present an updated study of this covariance when using the synthetic
datasets described above. 
The correlation coefficients are smaller than 2\% for all combinations,
probing the assumption that the measurements are independent.

%
%
\section{Fits to the data}
\label{section::Fit_to_the_data}

This section presents the results of the fits to the
data\footnote{We use the minimizer
  \url{https://github.com/scikit-hep/iminuit}}.
We fit the two auto-correlation functions, \lyalyalyalya~and \lyalyalyalyb,
and the two cross-correlation functions, \lyalyaq~and \lyalybq.
Table~\ref{table::best_fit_parameters_data} lists the parameters
used to fit the correlation functions and their best-fit values.
Various combinations of the four functions have also been fit
and are listed in the table.

\subsection{The correlation functions}
\label{section::The_correlation_functions}

The correlation functions are fit for separations
$r \in [10,180] \, \hMpc{}$, and for all directions,
$\mu \in [0,1]$ for the auto-correlations and $\mu \in [-1,1]$
for the cross-correlations.
This gives $1590$ bins for the two auto-correlations and twice as many,
$3180$, for both cross-correlations.
The different best fit models differ in their number and nature of
the free parameters.
However, they all share the two BAO parameters,
$(\alpha_{\parallel},\alpha_{\perp})$, that are the focus of this study.

The auto-correlations baseline model is composed of the Kaiser
model for biased matter density tracers,
the effect of correlated sky residuals,
the Kaiser model for metal-induced correlations,
and finally the effect of HCDs.
The model has $13$ free parameters
 when fitting separately the \lyalyalyalya~or \lyalyalyalyb~functions.
 The simultaneous fit for the two functions
 has 15 free parameters because the two sky subtraction parameters
 are independent for the two function while
 the other parameters are common to them.

The cross-correlation baseline model is composed of the Kaiser
model for quasars, Ly$\alpha$ and the contamination by metals, along with
the modeling of the quasar systemic and statistical redshift errors.
Because the RSD parameter of quasars, $\beta_{\mathrm{QSO}}$,
is highly correlated to the bias and beta parameters of Ly$\alpha$,
we fix it following
eqns. \ref{equation::quasarbias} and \ref{equation::quasarbeta}
at the effective redshift of the fit.
  We fix the HCD parameters at the values determined by the combined
(auto + cross)
  fits since they are highly correlated with other parameters, especially
  the quasar velocity offset.
  We also fix the quasar proximity-effect parameters to those determined
  by the combined fit.
This results in a best fit model with $10$ parameters, for
both cross-correlations. Since all these parameters are shared between
Ly$\alpha$(Ly$\alpha$) $\times$ quasar and
Ly$\alpha$(Ly$\beta$) $\times$ quasar, the combined fit is composed also
of $10$ free parameters.

The combined fit to the four measured correlation functions,
two auto and two cross-correlations, is composed of the different models
cited in the two previous paragraphs.
Running a combined fit to the auto and the cross-correlation allows us
to break degeneracies and thus allows us to free $\beta_{\mathrm{QSO}}$,
to fit for HCDs in the cross-correlation and take into account the
proximity effect of quasars onto Ly$\alpha$ absorption.
The resulting best fit model has $19$ parameters.

When fitting simultaneously more than one of the four
correlation functions, we neglect the
cross-covariance between them.
The subsampling covariance of
Sect.~\ref{subsection::the_cross_covariance_between_the_correlations}
were at the level of $1\%$ and this was confirmed by the mock-to-mock
variations shown in Fig.~\ref{figure::compare_covariance_data_mocks}.
The expected low correlation between BAO parameters measured
with different correlation functions was confirmed by
the mock-to-mock variations
discussed in
Sect.~\ref{section::Validation_of_the_analysis_with_mocks}.

We follow the prescriptions of \deSainteAgathe~(their Appendix~A) and
compute the effective redshift of the measurement to be
$z_{\mathrm{eff.}} = \zeff{}$, when running the combined fit to all the
four correlations. This effective redshift is defined to be the pivot
redshift for the two BAO parameters $\alpha_{\parallel}$ and
$\alpha_{\perp}$.
  All parameters are assumed to be redshift-independent
  except for the bias parameters
  (Section~\ref{section::bias_parameters})
  which the fitter returns at the effective redshift.

The best fit model of the individual fit to the auto-correlation
Ly$\alpha$(Ly$\alpha$) $\times$ Ly$\alpha$(Ly$\alpha$) is given in 2D
in the right panel of Figure~\ref{figure::auto_2d}
and in four wedges of $|\mu|$ in the top four panels of
Figure~\ref{figure::auto_4_wedges__cross_4_wedges}.
In a similar way the best fit model of the cross-correlation
Ly$\alpha$(Ly$\alpha$) $\times$ quasar
is given in 2D in the right panel of Figure~\ref{figure::cross_2d}
and for four wedges of $|\mu|$ at the bottom of
Figure~\ref{figure::auto_4_wedges__cross_4_wedges}.
Finally both auto and cross-correlation using the Ly$\beta$ region, 
Ly$\alpha$(Ly$\alpha$) $\times$ Ly$\alpha$(Ly$\beta$)
and Ly$\alpha$(Ly$\beta$) $\times$ quasar,
are given in
Figure~\ref{figure::autoLyaLyainLyb_4_wedges__crossLyainLyb_4_wedges}
of Appendix~\ref{section::more_plots}.

\begin{table*}
    \caption{
      Best fit parameters for seven different fits to the
      correlation functions.
}
    \scalebox{0.90}{
    \begin{tabular}{l | l | l | l | l || l | l | l}
    
    Parameter 
    & Ly$\alpha$(Ly$\alpha$)
    & Ly$\alpha$(Ly$\alpha$)
    & Ly$\alpha$(Ly$\alpha$)
    & Ly$\alpha$(Ly$\beta$)
    & full auto
    & full cross
    & all combined
    \\
     
    & $\,\,\,$ $\times$ Ly$\alpha$(Ly$\alpha$)
    & $\,\,\,$ $\times$ Ly$\alpha$(Ly$\beta$)
    & $\,\,\,$ $\times$ quasar
    & $\,\,\,$ $\times$ quasar
    &
    &
    &
    \\

\noalign{\smallskip} \hline \hline \noalign{\smallskip}

 $\alpha_{\parallel}$ &  $1.047 \pm 0.035$ & $1$ & $1.059 \pm 0.039$ & $1$ & $1.038 \pm 0.032$ & $1.056 \pm 0.039$ & $1.045 \pm 0.023$  \\
 $\alpha_{\perp}$     &  $0.980 \pm 0.052$ & $1$ & $0.932 \pm 0.047$ & $1$ & $0.959 \pm 0.049$ & $0.952 \pm 0.042$ & $0.956 \pm 0.029$  \\
 
 \noalign{\smallskip}
 \noalign{\smallskip}

 $b_{\eta, \mathrm{Ly\alpha}}$ &  $-0.2009 \pm 0.0039$ & $-0.2045 \pm 0.0065$ & $-0.225 \pm 0.010$ & $-0.202 \pm 0.024$ & $-0.201 \pm 0.0034$ & $-0.2222 \pm 0.0093$ & $-0.2014 \pm 0.0032$  \\
 $\beta_{\mathrm{Ly}\alpha}$ &  $1.657 \pm 0.088$ & $1.74 \pm 0.16$ & $1.95 \pm 0.14$ & $1.67 \pm 0.30$ & $1.669 \pm 0.077$ & $1.92 \pm 0.13$ & $1.669 \pm 0.071$  \\
 $10^{3} b_{\eta, \mathrm{SiII(119)}}$ &  $-2.96 \pm 0.50$ & $-2.85 \pm 0.96$ & $-4.5 \pm 1.2$ & $3.0 \pm 3.1$ & $-2.94 \pm 0.45$ & $-3.6 \pm 1.1$ & $-2.62 \pm 0.39$  \\
 $10^{3} b_{\eta, \mathrm{SiII(119.3)}}$ &  $-2.08 \pm 0.50$ & $-1.73 \pm 0.93$ & $2.1 \pm 1.2$ & $-3.6 \pm 3.1$ & $-2.02 \pm 0.44$ & $1.4 \pm 1.1$ & $-1.21 \pm 0.39$  \\
 $10^{3} b_{\eta, \mathrm{SiII(126)}}$ &  $-2.2 \pm 0.63$ & $-4.1 \pm 1.1$ & $-1.81 \pm 0.77$ & $-1.7 \pm 1.9$ & $-2.68 \pm 0.55$ & $-1.8 \pm 0.72$ & $-2.29 \pm 0.43$  \\
 $10^{3} b_{\eta, \mathrm{SiIII(120.7)}}$ &  $-4.54 \pm 0.51$ & $-4.34 \pm 0.95$ & $-0.98 \pm 0.96$ & $-0.5 \pm 2.4$ & $-4.5 \pm 0.45$ & $-0.92 \pm 0.89$ & $-3.72 \pm 0.40$  \\
 $10^{3} b_{\eta, \mathrm{CIV(eff)}}$ &  $-5.2 \pm 2.6$ & $-5.1 \pm 2.6$ & $-4.8$ & $-4.8$ & $-5.2 \pm 2.7$ & $-4.8$ & $-4.9 \pm 2.6$  \\
 $b_{\mathrm{HCD}}$ &  $-0.0522 \pm 0.0044$ & $-0.0556 \pm 0.0078$ & $-0.0501$ & $-0.0501$ & $-0.0523 \pm 0.0039$ & $-0.0501$ & $-0.0501 \pm 0.0036$  \\
 $\beta_{\mathrm{HCD}}$ &  $0.610 \pm 0.083$ & $0.549 \pm 0.087$ & $0.703$ & $0.703$ & $0.646 \pm 0.081$ & $0.703$ & $0.704 \pm 0.080$  \\
 $\beta_{\mathrm{QSO}}$ &  & & $0.2602$ & $0.2602$ & & $0.2602$ & $0.2601 \pm 0.0059$  \\
 $\Delta r_{\parallel, \mathrm{QSO}}$ ($\hMpc$) &  & & $0.23 \pm 0.13$ & $0.39 \pm 0.33$ & & $0.25 \pm 0.12$ & $0.10 \pm 0.11$  \\
 $\sigmav$ ($\hMpc$) &  & & $7.73 \pm 0.44$ & $7.9 \pm 1.1$ & & $7.77 \pm 0.41$ & $6.86 \pm 0.27$  \\
 $\xi_{0}^{\mathrm{TP}}$ &  & & $0.739$ & $0.739$ & & $0.739$ & $0.739 \pm 0.092$  \\
 $10^{2} A_{\mathrm{sky}, auto}$ &  $0.941 \pm 0.060$ & & & & $0.940 \pm 0.058$ & & $0.930 \pm 0.058$  \\
 $\sigma_{\mathrm{sky}, auto}$ &  $31.4 \pm 1.7$ & & & & $31.4 \pm 1.7$ & & $31.5 \pm 1.7$  \\
 $10^{2} A_{\mathrm{sky}, autoLyb}$ &  & $1.32 \pm 0.10$ & & & $1.332 \pm 0.095$ & & $1.323 \pm 0.094$  \\
 $\sigma_{\mathrm{sky}, autoLyb}$ &  & $34.2 \pm 2.4$ & & & $34.1 \pm 2.3$ & & $34.2 \pm 2.3$  \\
 
\noalign{\smallskip}
\hline \hline
\noalign{\smallskip}

 $N_{\mathrm{bin}}$ &  $1590$ & $1590$ & $3180$ & $3180$ & $3180$ & $6360$ & $9540$  \\
 $N_{\mathrm{param}}$ &  $13$ & $11$ & $10$ & $8$ & $15$ & $10$ & $19$  \\
 $\chi^{2}_{\min{}}$ &  $1604.79$ & $1583.61$ & $3238.96$ & $3193.29$ & $3190.88$ & $6436.80$ & $9654.56$  \\
 $probability$ &  $0.31$ & $0.46$ & $0.19$ & $0.39$ & $0.37$ & $0.22$ & $0.17$  \\

 \noalign{\smallskip}

 $\chi^{2}_{\min,auto}$ &  & & & & $1603.74$ & & $1608.52$  \\
 $\chi^{2}_{\min,cross}$ &  & & & & & $3235.31$ & $3258.41$  \\
 $\chi^{2}_{\min,autoLyb}$ &  & & & & $1584.48$ & & $1585.48$  \\
 $\chi^{2}_{\min,crossLyb}$ &  & & & & & $3196.39$ & $3197.02$  \\
 
 \noalign{\smallskip}
 
 $\Delta \chi^{2}(\alpha_{\parallel}=\alpha_{\perp}=1)$ &  $2.00\,(0.81\,\sigma)$ & $-$ & $3.94\,(1.33\,\sigma)$ & $-$ & $1.72\,(0.77\,\sigma)$ & $3.22\,(1.16\,\sigma)$ & $4.62\,(1.55\,\sigma)$  \\
 $\mathrm{free}\,A_{\mathrm{BAO}}$                     &  $1.37 \pm 0.25 $ & $1.30 \pm 0.41$ & $1.08 \pm 0.20$ & $1.25 \pm 0.58 $ & $1.32 \pm 0.22 $ & $0.92 \pm 0.19 $ & $1.10 \pm 0.15 $  \\
                      &  $(5.48\,\sigma)$ & $(3.17\,\sigma)$ & $(5.4\,\sigma)$ & $(2.16\,\sigma)$ & $(6.0\,\sigma)$ & $(4.84\,\sigma)$ & $(7.33\,\sigma)$  \\

    \end{tabular}
    }
\tablecomments{
    The first four columns give fit to each measurement independently,
    the last three gives the results when running combined fits.
    All parameters  are given at $z_{\mathrm{eff}} = \zeff{}$.
    The first two lines give the BAO parameters with error bars giving
    $68.27\%$ of trials as estimated with fastMC, as explained in the text.
    Error bars of other best fit parameters are given by
    minuit.
    Quantities without error bars are fixed in the fit.
    The second section gives the attributes of the best fit, along with
    the significance of the shift with respect to the \Planck~cosmology
    and the significance of the BAO peak.
    }
    \label{table::best_fit_parameters_data}

\end{table*}

The results of the seven different fits are given in
Table~\ref{table::best_fit_parameters_data}. The first four columns give
the results for each individual measured correlations.
The last three columns give combined fits:
``full auto'', ``full cross'', and ``all combined''
\footnote{\label{picca_fit} Best-fit~values~and~likelihoods~are~also~given~in\\
\href{https://github.com/igmhub/picca/blob/master/data/duMasdesBourbouxetal2020}{picca/tree/master/data/duMasdesBourbouxetal2020/fits}}.
In the first section, the first two lines give the BAO parameters $\alpha_{\parallel}$ and
$\alpha_{\perp}$, follows the different parameters that describe the overall
shape of the correlation functions.
The second section gives the attributes of the fit.
The last two lines give properties of two different models.
The second to last line is for
a model where the BAO parameters are fixed to the cosmology of
\Planck. The first number is the difference of $\chi^{2}$ and the
second number in parenthesis gives the conversion to the significance
in number of trials, as computed using Monte Carlo realizations
(Table~\ref{table::confidence}).
The last line gives the significance of the BAO peak.

  In Table~\ref{table::best_fit_parameters_data}, parameters
  without error bars are fixed in the fit.
  This is the case for the BAO parameters using only the \lyb~region
  where the BAO peak is not detected with high significance
  and we choose to fix $(\apar,\aperp)=(1,1)$.
  In fits for the cross-correlation we fix the values of
  some parameters to the values found in the combined
  fit.  For
  the HCD bias parameters, this is motivated by their
  high correlation with the quasar velocity distribution
  parameter $\sigma_v$.
  The parameters  $b_{\rm \eta,CIV(eff)}$, $\beta_{\rm QSO}$ and
  $\xi_0^{\rm TP}$ are fixed because they
  are not significantly constrained by the cross-only fits. 

We can notice that both widths, $\sigma_{\mathrm{sky}}$, of the model of
the correlated calibration residuals are of the same order of magnitude
as that found for the MgII(MgII(11)) $\times$ MgII(MgII(11)) auto-correlation
in Section~\ref{contribution_of_sky_residuals}. The strength, $A_{\mathrm{sky}}$,
is however one magnitude stronger, due to the difference in effective
redshift of the correlation functions between Ly$\alpha$ and MgII.

In another aspect we can note that the parameter describing the systematic
quasar redshift error,
$\Delta r_{\parallel, \mathrm{QSO}} = 0.31 \pm 0.11 \, \hMpc{}$,
is compatible with zero at the $3~\sigma$ level.
Using DR12 data, \duMasdesBourboux~measured this parameter to be
$-0.79 \pm 0.13 \, \hMpc{}$, which was $6~\sigma$ from zero.
This difference is linked
to the change in the quasar redshift estimator
(Section~\ref{subsection::quasar_redshifts}).
Other estimators give even larger values, as shown in
Table~\ref{table::drp_different_estimator}.
Though this suggests
that our new quasar redshift estimator is less biased, it is not possible to
draw any definitive conclusion. Indeed \Blomqvist~showed that this
parameter is highly correlated with models for
relativistic effects
\citep{2020JCAP...04..006L}.

\subsection{Measurement of the BAO parameters}
\label{section::BAO_measurement}

Combining the two measurements of the Ly$\alpha$ auto-correlation, yields:
\begin{equation}
    \left\{
    \begin{array}{ll}
        D_{H}(z=\zeff{})/r_{d} = 8.93 \;_{-0.27}^{+0.28} \;_{-0.63}^{+0.64} \\[4.pt]
        D_{M}(z=\zeff{})/r_{d} = 37.6 \;_{-1.9}^{+1.9} \;_{-4.0}^{+4.2} \\[4.pt]
        \rho \left( D_{H}(z)/r_{d},D_{M}(z)/r_{d} \right) = -0.49 \\
    \end{array}
    \right. \;.
    \label{equation::autoresults}
\end{equation}
Combining the two measurements of the Ly$\alpha$ $\times$ quasar cross-correlation yields:
\begin{equation}
    \left\{
    \begin{array}{ll}
        D_{H}(z=\zeff{})/r_{d} = 9.08 \;_{-0.34}^{+0.34} \;_{-0.60}^{+0.61} \\[4.pt]
        D_{M}(z=\zeff{})/r_{d} = 37.3 \;_{-1.6}^{+1.7} \;_{-3.6}^{+4.2} \\[4.pt]
        \rho \left( D_{H}(z)/r_{d},D_{M}(z)/r_{d} \right) = -0.43 \\
    \end{array}
    \right. \;.
    \label{equation::crossresults}
\end{equation}

Finally, combining the two auto-correlations along with the two cross-correlations yields:
\begin{equation}
    \left\{
    \begin{array}{ll}
        D_{H}(z=\zeff{})/r_{d} = 8.99 \;_{-0.19}^{+0.20} \;_{-0.38}^{+0.38} \\[4.pt]
        D_{M}(z=\zeff{})/r_{d} = 37.5 \;_{-1.1}^{+1.2} \;_{-2.3}^{+2.5} \\[4.pt]
        \rho \left( D_{H}(z)/r_{d},D_{M}(z)/r_{d} \right) = -0.45 \\
    \end{array}
    \right. \;.
    \label{equation::combinedresults}
\end{equation}

\begin{figure}
    \centering
    \includegraphics[width=0.98\columnwidth]{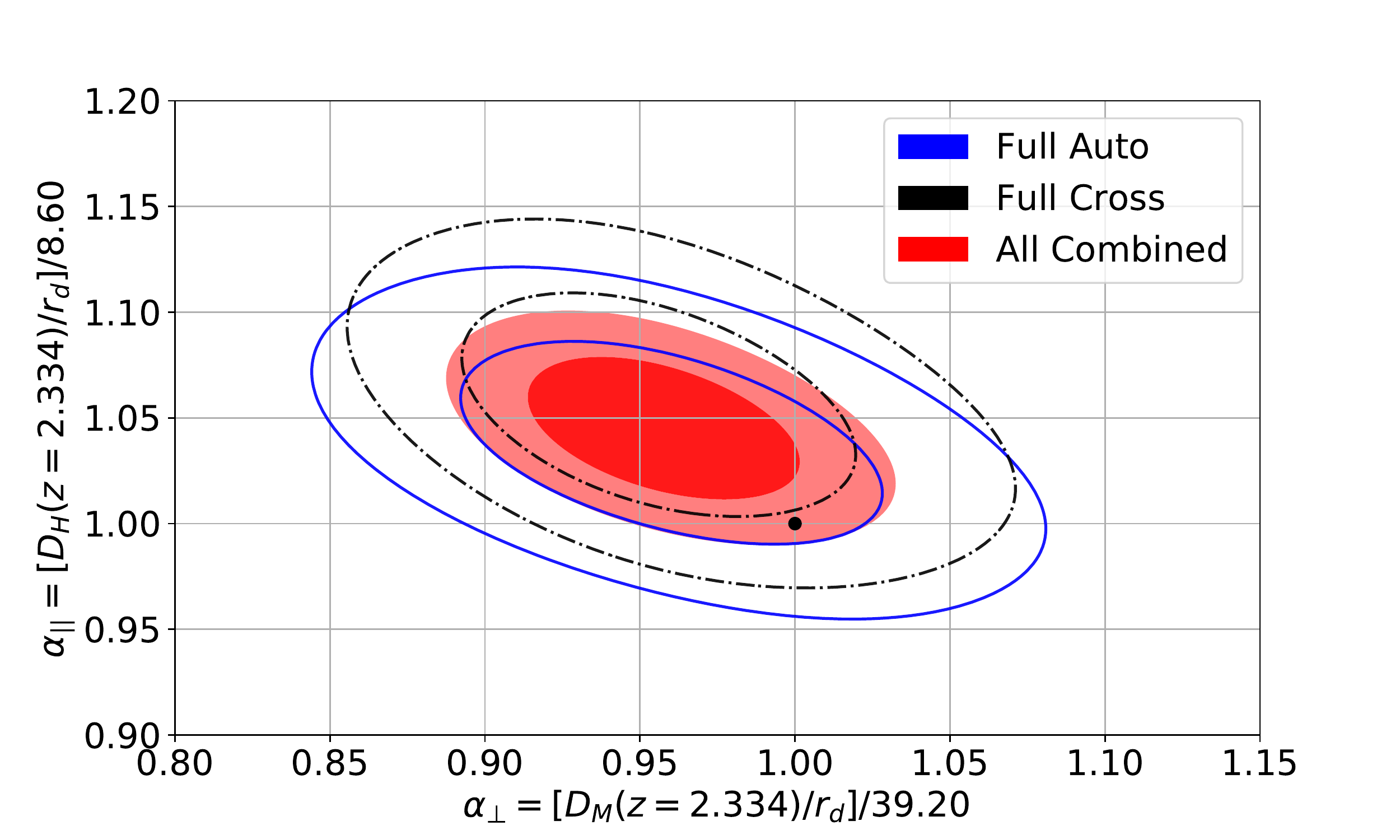}
    \caption{
      The 68\% and 95\% confidence level contours in the
      $(\apar,\aperp)$ plane from fits to the
      auto-correlation and cross-correlation functions and from the
      combined fit.
      The black dot shows the position of the flat-\lcdm~model of \Planck.
    }
    \label{figure::BAO_contours}
\end{figure}

These three results are presented in Figure~\ref{figure::BAO_contours}.
The black point gives the \Planck~cosmology, the contours
give the $1$ and $2$ confidence levels and corresponds to $(68.27, 95.45)\%$.

As first shown in  \duMasdesBourboux, the BAO parameters
$(\alpha_{\parallel}, \alpha_{\perp})$ have non-Gaussian uncertainties.
This
means that e.g. $\Delta \chi^{2} = 1$ does not correspond to $68\%$ of trials
for one degree of freedom.
To estimate the mapping between $(68.27, 95.45)\%$ confidence levels
to $\Delta \chi^2$ values,
  we generate $1000$ fast Monte-Carlo (fastMC) realizations
of our seven fits.
Each fit is reduced to the simple Kaiser model, in order to build enough
statistics given the available computation time. A complete model would
give similar results, as shown by \duMasdesBourboux.
In each realization, we generate a random noisy measurement of the
best fit model, given the covariance matrix, with BAO parameters
set to $1$. Then the fit is performed
in four different combinations:
for the BAO parameters free, another for both of them fixed,
then two more for one fixed and the other free.
This allows for each of the seven results to give the error bars
for $\alpha_{\parallel}$ and $\alpha_{\perp}$ independently.
Table~\ref{table::confidence}
gives the results of the estimation of the error bars.
The BAO error bars given in Table~\ref{table::best_fit_parameters_data}
are the results of this fastMC correction.

\begin{table}[tb]
\centering
\caption{
  Values of $\Delta\chi^{2}$ corresponding to $CL=(68.27,95.45\%)$.
  }
\label{table::confidence}
\begin{tabular}{l c c}
\hline
\hline
\noalign{\smallskip}
Parameter & $\Delta\chi^{2}$ $(68.27\%)$ & $\Delta\chi^{2}$ $(95.45\%)$ \\
\noalign{\smallskip}
\hline
\noalign{\smallskip}
\lyalyalyalya \\
 $\apar$ & $  1.06 \pm 0.06$ & $ 5.06 \pm 0.35 $\\
 $\aperp$ & $ 1.24 \pm 0.08 $& $ 4.51 \pm 0.25 $\\
 ($\apar$,$\aperp$) & $ 2.64 \pm 0.09 $& $ 7.07 \pm 0.33 $\\
\noalign{\smallskip}
\hline
\noalign{\smallskip}
\lyalyalyalyb \\
 $\apar$ & $ 1.25 \pm 0.09 $& $ 5.1 \pm 0.41 $\\
 $\aperp$ & $ 1.39 \pm 0.07 $& $ 5.03 \pm 0.26 $\\
 ($\apar$,$\aperp$) & $ 2.91 \pm 0.11 $& $ 7.9 \pm 0.4 $\\
\noalign{\smallskip}
\hline
\noalign{\smallskip}
\lyalyaq \\
 $\apar$ & $ 1.21 \pm 0.06 $& $ 4.4 \pm 0.22 $\\
 $\aperp$ & $ 1.22 \pm 0.08 $& $ 4.98 \pm 0.26 $\\
 ($\apar$,$\aperp$) & $ 2.59 \pm 0.12 $& $ 6.79 \pm 0.31 $\\
\noalign{\smallskip}
\hline
\noalign{\smallskip}
\lyalybq \\
 $\apar$ & $ 1.43 \pm 0.07 $& $ 4.5 \pm 0.18 $\\
 $\aperp$ & $ 1.41 \pm 0.07 $& $ 4.74 \pm 0.2 $\\
 ($\apar$,$\aperp$) & $ 2.93 \pm 0.13 $& $ 6.69 \pm 0.22 $\\
\noalign{\smallskip}
\hline
\noalign{\smallskip}
auto all \\
 $\apar$ & $ 1.06 \pm 0.07 $& $ 4.83 \pm 0.46 $\\
 $\aperp$ & $ 1.14 \pm 0.08 $& $ 4.86 \pm 0.29 $\\
 ($\apar$,$\aperp$) & $ 2.42 \pm 0.1 $& $ 7.16 \pm 0.33 $\\
\noalign{\smallskip}
\hline
\noalign{\smallskip}
cross all \\
 $\apar$ & $ 1.23 \pm 0.06 $& $ 4.3 \pm 0.19 $\\
 $\aperp$ & $ 1.11 \pm 0.09 $& $ 5.02 \pm 0.41 $\\
 ($\apar$,$\aperp$) & $ 2.64 \pm 0.11 $& $ 6.97 \pm 0.36 $\\
\noalign{\smallskip}
\hline
\noalign{\smallskip}
combined \\
 $\apar$ & $ 1.05 \pm 0.05 $& $ 4.07 \pm 0.37 $\\
 $\aperp$ & $ 1.05 \pm 0.08 $& $ 4.37 \pm 0.27 $\\
 ($\apar$,$\aperp$) & $ 2.4 \pm 0.07 $& $ 6.4 \pm 0.27 $\\
\noalign{\smallskip}
\hline
\end{tabular}
\tablecomments{
  Values are derived from 1000 Monte Carlo simulations of the correlation function that are fit using the model containing only Ly$\alpha$ absorption. Confidence levels are the fractions of the generated data sets that have best fits below the $\Delta\chi^{2}$ limit. The uncertainties are statistical and estimated using
  the bootstrap technique.}
\end{table}

The distribution of $\Delta \chi^{2}(\alpha_{\parallel}=\alpha_{\perp}=1)$
in the $1000$ fastMC allows us to compute the confidence levels in two
dimensions. Furthermore, it allows us to estimate how significant the shift of our best fit
measurement is against the \Planck~cosmology. This result is given
in Table~\ref{table::best_fit_parameters_data} in the second to last line.
Our final measurement, ``all combined'', is $1.5\,\sigma$ from the
cosmology of \Planck.

The significance of the BAO detection is estimated by leaving free the parameter
$A_{\mathrm{BAO}}$ that describes the size of the BAO peak relative to what
is expected according to the \Planck~cosmology.
We give the results for this extension to the baseline model in the last line
of Table~\ref{table::best_fit_parameters_data}.
Using fast Monte-Carlo realizations, we verify that this parameter is linear
in its mapping from $\Delta \chi^{2}$ to confidence levels.
Using $1000$ fast Monte-Carlo we verify this property up to $3\,\sigma$,
and then assume that it holds true for higher confidence levels.

Our measurement of the auto-correlation, \lyalyalyalya,
and cross-correlation \lyalyaq, have both
a BAO peak that has a significance of higher than $4\,\sigma$.
It is barely $3\,\sigma$ for the auto-correlation
\lyalyalyalyb, and barely
$2\,\sigma$ for the \lyalybq.
Because of this low significance of the BAO measurement in these two
extra correlations, the $2\,\sigma$ error bars are very non-linear,
with \lyalybq~ having none.
We thus choose not to give the BAO measurement for
these two correlations because it should
not be used alone, but only in combination with other measurements.
The final BAO measurement that combines all measured correlation functions
has a BAO significance of more than $7\,\sigma$.

Our measured $(\alpha_{\parallel}, \alpha_{\perp})$ are marginally
correlated to other parameters but significantly correlated,
$\rho\sim -50\%$,  with one another.
The only other parameter that is correlated with $(\apar,\aperp)$
is  the bias parameter of the SiII(126) metal line.
Indeed
the correlations of absorption by this metal and \lya~absorption
at the same physical position generates an apparent peak
in the reconstructed correlation function at
$(r_{\parallel}, r_{\perp}) \sim (+111,0) \, \hMpc{}$
(Table~\ref{table::metals_in_mocks}).
This correlation is  $\approx-22\%$ with $\alpha_{\parallel}$
and $+9\%$ with $\alpha_{\perp}$ in the case of the ``full auto''.
Since in the cross-correlations we have access to positive and negative
distances along the line-of-sight, this correlation is reduced to 
$-14\%$ with $\alpha_{\parallel}$ and $+5\%$ with $\alpha_{\perp}$
for the ``full cross''.
Combining all results give a correlation of
$-19\%$ with $\alpha_{\parallel}$ and $+8\%$ with $\alpha_{\perp}$
for the ``all combined'' fit.
All other parameters are less than $\pm 3\%$ correlated with the BAO
parameters.

In Appendix~\ref{section::Systematic_tests_on_BAO}, we test for systematic
errors in our measurement of the BAO parameters, along with changes in the
estimation of their error bars. We either modify the best fit model
in Table~\ref{table::best_fit_BAO_data_different_model} or change
some aspects of the analysis or study data splits in
Table~\ref{table::best_fit_BAO_data_systematic_tests}.
In Table~\ref{table::best_fit_BAO_data_different_model} the strongest effect
on both the recovered BAO best fit value and error bars happens when
taking the effect of metals into account
  which results in a $0.5\sigma$ shift in $\apar$.
As previously discussed,
the SiII(126) absorption line produces a scale along the line-of-sight
of $+111\,\hMpc{}$. The change in the error bars is explained by the effect
of preventing the best fit model to build false BAO significance
from this line.

Of the fits in
Table~\ref{table::best_fit_BAO_data_different_model}, those
that add polynomial ``broadband'' curves to the model are of special
importance because they  test the sensitivity of the BAO parameters to
unidentified systematic errors in the model.
We performed two fits, allowing the broadband curves different
amounts of freedom in each.
The first placed
``physical priors'' on $(\blya,\betalya,\bhcd)$ in the form of
a Gaussian of mean and width of the fit without broadband terms.
Such priors ensured that the broadband terms were relatively small
perturbations to the physical model.
The second type of fit placed no priors on $(\blya,\betalya,\bhcd)$.
The results of these fits are given in
Table~\ref{table::best_fit_BAO_data_different_model}.
We see that the addition of such terms does not change significantly
the values of $(\apar,\aperp)$ in any of the fits.

%
%

\section{Comparison with previous \lya\ BAO studies}
\label{section::Comparison}

The final \lya\ BAO analyses from BOSS, using data from DR12, were presented
in \citet{2017A&A...603A..12B} (\Bautista) and \citet{2017A&A...608A.130D} 
(\duMasdesBourboux).
When combined, the measurement of the BAO scale was in mild 
($\approx 2.3\sigma$) tension with the predictions from the best
fit \lcdm~cosmology from the {\it Planck} satellite (\duMasdesBourboux).
In the present study, combining all data from BOSS and eBOSS, the tension
has now reduced to only $\approx 1.5\sigma$.

\begin{figure}
    \centering
    \includegraphics[width=0.98\columnwidth]{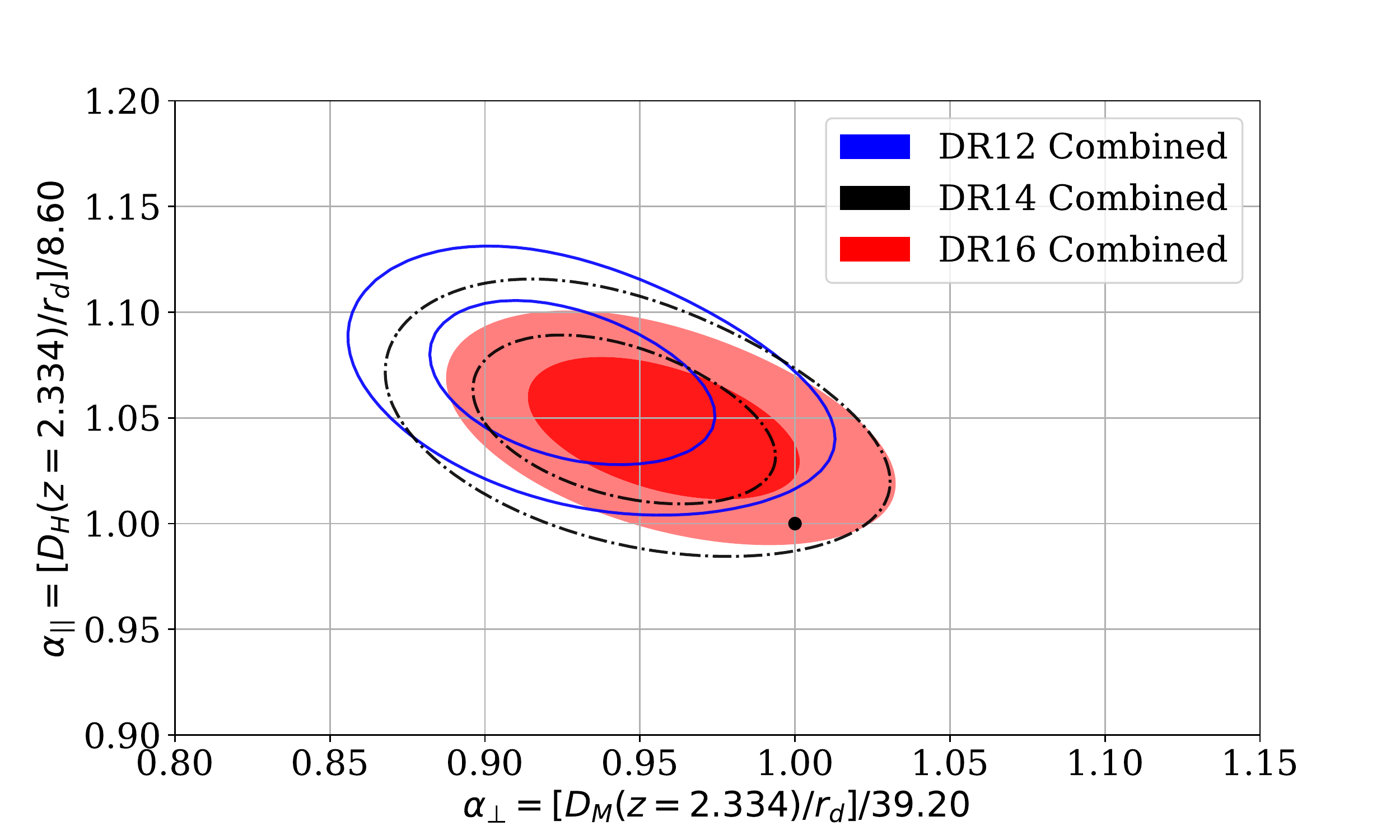}
    \caption{
      The 68\% and 95\% confidence level contours for the BAO parameters
      $(\apar,\aperp)$ for the combined \lya\ results from different
      data releases of SDSS: DR12 (\Bautista, \duMasdesBourboux, in blue),
      DR14 (\deSainteAgathe, \Blomqvist, in black) and DR16 (this work, in red).
      The black dot shows the position of the flat-\lcdm~model of \Planck.
    }
    \label{figure::history_lya_bao}
\end{figure}

\begin{figure}
    \centering
    \includegraphics[width=0.98\columnwidth]{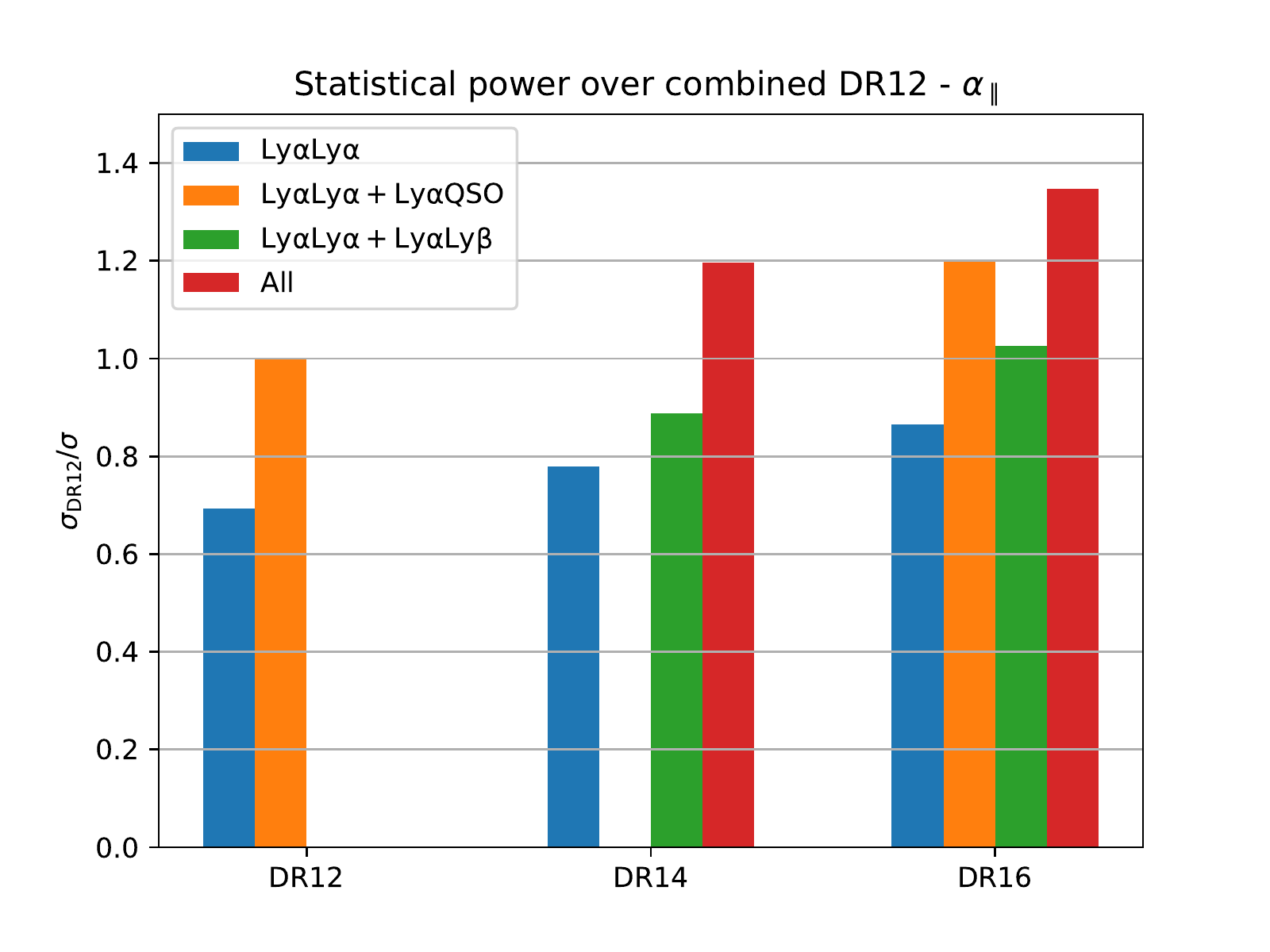}
    \includegraphics[width=0.98\columnwidth]{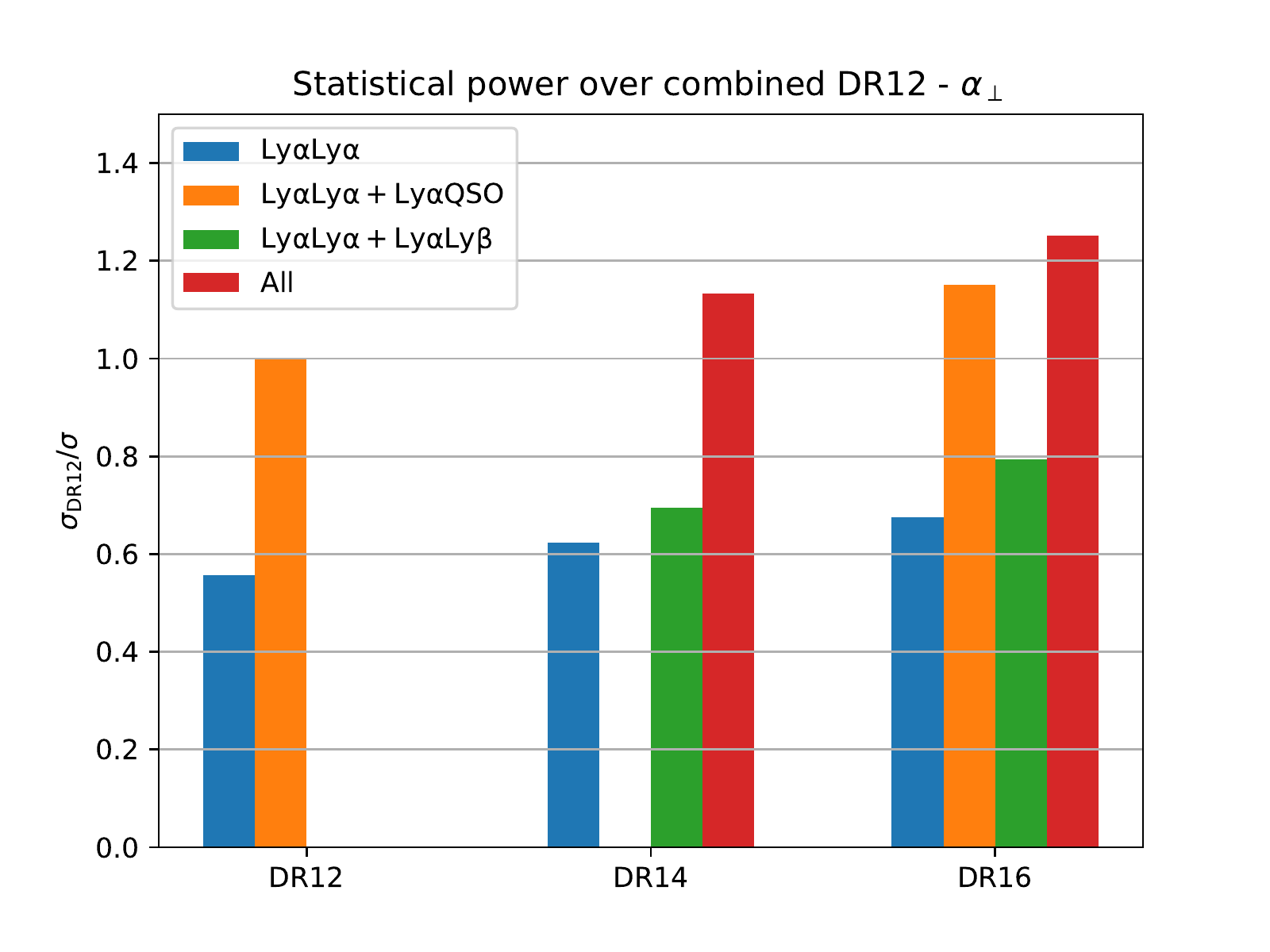}
    \caption{
      Gain in statistical power on BAO parameters over that in DR12,
      for $\apar$ (top) and $\aperp$ (bottom).
      The blue bars show the constraining power of the \lya\ auto-correlation
      in the \lya\ region; 
      orange bars add the cross-correlation with quasars, still using the 
      \lya\ region only (this result was not published for DR14); 
      green bars show the auto-correlation using both the \lya\ and the \lyb\
      region (not computed in DR12);
      red bars show the combination of auto-correlation and cross-correlation
      with quasars, using both the \lya\ and the \lyb\ regions 
      (not computed in DR12).
    }
    \label{figure::history_stat_power}
\end{figure}

In Figure \ref{figure::history_lya_bao} we show the BAO contours for the
combined \lya\ BAO studies from different SDSS data releases:
DR12 (\Bautista, \duMasdesBourboux), DR14 (\deSainteAgathe, \Blomqvist)
and DR16 (this work).
The contours have shifted towards the fiducial cosmology 
($\apar=1$, $\aperp=1$), and the area of the contours have shrunk
by roughly 25\% between DR12 and DR16.
It is important to remember, though, that the BAO uncertainty is itself a
random variable, and it varies significantly from realization to realization.

  In this section, we present first a comparison of the expected statistical
  power of the DR12 and DR16 datasets.
  We then discuss possible origins for the evolution
  of the BAO results from DR12 to DR16.
  The possibility that changes in the analysis pipeline induced
  significant changes is studied and rejected.
  We then investigate the effects of statistical changes, both
  in the enlarged data set and in the changed catalog of quasar
  redshifts.
  We will show that these two purely statistical effects are sufficient
  to explain the DR12 to DR16 evolution.


We would like to have a comparison of the statistical power of the different
datasets that does not suffer from this randomness. 
In order to do so, we generate Monte Carlo
measurements of the different correlations
that have the same covariance matrix as the real data, but where the measured
correlations have been set to their theoretical prediction obtained using the
combined best fit presented in the last column of 
Table \ref{table::best_fit_parameters_data}.
These mock correlations are then fitted using the same configuration as
our main analyses, resulting in a fit that is clearly too good ($\chi^2=0$),
but with parameter uncertainties that should be equivalent to those obtained
in a Fisher forecast.

In Figure \ref{figure::history_stat_power} we show this gain in statistical
power with respect to DR12 combined results, quantified as the ratio of
uncertainties expected for the BAO parameters $\apar$ (top) and $\aperp$
(bottom).
From this plot we can read that the statistical power in the \lya\ 
auto-correlation (blue) of DR16 is 20-25\% larger than that in DR12 for both
BAO parameters.
These numbers are consistent with the average ratio of the errorbars
in the measured correlations.
They are also comparable to the 27-30\% gain forecasted for eBOSS in
\citet{2016AJ....151...44D}, where we have used their forecasted values for a
4 years survey with an area of 4500 squared degrees, similar to the final
eBOSS area.
The gain is a bit smaller for the combination of the auto- and
cross-correlation (orange), where the gain is roughly 15-20\% from DR12 to DR16.
When we also include measurements in the \lyb\ region, the gain in statistical
power increases to 25-30\%.
This gain corresponds to the average gain that we would see in a large ensemble
of survey realizations, but each realization will have a different value. 
As seen in Figure \ref{figure::history_lya_bao}, the statistical gain between
DR12 and DR16 in the actual realization observed is only 12 \%.



\begin{table*}
    \caption{
      Comparison of \lya\ BAO results using DR12. }
    \centering

    \begin{tabular}{lllll}

    DR12 analysis
    &       $\alpha_{\parallel}$
    &       $\alpha_{\perp}$
    &       $\chi^{2}_{\mathrm{\min{}}}/DOF, \, proba$
    \\
    \noalign{\smallskip} \hline \hline \noalign{\smallskip}
    Ly$\alpha$(Ly$\alpha$) $\times$ Ly$\alpha$(Ly$\alpha$): & & & \\
    \noalign{\smallskip}

    Bautista et al. (2017) &  $1.053 \pm 0.036$ & $0.965 \pm 0.055$ & $1556.5/(1590-13), p=0.639$  \\
    Re-analysis (this work) & $1.055 \pm 0.036$ & $0.987 \pm 0.051$ & $1582.59/(1590-13), p=0.456$ \\

    \noalign{\smallskip} \hline \hline \noalign{\smallskip}
    Ly$\alpha$(Ly$\alpha$) $\times$ quasar: & & & \\ 
    \noalign{\smallskip}

    du Mas des Bourboux et al. (2017) & $1.077 \pm 0.038$ & $0.898 \pm 0.038$ & $2576.3/(2504-15), p=0.11$ \\
    Re-analysis (this work, $10<r<160~\hMpc$ ) & $1.080 \pm 0.037$ & $0.896 \pm 0.036$ & $2544.58/(2504-15), p=0.214$ \\
    Re-analysis (this work, $10<r<180~\hMpc$ ) & $1.078 \pm 0.037$ & $0.893 \pm 0.035$ & $3292.87/(3180-10), p=0.063$ \\

    \noalign{\smallskip} \hline \hline \noalign{\smallskip}
    all\,combined: & & & \\
    \noalign{\smallskip}

    du Mas des Bourboux et al. (2017) & $1.069 \pm 0.027$ & $0.920 \pm 0.031$ & $3833.16/(3756-17), p=0.14$  \\
    Re-analysis (this work, $10<r<160~\hMpc$) & $1.074 \pm 0.026$ & $0.926 \pm 0.030$ & $3810.71/(3756-17), p=0.203$ \\
    Re-analysis (this work, $10<r<180~\hMpc$) & $1.066 \pm 0.026$ & $0.928 \pm 0.031$ & $4883.70/(4770-17), p=0.091$ \\

    \end{tabular}
\tablecomments{
    The results published in \Bautista\ and \duMasdesBourboux\ are compared
      to a re-analysis of the measurement using the data analysis pipeline used
      in this work.
      Top section: results for the \lya\ auto-correlation; 
      middle section: results for the cross-correlation;
      bottom section: combined results.
      Note that all analyses exclude the \lyb\ region, as done in the DR12
      publications.
      \duMasdesBourboux\ used a shorter range of separations in the fits 
      ($10< r < 160~\hMpc$), resulting in a different number of degrees
      of freedom.
      We show a reanalysis for both ranges of separations with the value
      of $DOF$ in the last column specifying the range.
      BAO errors correspond to $\Delta \chi^2 = 1$.
    }
    \label{table::redo_dr12}
\end{table*}

  We now investigate the possibility that changes in the analysis
  pipeline could explain the evolution of the BAO results.
The DR12 and DR16 results presented in Figure \ref{figure::history_lya_bao}
used slightly different software to measure the correlations, and different
models to fit BAO. 
In order to address whether the shift is caused by the addition of new data,
or by changes in the analysis pipeline, in Table \ref{table::redo_dr12} we
compare the DR12 results published with our re-analysis of the DR12 data set,
using the current analysis software \picca, and the modeling described in the
sections above.
In order to better compare the analyses, we have used the exact DR12 quasar
catalog and  the $Z_{\rm VI}$ redshifts that were used in the original
DR12 analyses, as well as the DR12 DLA catalog.
Similarly, we do not include \lya\ information from the \lyb\ region, as was
done in the DR12 analyses.

Even though there are many differences in the analysis pipeline, the BAO
results are in very good agreement, highlighting again the robustness of these
measurements with respect to choices in the modeling (see Table 
\ref{table::best_fit_BAO_data_systematic_tests}).

  We now turn to the question of whether the statistical changes
  due to the addition of new quasars and forests
  can explain the DR12 to DR16 evolution.
We estimated the expected changes in the BAO parameters using the
fastMC technique described in Sect. \ref{section::BAO_measurement}.
We first 
created 100 DR12 mock correlation functions, $\xi_{12}$,
Gaussian distributed about the best-fit DR16 model
according to the observed DR12 covariance matrix, $C_{12}$.
For each  DR12 correlation function,
we created 100 correlation functions for eBOSS-only data 
using the noise-dominated covariance
$C_{16-12} = (C_{16}^{-1} - C_{12}^{-1})^{-1}$, where $C_{16}$ is the observed
DR16 covariance matrix,.
Mock DR16 correlation functions were then created by adding the
DR12 and e-BOSS-only functions:
$ \xi_{16}= C_{16}(C_{12}^{-1}\xi_{12} + C_{16-12}^{-1}\; \xi_{16-12})$.
The DR12-DR16 mock pairs were then fit for the BAO  parameters.
Approximately 30\% of the mock pairs had changes in the BAO parameters
that were greater than that observed for the combined fit in the data.
We can then conclude that the observed changes in BAO parameters are
consistent with those expected from statistical fluctuations.

  A second source of statistical differences between DR12 and DR16
  is our change in quasar redshift estimator from \zvi~
to \zlyawg, made necessary by the lack of inspection for all
DR16 quasars.
This leads to random migration of quasar-pixel pairs in $\rpar$ space
due to random differences between the two estimators.
The statistical effect of this on the BAO parameters for the
cross-correlation is studied in Appendix~\ref{section::quasar_redshifts}
with the conclusion that it has a non-negligible impact on
DR12-DR16 evolution of BAO parameters for the cross-correlation.

%
%

\section{Summary and conclusions}
\label{section::Summary_and_conclusions}

This paper presents the
measurement of baryonic acoustic oscillations
with Lyman-$\alpha$ absorption and quasars from
BOSS and eBOSS.
It benefits from ten years of SDSS observations,
from 2009 to 2019.
Since the first measurement of the
large scale 3D Ly$\alpha$ auto-correlation and 3D Ly$\alpha$-quasar
cross-correlation,
many improvements
have been made in the analysis:

\begin{itemize}

    \item \citet{2011JCAP...09..001S} measured for the first time the 3D large scale
    auto-correlation of Ly$\alpha$, from $\sim \num{10000}$ spectra,
    from the first few months of BOSS data.
    The auto-correlation was measured in the $(r,\mu)$
    plane up to separations $100 \, \hMpc{}$.
    The analysis also presented a set of mocks \citep{2012JCAP...01..001F} allowing
    tests of the measurement with the different astrophysical and
    observational effects: Poisson noise, quasar continuum, high absorption
    systems, metals.
    The first discussion of distortion and of the Wick expansion were presented.

    \item \citet{2013JCAP...05..018F} measured for the first time the 3D large scale
      cross-correlation of Ly$\alpha$ and quasars using the techniques
      introduced by \citet{2012JCAP...11..059F} applied to
       $\sim \num{60000}$ DR9 spectra.
    The cross-correlation was measured in the $(r_{\parallel},r_{\perp})$
    plane up to separations $80 \, \hMpc{}$.
    The analysis estimated the bias of quasars to agree with other studies.

  \item The DR9 auto-correlation measurement
    \citep{2013A&A...552A..96B, 2013JCAP...04..026S, 2013JCAP...03..024K}
    made the first detection and measurement of the BAO scale in
    the Ly$\alpha$ auto-correlation with DR9 and $\sim \num{50000}$ spectra.
    Two different studies measured this correlation either in the
    Cartesian coordinates $(r,\mu)$ or in the observable coordinates
    $(\Delta \log \lambda, \theta, z)$.
    Without any models for the distortion of the correlation, from the
    fit of the quasar continuum, the correlation was fit using broad-band
    functions in addition to the Kaiser model.

  \item The DR11 quasar-cross-correlation measurement \citep{2014JCAP...05..027F}
    was the first to observe the BAO peak in cross-correlation.
    \citet{2015A&A...574A..59D} updated the auto-correlation measurement
    and  combined constraints on BAO parameters were given.

  \item The DR12 auto-correlation (\Bautista) and
    cross-correlation (\duMasdesBourboux) measurements
    used the complete BOSS data set.
    They performed the important breakthrough
    of modeling the effect of the distortion of the correlation from the
    fit of the quasar continuum, allowing for the first physical
    fit of the 3D correlation.
    Broadband functions were kept
    to test for systematics in the measurement of BAO but no longer
    played a main role in the analysis.
    In addition to improvements in the data analysis, a main effort
    was dedicated in improving the realism of the mock spectra \citep{2015JCAP...05..060B}
    for metals and HCDs.
    The first mocks with realistic quasar-forest correlations were
    produced \citep{2011A&A...534A.135L}.
    The first combined fit between the auto and
    the cross-correlation was performed, breaking parameter degeneracies.
    Covariances were better understood through the study of Wick expansions.

  \item The DR14 auto-correlation (\deSainteAgathe) and
    cross-correlation (\Blomqvist) measurements
    were the first to use eBOSS data.
    Development were made to benefit from Ly$\alpha$
    absorption blueward of the Ly$\beta$+OVI emission line
    in $\sim \num{60000}$ spectra.
    The analysis used all observations of the same quasar, instead of
    the best one as was done in previous analyses.
    Furthermore, the analysis investigated different models for the
    effect of HCDs onto the auto-correlation function.
    In addition, the analysis developed
    a new way of splitting the sample in order to measure the auto-correlation
    in two redshift bins, while limiting the cross-covariance to sub percent levels.

    \item The DR16 study presented here is the first to use
    the complete BOSS and eBOSS datasets. It uses $\sim \num{210000}$ spectra for the
    Ly$\alpha$ spectral region and $\num{70000}$ for the Ly$\beta$ region.
    The analysis gives BAO for both the auto and cross-correlation.
    This study focus on:
    improving the understanding of the effects of the calibration from sky and
    standard stars fibers, expanding the Ly$\beta$ spectral region,
    better understanding the effects of the quasar redshift estimator,
    and continue the development of realistic 3D Gaussian random field
    mocks
    \citep{2020JCAP...03..068F, SaclayMocks}.

\end{itemize}

With the beginning of the Dark Energy Spectroscopic Instrument
\citep{2016arXiv161100036D} and WEAVE \citep{2016sf2a.conf..259P} projects,
the measurement of BAO
with Ly$\alpha$ forest will continue.
In order to profit from the improved statistical precision,
it will be useful to continue to improve various parts of the analysis, among
which are:
better understanding of quasar redshift estimators
(Sect.~\ref{subsection::quasar_redshifts}), and 
improved mock spectra through improved modeling 
of quasar spectral diversity and  absorption by metals
(Sect.~\ref{section::Validation_of_the_analysis_with_mocks}).
Improved modeling of the correlation function is essential in light
of the fact that
including additional ad hoc polynomial broadband terms improves the fit,
though without affecting
the BAO parameters.
Improvements here will  allow us to profit fully from the full shape
of the correlation function and yield a reliable measurement of
the growth rate of cosmological structure \citep{dumasdesbourboux:tel-01587743}.

The measurement of BAO at redshift $z\approx2.4$ has had an important
impact on cosmology.
While these data do not significantly constrain simple models
when CMB \citep{2016A&A...594A..13P}
and galaxy BAO data \citep{2017MNRAS.470.2617A}
are already used, the \lya~BAO data are
essential for providing a measurement of \lcdm~parameters using
only low-redshift data \citep{2015PhRvD..92l3516A}.
They thus provide an independent check on the CMB-inspired flat \lcdm~model.
The cosmological constraints from the DR16 data presented here
are given in a companion paper \citep{eBOSS_Cosmology}.

%
%

\begin{acknowledgements}

  \textit{Acknowledgements.}
It is a pleasure to thank Nicolas Busca for 
spearheading this analysis over an 8-year period, during which he
energetically provided creative solutions to innumerable issues.

The work of H\'elion du Mas des Bourboux, and Kyle Dawson was supported in
part by U.S. Department of Energy, Office of Science,
Office of High Energy Physics, under Award Number DESC0009959.
Andreu Font-Ribera acknowledges support by an STFC Ernest Rutherford
Fellowship, grant reference ST/N003853/1,
and  by FSE funds trough the program Ramon y Cajal (RYC-2018-025210) of
the Spanish Ministry of Science and Innovation.

This work was supported by the A*MIDEX project (ANR-11-IDEX-0001-02)
funded by the “Investissements d'Avenir” French Government program,
managed by the French National Research Agency (ANR),
and by ANR under contracts ANR-14-ACHN-0021 and ANR-16-CE31–0021.

Funding for the Sloan Digital Sky Survey IV has been provided by the Alfred P. Sloan Foundation, the U.S. Department of Energy Office of Science, and the Participating Institutions. SDSS acknowledges support and resources from the Center for High-Performance Computing at the University of Utah. The SDSS web site is www.sdss.org.
SDSS is managed by the Astrophysical Research Consortium for the Participating Institutions of the SDSS Collaboration including the Brazilian Participation Group, the Carnegie Institution for Science, Carnegie Mellon University, the Chilean Participation Group, the French Participation Group, Harvard-Smithsonian Center for Astrophysics, Instituto de Astrofísica de Canarias, The Johns Hopkins University, Kavli Institute for the Physics and Mathematics of the Universe (IPMU) / University of Tokyo, the Korean Participation Group, Lawrence Berkeley National Laboratory, Leibniz Institut f\"ur Astrophysik Potsdam (AIP), Max-Planck-Institut f\"ur Astronomie (MPIA Heidelberg), Max-Planck-Institut f\"ur Astrophysik (MPA Garching), Max-Planck-Institut f\"ur Extraterrestrische Physik (MPE), National Astronomical Observatories of China, New Mexico State University, New York University, University of Notre Dame, Observatório Nacional / MCTI, The Ohio State University, Pennsylvania State University, Shanghai Astronomical Observatory, United Kingdom Participation Group, Universidad Nacional Aut\'onoma de M\'exico, University of Arizona, University of Colorado Boulder, University of Oxford, University of Portsmouth, University of Utah, University of Virginia, University of Washington, University of Wisconsin, Vanderbilt University, and Yale University.

In addition, this research relied on resources provided to the eBOSS
Collaboration by the National Energy Research Scientific Computing
Center (NERSC).  NERSC is a U.S. Department of Energy Office of Science
User Facility operated under Contract No. DE-AC02-05CH11231.

\end{acknowledgements}


\bibliographystyle{aa}

\bibliography{lyaDR16}

\begin{appendix}

\section{Effect of the assumed cosmology}
\label{section::physical_model_to_the_broadband_of_the_correlation_functions}

The BAO parameters $(\alpha_{\parallel}, \alpha_{\perp})$ are given in
Table~\ref{table::best_fit_parameters_data} with respect to the assumed
cosmology of \Planck. For this reason they depend on the assumed
cosmology. However, as given in equation~\ref{equation::BAO_definition},
the parameters $(D_{H}(z)/r_{d}, D_{M}(z)/r_{d})$ are independent
of this assumption,
as long as the real cosmology is not ``too far'' from the \Planck~cosmology.
Here,
we demonstrate this property for the auto and for the cross-correlation
by replacing the fiducial cosmology of Table~\ref{table::cosmology_parameters}
with another model that yields $(\apar,\aperp)=(1,1)$.
This test has been performed on mock galaxy catalogs
\citep{2020MNRAS.494.2076C}, but this is the first test
using \lya~correlations.

We modify the \Planck~model by changing the values of $(h,\Omega_k)$
while maintaining the values of $(\Omega_ch^2,\Omega_bh^2$)
which are precisely determined by the CMB anisotropies,
independent of the curvature and dark-energy model.
This maintains the values of $r_d$ but changes the functions
$\DM(z)$ and $\DHub(z)$.
For a given $(\alpha_{\parallel}, \alpha_{\perp})$  found using the
\Planck~model
(Table~\ref{table::best_fit_parameters_data})
we can predict the values of
$(h, \Omega_{k})$ that allow us to find 
$(\alpha_{\parallel}, \alpha_{\perp}) = (1,1)$.
We do this separately for  the auto-
and cross-correlations.
As shown in Table~\ref{table::old_vs_new_cosmology},
with the new cosmologies we recover the same BAO parameters
$(D_{H}(z)/r_{d}, D_{M}(z)/r_{d})$.

\begin{figure*}
    \centering
    \includegraphics[width=0.48\columnwidth]{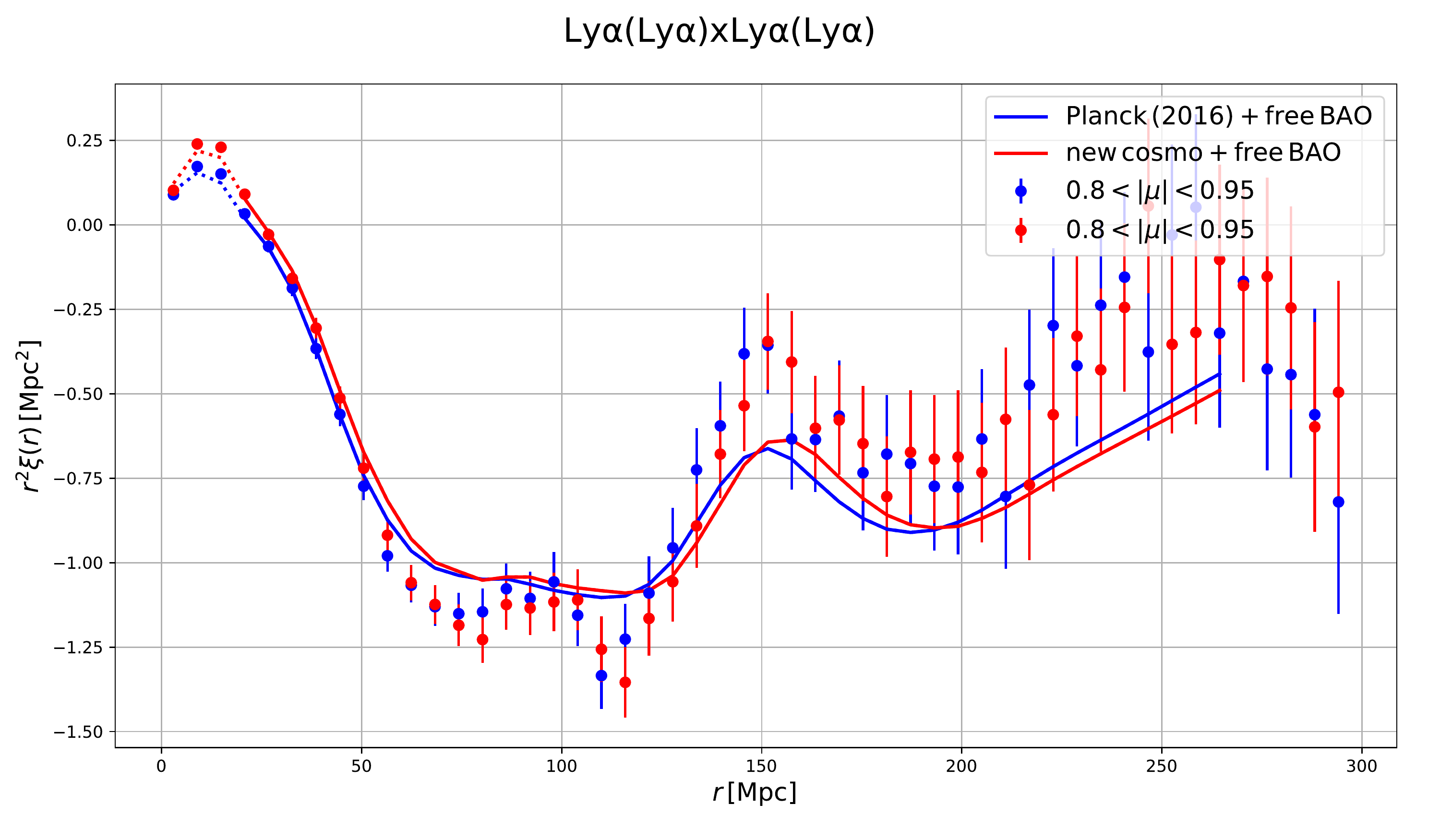}
    \includegraphics[width=0.48\columnwidth]{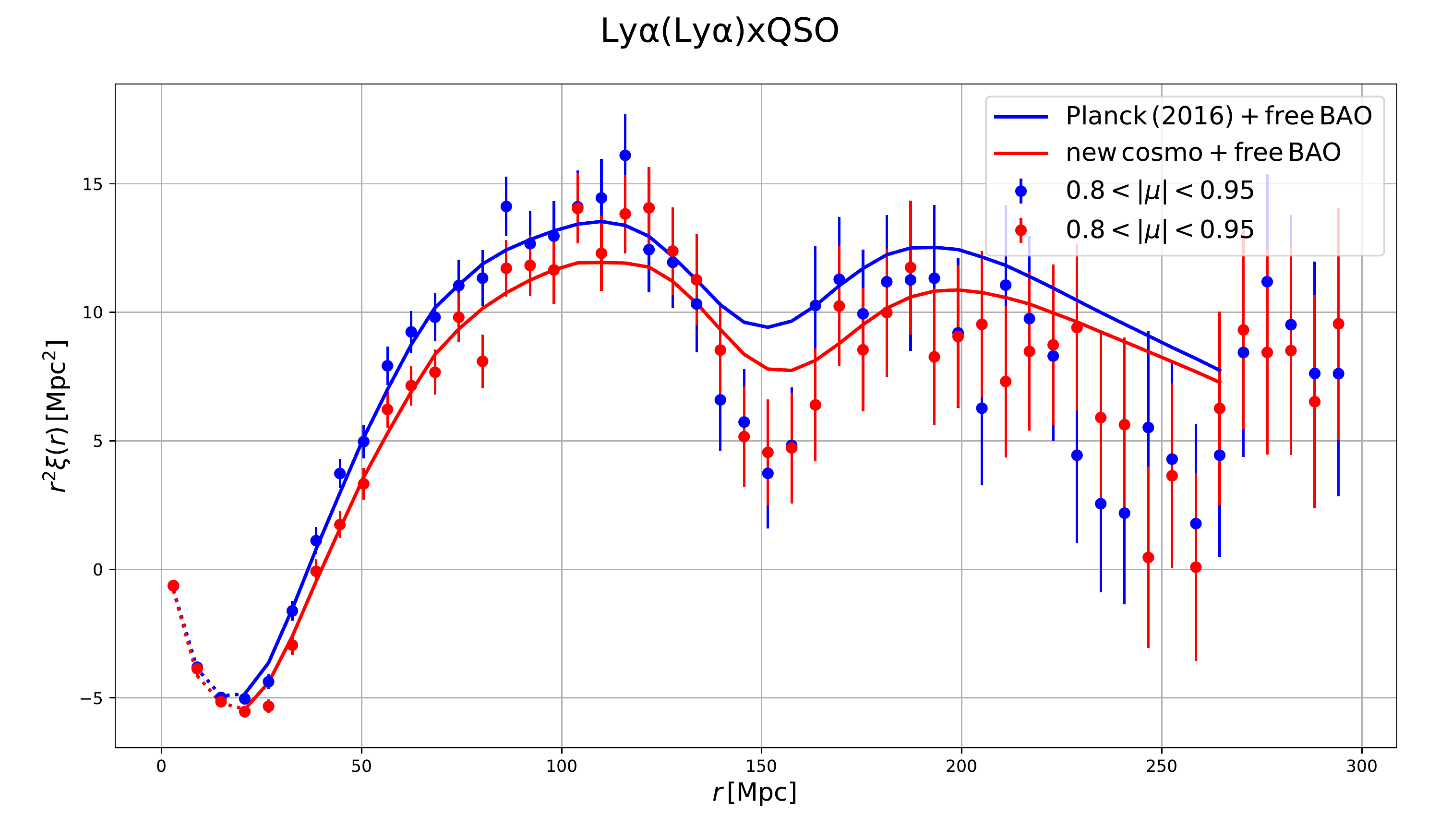}
    \caption{
    Auto-correlation Ly$\alpha$(Ly$\alpha$) $\times$
    Ly$\alpha$(Ly$\alpha$) (left) and cross-correlation
    Ly$\alpha$(Ly$\alpha$) $\times$ quasars (right) for the wedge
    $0.8 < |\mu| < 0.95$ as a function of the separation in $\mathrm{Mpc}$.
    The blue points and curves show these correlations assuming \Planck,
    Table~\ref{table::cosmology_parameters},
    while the red points and curves show them assuming the cosmology
    that measures $\alpha_{\parallel} = \alpha_{\perp} = 1$ for the auto-correlation
    on the left and the different cosmology that measures $\alpha_{\parallel} = \alpha_{\perp} = 1$
    for the cross-correlation on the right.
    }
    \label{figure::1_wedge_compare_new_cosmo_auto_cross}
\end{figure*}

The two 3D correlations, auto Ly$\alpha$(Ly$\alpha$) $\times$ Ly$\alpha$(Ly$\alpha$)
and cross Ly$\alpha$(Ly$\alpha$) $\times$ quasars are treated
independently in this section.
Using the \Planck~cosmology of Table~\ref{table::cosmology_parameters},
we compute $(D_{H}(z=\zeffshort)/r_{d}, D_{M}(z=\zeffshort)/r_{d})$
for different values of $(h, \Omega_{k})$. We find
that $(h,\Omega_{k}) = (0.70743,-0.09837)$ and
$(h,\Omega_{k}) = (0.80829,-0.12035)$ should reproduce the
auto and cross BAO results
(eqns.~\ref{equation::autoresults} and \ref{equation::crossresults}) but
with $(\alpha_{\parallel},\alpha_{\perp}) = (1,1)$.
Both 3D correlation, with their distortion matrix and their metal matrix
are computed using the updated new cosmological parameters
$(\Omega_{m}, \Omega_{r}, \Omega_{k})$ to map the observable coordinates
$(\Delta \lambda, \Delta \theta)$ to their respective Cartesian coordinates
$(r_{\parallel}, r_{\perp})$.
In order to fit these new correlations, we compute their respective matter
power-spectrum at $z=\zeffshort$ with CAMB, then using \picca~and
\texttt{nbodykit}, we extract the wiggle-less
power-spectrum of the side-band (sec.~\ref{section::Model_of_the_correlations}).
The resulting measured and best-fit correlation function is shown
for the wedge near the line-of-sight in Figure~\ref{figure::1_wedge_compare_new_cosmo_auto_cross}.
For these different cosmologies, the BAO scale, $r_{d}$, is unchanged
at better than $0.01$\%,
however the assumed values for the Hubble parameter are very different,
we thus show the separation on the x-axis in $\mathrm{Mpc}$ as opposed to $h^{-1}\,\mathrm{Mpc}$
as was done in Figure~\ref{figure::auto_4_wedges__cross_4_wedges}.
This allows us to present the BAO scale at nearly the
same position for all the different assumed
cosmologies.

Table~\ref{table::old_vs_new_cosmology} summarizes the best-fit BAO results.
As expected, the two customized cosmologies for the auto and
for the cross-correlation give $(\alpha_{\parallel},\alpha_{\perp}) = (1,1)$.
We also observe that the $\alpha_{i}$ parameters are dependent on the
assumed cosmology.
On the other hand, the BAO parameters
$(D_{H}(z=\zeffshort)/r_{d}, D_{M}(z=\zeffshort)/r_{d})$
are measured to be the same regardless of the assumed cosmology, furthermore
they are measured with the same error bars.

\begin{table*}
    \caption{
    BAO parameters for different assumed cosmologies}
    \centering
    
    \begin{tabular}{llllll}
    
    Cosmology
    &       $\apar$ 
    &       $\aperp$
    &       $D_{H}(z=\zeffshort)/r_{d}$ 
    &       $D_{M}(z=\zeffshort)/r_{d}$
    &       $\chi^{2}_{\mathrm{\min{}}}/DOF, \, proba$
    \\
    \noalign{\smallskip}
    \hline \hline
    \noalign{\smallskip}

  Ly$\alpha$(Ly$\alpha$) $\times$ Ly$\alpha$(Ly$\alpha$): &  &  & \\ 
  \noalign{\smallskip}

 $\mathrm{\,\,Planck~(2016,\,baseline)}$ &  $1.043 \pm 0.035$ & $0.986 \pm 0.051$ & $8.97 \pm 0.30$ & $38.6 \pm 2.0$ & $1599.55/(1590-14), p=0.33$   \\
 $\mathrm{\,\,new\,cosmo.\,auto}$        &  $0.997 \pm 0.035$ & $1.003 \pm 0.053$ & $8.94 \pm 0.32$ & $38.7 \pm 2.1$ & $1631.33/(1590-14), p=0.16$  \\

 \noalign{\smallskip}
 Ly$\alpha$(Ly$\alpha$) $\times$ quasar: &  &  & \\ 
 \noalign{\smallskip}

 $\mathrm{\,\,Planck~(2016,\,baseline)}$ &  $1.058 \pm 0.042$ & $0.929 \pm 0.053$ & $9.10 \pm 0.36$ & $36.4 \pm 2.1$ & $3232.82/(3180-10), p=0.21$  \\
 $\mathrm{\,\,new\,cosmo.\,cross}$        &  $1.002 \pm 0.040$ & $1.002 \pm 0.057$ & $9.11 \pm 0.36$ & $36.5 \pm 2.1$ & $3211.82/(3180-10), p=0.30$ \\

    \end{tabular}
   
    \tablecomments{
      The cosmologies are 
    ``Planck~(2016)'' the baseline of the study of this work,
    ``new\,cosmo.\,auto'' the custom cosmology for the
    Ly$\alpha$(Ly$\alpha$) $\times$ Ly$\alpha$(Ly$\alpha$) auto-correlation,
    and ``new\,cosmo.\,cross'' the custom cosmology for the
    Ly$\alpha$(Ly$\alpha$) $\times$ quasars cross-correlation.
    The error bars on the BAO parameters are given without fastMC corrections.
    The last column gives the goodness of the fit.
    }
    \label{table::old_vs_new_cosmology}
    
\end{table*}

\begin{table}
\caption{
        Characteristics of six quasar-redshift estimators as described
        in the text.}
    \centering
    
    \begin{tabular}{lllll}

    Redshift 
    & estimator -\zlyawg
    &       $\Delta r_{\parallel, \mathrm{QSO}}$
    &       $\sigmav$
    &       $\chi^{2}_{\mathrm{\min{}}}/DOF, \, proba$
    \\
    estimator & mean (r.m.s.) & ($\hMpc$) & ($\hMpc$) & \\
\noalign{\smallskip}
\hline \hline
\noalign{\smallskip}

 \zlyawg & 0 (0)     & $0.31 \pm 0.11 $   & $6.88 \pm 0.28$ & $9626.45/(9540-20), p=0.22$  \\
 \zpca   & -0.0022 (0.0077) & $-0.84 \pm 0.11 $ & $6.81 \pm 0.31$  & $9734.68 /( 9540 - 20 ),  p= 0.061$\\
 \zciv   & -0.0059 (0.0084) & $-4.31 \pm 0.12$     & $8.54 \pm 0.34$ & $9746.02/(9540-20), p=0.052$  \\
 \zciii  & -0.0066  (0.0113) & $-4.34 \pm 0.11$ & $6.90 \pm 0.33$ & $9786.63/(9540-20), p=0.027$  \\
 \zcomp  & $3\!\cdot\!10^{-6}$ (0.0076) &  $-0.783 \pm 0.099$ & $4.94 \pm 0.23$ & $9730.85/(9540-20), p=0.064$  \\
 \zvi~ (DR12)& 0.0003 (0.0087) & $-0.79 \pm 0.14$ & $4.46 \pm 0.30$ & $4853.35/(4770-18), p=0.15$  \\

    \end{tabular}
    
    \tablecomments{
    The second column gives the mean and standard deviation of the
        difference between
        the estimator and \zlyawg, the estimator used in this study.
        All mean and standard deviations are calculated
        for quasars in DR16Q with redshifts $>1.77$ and after eliminating
        outliers with $|\Delta z_q|>0.05$.
        The third and fourth columns give the
      best-fit values of the parameters $\Delta r_{\parallel, \mathrm{QSO}}$ and 
      $\sigmav$.
      The relatively large values of the means of
      \zciv-\zlyawg~ and \zciii-\zlyawg~ in column~2
      are correlated with the large values of
      $\Delta r_{\parallel, \mathrm{QSO}}$ in column~3, with
      the expected constant of proportionality, $\DHub(z_{\rm eff})=852~\hMpc$.
      The results for \zlyawg~differ slightly from those of the baseline fits
      (Table \ref{table::best_fit_parameters_data})
      because in this table the HCD model of \Bautista~was used. 
   }
    \label{table::drp_different_estimator}
    
\end{table}

\section{Quasar Redshifts}
\label{section::quasar_redshifts}

Quasar redshifts enter our study in two ways.
For the auto-correlation, they are used only to define
the continuum model over the \lya~and \lyb~spectral ranges .
Errors on quasar redshifts therefore only add an
effective quasar spectral diversity by moving features in
the spectral template randomly in wavelength.  This increases
the noise in the measurement of the transmission field
but does not systematically shift the $(\rpar,\rperp)$
of pixel pairs.
On the other hand, for the cross-correlation,
random errors in quasar redshift estimates smear the correlation
function in the $\rpar$ direction while  a systematic under- or over-estimate
of redshifts
would produce an asymmetry between positive and negative $\rpar$.
Both of these effects are accounted for in the fits:
the smearing via the parameter $\sigmav$
(eqn. \ref{equation::FNL_cross}) and
the asymmetry via the parameter $\Delta\rpar$.
As such, we do not expect quasar redshift errors to
systematically bias BAO parameters.

In this appendix we quantify the effects of redshift uncertainties on
several redshift estimators by using  mock and real spectra.
The different quasar redshift estimates that are included in the DR16Q catalog
are described in
\citet{2020ApJS..250....8L}
The important estimators are listed in
Table \ref{table::drp_different_estimator} with their best-fit values of
$\Delta r_{\rm \parallel, QSO}$ and $\sigma_v$.
The \zlyawg, \zpca, \zciv~and \zciii~estimates are derived by
the code \texttt{redvsblue}
as described below.
A composite redshift estimate, called \texttt{Z}, was
derived primarily from the visual inspection redshift,
\zvi, and
the pipeline redshift, \zpipe, using
a prior from the neural network
\texttt{quasarNET}\footnote{\url{https://github.com/ngbusca/QuasarNET}},
\citep{2018arXiv180809955B}.
\zpipe~was derived by fitting a
library of spectral templates based on
Principal Component Analysis (PCA).
\zvi~was 
estimated on all quasar spectra in the DR12 sample
\citep{2017A&A...597A..79P}.  Visual inspections were performed for only
a subset
of the spectra obtained during eBOSS observations to reduce the rate of
false detections
\citep{2018A&A...613A..51P}.

The DR12 analyses (\Bautista, \duMasdesBourboux) used \zvi~as
the redshift estimator while the
DR14 analyses (\deSainteAgathe, \Blomqvist) used its
closest descendant, \zcomp.
All eBOSS DR16 BAO analyses including this one use PCA-derived redshifts
because the lack of systematic visual scanning
makes the \zcomp~estimate inhomogeneous over the whole sample.

A known shortcoming of the pipeline redshift is the lack of redshift
evolution of
Ly$\alpha$ absorption in the spectral templates.
To mitigate the biases resulting from incorrect models of the quasar
spectrum at blue wavelengths,
we developed the  publicly available code ``redshift versus blueshift'':
\texttt{redvsblue}\footnote{\url{https://github.com/londumas/redvsblue}}.
As with the \texttt{Z\_PIPE} redshifts, this code performs redshift
estimates
using four PCA eigenvectors and three nuisance terms represented by
Legendre polynomials.
  However, \texttt{redvsblue} can correct the eigenvectors
  for the redshift evolution of the mean
Ly$\alpha$ optical depth.
The fits can also be restricted to a  chosen redshift range informed by
the \texttt{Z} estimate.
The estimates \zciv~and \zciii~restrict the range to a given emission line.
The estimate \zlyawg~used in this study, restricts the range  to
restframe wavelengths greater than $135.6$~nm, i.e.,
near the beginning of rise of the spectrum toward the \lya-emission peak.
The estimate \zpca~does not make this wavelength restriction  but
corrects the model for the \lya~optical depth of \citet{2012MNRAS.422.3019C}.

\begin{figure}
    \centering
\includegraphics[width=0.5\columnwidth]{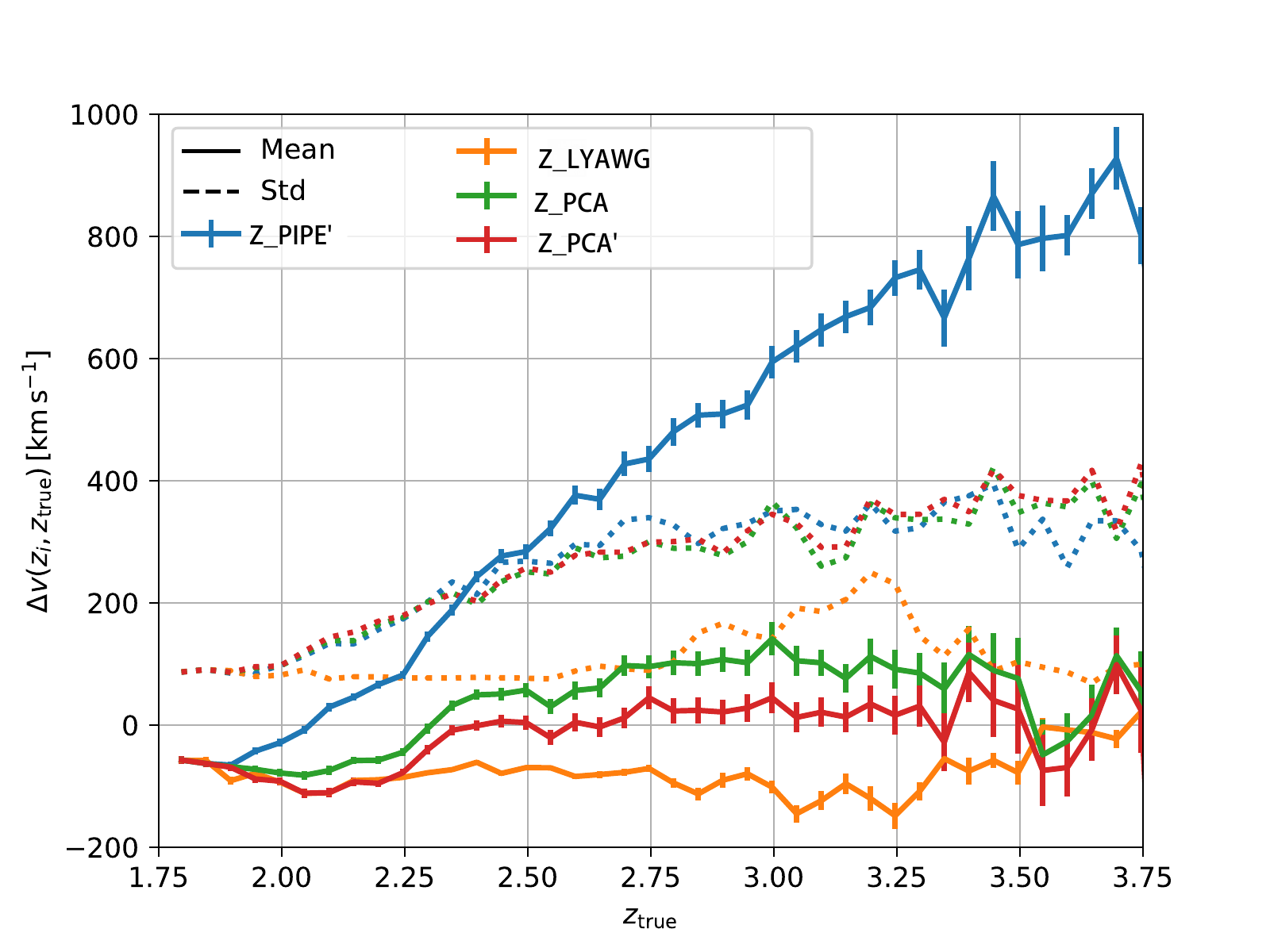}
\caption{
    Velocity difference between the estimated quasar redshift and the
    true input redshift of mocked quasar spectra as a function of the true
    redshift. Solid curves give the
    mean of the distribution, dashed curves give the standard deviation.
    The four \texttt{redvsblue} redshift estimators are:
    \texttt{Z\_PIPE'} (no correction for redshift
    evolution of Ly$\alpha$ optical depth);
    \zlyawg~ (masking of Ly$\alpha$ emission and forest
    wavelengths);
    \zpca~ (correcting for redshift evolution of  Ly$\alpha$ optical depth
    using the model of \citet{2012MNRAS.422.3019C});
    and \texttt{Z\_PCA'}
 using the model of \citet{2020ApJ...892...70K}.
  }
    \label{figure::dv_vs_ztrue_PCA_ZPCA}
\end{figure}

We  assessed the accuracy and precision of the \texttt{redvsblue}
estimates
using the mock spectra of
Sect.~\ref{section::Validation_of_the_analysis_with_mocks}.
Fig.~\ref{figure::dv_vs_ztrue_PCA_ZPCA} shows
the standard deviation (dashed curves)
and mean difference (solid curves) between the input
redshift and the estimated redshift for four different configurations of \texttt{redvsblue}.
The four configurations labeled in the figure are
1) ``noLy$\alpha$Corr'' which is nearly identical to \texttt{Z\_PIPE};
2) ``Ly$\alpha$Mask'' (\zlyawg);
3) ``Ly$\alpha$Corr'' (\zpca); and
4) ``Ly$\alpha$Corr, diff model'' also correcting for
the evolution of the \lya~optical depth but using the model of
\citet{2020ApJ...892...70K}.

Figure \ref{figure::dv_vs_ztrue_PCA_ZPCA} shows that
not correcting for the Ly$\alpha$ absorption in the spectral models
leads to systematic errors and statistical errors that increase with
redshift.
The bias in redshift estimates is reduced to values
less than $100~{\rm km\,s^{-1}}$
when using either correction for the Ly$\alpha$ absorption, albeit with
slightly different signatures of the systematic errors.
Of the four methods, the statistical errors are minimized when masking
the spectra around the Lyman-$\alpha$ emission line and at shorter
wavelengths.
This improvement can likely be explained by variance in large scale
structure variance and absorption from structures proximate to the
quasar that lead to distortions of the Lyman-$\alpha$ emission line that
are difficult to model.
These results motivated our choice of \zlyawg~for the DR16 BAO analysis
presented in this paper.

Table~\ref{table::drp_different_estimator} shows the
effect of using different quasar redshift estimators on the
nuisance variables $\Delta r_{\parallel,{\rm QSO}}$ and $\sigmav$.
The largest $\rpar$ asymmetries, reflected in $\Delta r_{\parallel,{\rm QSO}}$,
appears in the \texttt{Z\_CIV} and \texttt{Z\_CIII}
estimates.
The values of  $\Delta r_{\parallel,{\rm QSO}}$ relative to that
for \zlyawg~are correlated with the difference
between the estimator and \zlyawg~ (column 2) with
the expected factor of proportionality, $\DHub(z_{\rm eff})=852~\hMpc$.

  The best-fit values of $(\apar,\aperp)$ for four alternative estimators
  along with those for the adopted \zlyawg~
are listed in 
Table~\ref{table::best_fit_BAO_data_systematic_tests}.
We do not expect the choice of quasar-redshift
estimator to systematically change the BAO parameters
in one way or another
because any
small mean shift in quasar redshifts would be absorbed into the
nuisance parameter $\Delta\rpar$. 
Pairs of estimators do, however,
have a r.m.s. differences of $\Delta z_q\approx0.008$. Use of different
estimators would randomly
shift quasar-forest pixel pairs in $\rpar$ space with
$\Delta\rpar=0.008\DHub\approx7~\hMpc$, corresponding to $\approx$two bins
in $\rpar$.
This effect results in
random changes in the cross-correlation function, resulting in
random changes in the BAO parameters.

  We estimated the range of expected differences in BAO parameters
  when using different quasar-redshift estimators by using
  twenty Saclay mock sets
  (Sect. \ref{section::Validation_of_the_analysis_with_mocks}).
  For each mock set, a pair of new sets were created with the only
  change being in the quasar redshift: for each quasar the mean
  of the two redshifts was equal to the original redshift and
  the difference of the two redshifts was drawn randomly
  from the observed distribution of \zcomp-\zlyawg~
in the DR16Q catalog (r.m.s = 0.0076 from
Table~\ref{table::drp_different_estimator}, column 2).
The mock pairs were then fit for $\apar$ and $\aperp$.
The r.m.s. differences between pairs were $0.49\sigma$
and  $0.35\sigma$ for $\apar$ and $\aperp$ respectively.
A $\chi^2$ statistic characterizing the overall
change in BAO parameters was calculated.
Of twenty mock pairs,
eight
had changes greater than the
observed difference between the use of \zcomp~and \zlyawg.
DR12 (Table~\ref{table::best_fit_BAO_data_systematic_tests}).
This indicates that the difference observed for different estimators
are statistical in origin.
Furthermore four of the mock pairs
had changes greater than the
observed change in the \lyalyaq~ BAO parameters going from
DR12 (Table~\ref{table::redo_dr12})
to DR16 (Table~\ref{table::best_fit_parameters_data}).
This indicates that the change in redshift estimator had
a non-negligible impact on the evolution from DR12 to DR16.

\section{Tests of the covariance matrix}
\label{section::covariance_matrix_tests}

The covariance matrix for the correlation function was calculated
in Sect. \ref{section::Measurement_of_the_auto_and_cross_correlations}
by the sub-sampling methods.
The accuracy of this technique was verified  with mock spectra
(Sect. \ref{section::Validation_of_the_analysis_with_mocks})
by comparing the sub-sampling covariance with the mock-to-mock
variations of the correlation functions.
Here, we present two other techniques to estimate the covariance,
the ``Wick expansion'' and ``shuffling''.

\begin{figure*}
    \centering
    \includegraphics[width=.95\textwidth]{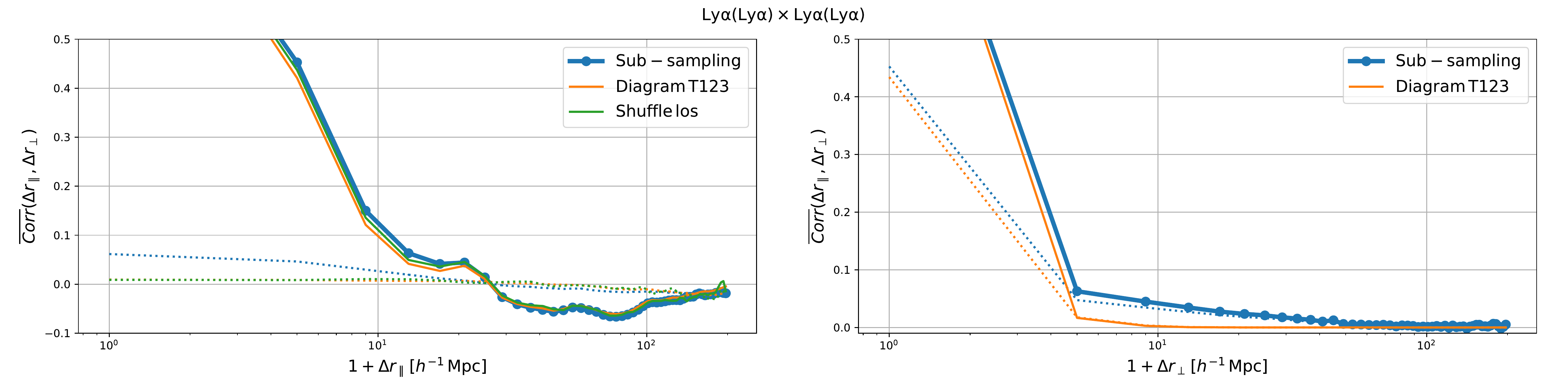}\\
    \includegraphics[width=.95\textwidth]{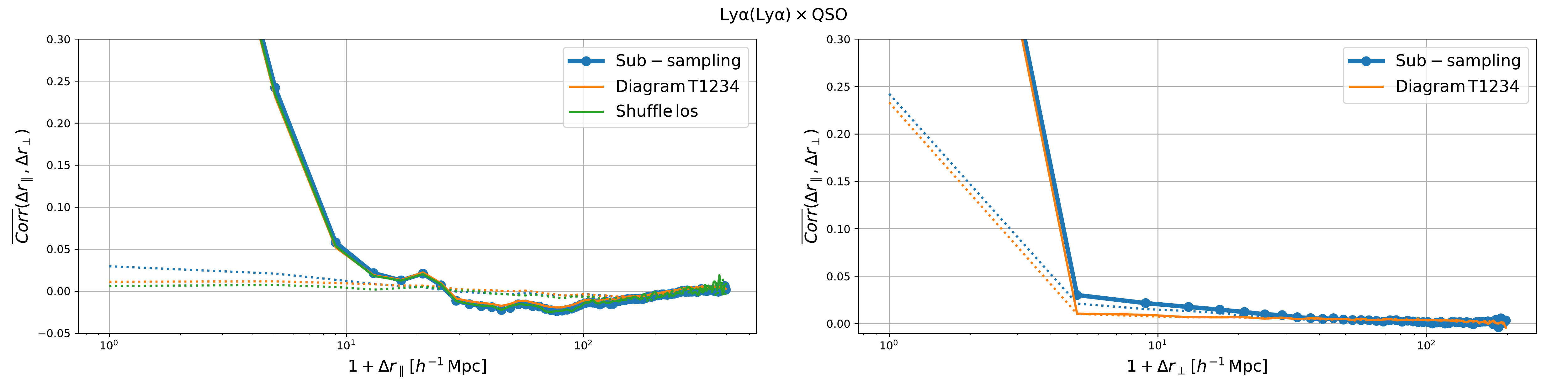}\\
    \caption{
    The normalized covariance matrix, i.e. the correlation matrix
    (eqn.~\ref{equation::correlation_matrix_estimator_auto}), for the auto-correlation,
    Ly$\alpha$(Ly$\alpha$) $\times$ Ly$\alpha$(Ly$\alpha$), top two panels,
    and for the cross-correlation, Ly$\alpha$(Ly$\alpha$) $\times$ quasar,
    bottom two panels.
    In each figure, the blue points and curves give the sub-sampling correlation matrix
    (eqn.~\ref{equation::covar_xi_estimator_subsampling_auto}), the orange
    curve the direct computation
    (eqns.~\ref{equation::covar_xi_estimator_wick_auto} and
    \ref{equation::covar_xi_estimator_wick_cross}),
    and the green curve the shuffle
    estimator (eqn.~\ref{equation::covar_xi_estimator_subsampling_auto}).
    The two figures on the left (resp. right) panel give the correlation 
    as a function of $\Delta r_{\parallel} = |r^{A}_{\parallel}-r^{B}_{\parallel}|$
    (resp. $\Delta r_{\perp} = |r^{A}_{\perp}-r^{B}_{\perp}|$).
    The solid curves are for $\Delta r_{\perp}=0\,\hMpc{}$
    (resp. $\Delta r_{\parallel}=0\,\hMpc{}$).
    The dashed curves are for
    $\Delta r_{\perp}=4\,\hMpc{}$
    (resp. $\Delta r_{\parallel}=4\,\hMpc{}$).
    }
    \label{figure::correlation_matrix_app}
\end{figure*}

The Wick method
(\citealt{2015A&A...574A..59D}, \Bautista, \duMasdesBourboux)
uses the expansion of the
four-point function in terms of products of two-point functions:\footnote{
  This is strictly true only for
  Gaussian fields but it works well in our context because intra-forest
  correlations are much larger than inter-forest correlations.}
\begin{equation}
    \frac{1}{W_A W_B}
    \sum\limits_{(i,j) \in A}
    \sum\limits_{(k,l) \in B}
    w_{i}w_{j}w_{k}w_{l}
    \langle \delta_{i}\delta_{j}\delta_{k}\delta_{l} \rangle
    \; \approx \; \frac{1}{W_AW_B}
    \sum\limits_{(i,j) \in A}
    \sum\limits_{(k,l) \in B}\;
    \sum\limits_{(\alpha,\beta,\mu,\nu) \in (i,j,k,l)}
    w_{\alpha}w_{\beta}w_{\mu}w_{\nu}
    \langle \delta_{\alpha}\delta_{\beta} \rangle
    \langle \delta_{\mu}\delta_{\nu} \rangle
    \; ,
    \label{equation::covar_xi_estimator_wick_auto}
\end{equation}
where $A$ and $B$ are two bins of the auto-correlation
and $W_{A,B}=\sum_{(i,j) \in A,B}w_iw_j$.
The covariance, $C_{AB}$, is the l.h.s. minus the
$(i,j,k,l)=(\alpha,\beta,\mu,\nu)$ term on the r.h.s.

Calculation of all terms in
eqn.~\ref{equation::covar_xi_estimator_wick_auto}
is computationally expensive.
Fortunately, the terms involving only two forests dominate, in which
case the covariance is 
\begin{equation}
C_{AB} =
\frac{1}{W_AW_B}
\sum_{ij\in A}\sum_{kl\in B} w_iw_j w_kw_l 
\xioned(\lambda_i/\lambda_k)\xioned(\lambda_j/\lambda_l)
\; ,
\label{equation::autocov_from_xi1d}
\end{equation}
where $\xioned$ is the intra-forest correlation function.

The resulting correlation matrix is presented in
Figure~\ref{figure::correlation_matrix_app}, by the orange curve. There,
``Diagram T123'' refers to the decomposition of
equation~\ref{equation::covar_xi_estimator_wick_auto} into different configuration
diagrams, as presented in Figure~A.1 of \citet{2015A&A...574A..59D}, and where T1, T2, and T3 are
two forests diagrams.
The figure shows that the direct computation estimator of the correlation
matrix agrees very well with the sub-sampling estimation.
The small discrepancies
come from taking into account only intra-forest correlations.

The direct computation of
equation~\ref{equation::covar_xi_estimator_wick_auto} is slightly different
for the cross-correlation, since there only one delta is used in each pair:
\begin{equation}
    C_{A B} = \frac{
    \sum\limits_{(i,j) \in A}
    \sum\limits_{(k,l) \in B}
    w_{i}w_{j}w_{k}w_{l}
    \langle \delta_{i}\delta_{k} \rangle
    }{
	W_{A} W_{B}
    }
    \; .
    \label{equation::covar_xi_estimator_wick_cross}
\end{equation}
In this equation, $\delta_i$ and $\delta_k$ are in the pairs $A$ and $B$,
and $j$ and $l$ are the quasars of pairs $A$ and $B$.
We show in Figure~\ref{figure::correlation_matrix_app} that the direct computation,
``Diagram T1234'' (see Figure~A.1 of \duMasdesBourboux), describes
to good approximation the correlation matrix estimated via sub-sampling.

In the shuffling method,
we shuffle the angular distribution of the line-of-sights and compute for each
realization the auto-correlation function of
equation~\ref{equation::xi_estimator_auto}.
This allows to have multiple measured auto-correlations, on the same footprint
and with the same redshift distribution. However, since we shuffled the angular
position of the line-of-sight, the expected value of the auto-correlation
is zero, $\langle \xi_{\mathrm{shuffle}} \rangle = 0$. The covariance matrix is then
given by the same equation as the one defined in
equation~\ref{equation::covar_xi_estimator_subsampling_auto}, where $s$
is one of the shuffle realization and where $W_{A}^{s}=1$.
Because of the shuffling, any correlations from the 3D distribution of the
line-of-sights is lost, however any correlations linked to pixels from the
same line-of-sight is preserved.

Figure~\ref{figure::correlation_matrix_app} shows in the top two panels, in
green, the estimated correlation matrix from shuffling the line-of-sights.
It has more noise than other estimators, for the simple reason that
we used only $100$ realizations.
This correlation matrix agrees very well with the one from sub-sampling.
We observe some differences of less than $5$\% of correlation in the
$\Delta r_{\perp}$ direction, linked to the lack of 3D correlation.

\begin{figure*}
    \centering
    \includegraphics[width=\textwidth]{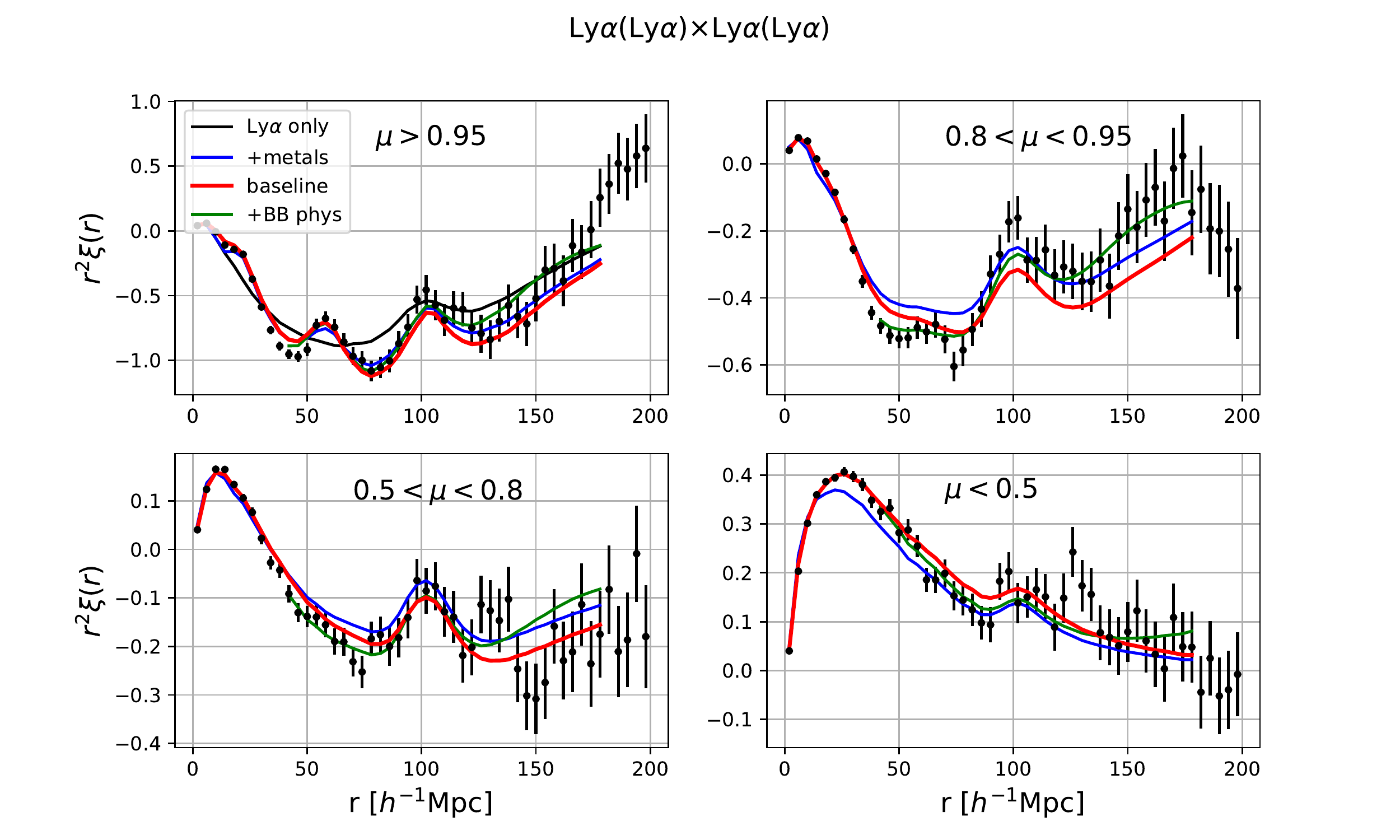} \\
    \includegraphics[width=\textwidth]{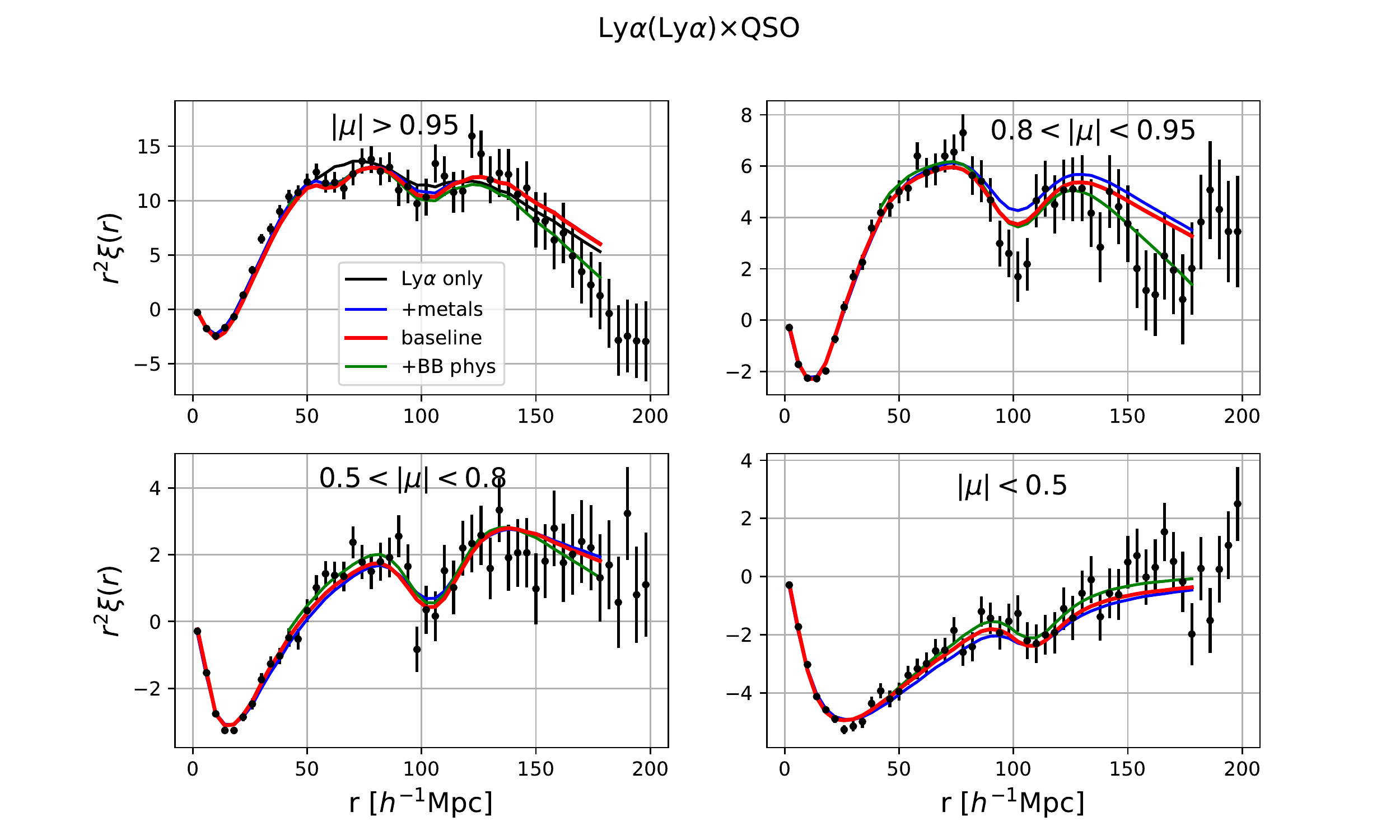} \\
    \caption{
      The \lyalyalyalya~(top) and \lyalyaq~(bottom) correlation functions in
      four ranges of $\mu$.
      The different lines correspond to models with an increasing
      number of components as shown in
      Table \ref{table::best_fit_BAO_data_different_model}:
      \lya~only (including $z_q$ smearing for \lyalyaq),
      addition of metals (and the sky-subtraction model for
      \lyalyalyalya),
      addition of HCDs
      (i.e. the baseline model) and the addition of a polynomial broadband with
      physical priors.
  }
    \label{figure::four_models}
\end{figure*}

\begin{table*}
    \caption{
      Evolution of the BAO and \lya~bias parameters with the addition of
      components of the model,  when analyzing the correlation functions
      measured in the main analysis. 
    }
    \centering
    
    \begin{tabular}{llllll}
    
    Models 
    &       $\alpha_{\parallel}$ 
    &       $\alpha_{\perp}$
    &       $b_{\eta,\mathrm{Ly}\alpha}$ 
    &       $\beta_{\mathrm{Ly}\alpha}$ 
    &       $\chi^{2}_{\mathrm{\min{}}}/DOF, \, proba$
    \\
\noalign{\smallskip}
\hline \hline
\noalign{\smallskip}

 Ly$\alpha$(Ly$\alpha$) $\times$ Ly$\alpha$(Ly$\alpha$): &  &  & & &\\ 
\noalign{\smallskip}

 $\mathrm{{Ly\alpha\;only}}$ &  $1.028 \pm 0.028$ & $1.002 \pm 0.048$ & $-0.1956 \pm 0.0030$ & $1.207 \pm 0.034$ & $2241.84/(1590-4), p=0$  \\
 $\mathrm{{\,\,\,\,+Sky}}$ &  $1.028 \pm 0.029$ & $1.001 \pm 0.048$ & $-0.1767 \pm 0.0032$ & $1.055 \pm 0.032$ & $1942.24/(1590-6), p=1.3\times10^{-9}$  \\
 $\mathrm{{\,\,\,\,+Metals}}$ &  $1.043 \pm 0.034$ & $0.991 \pm 0.050$ & $-0.181 \pm 0.0033$ & $1.098 \pm 0.035$ & $1766.05/(1590-11), p=6.5\times10^{-4}$  \\
baseline   & $ 1.047 \pm 0.034 $  & $  0.98\pm 0.042 $ & $ -0.2009\pm 0.0039 $  &  $ 1.657\pm 0.088 $  &  $ 1604.79 /( 1590 - 13 ),  p= 0.307$ \\
BBphysical   & $ 1.048 \pm 0.031  $ & $  0.988\pm 0.04 $ & $ -0.2113\pm 0.0052 $  & $  1.526\pm 0.095 $  &  $ 1488.95 /( 1515 - 25 ),  p= 0.503$ \\
BB           & $ 1.046 \pm 0.031 $  &  $ 0.989 \pm0.041 $ & $ -0.2382 \pm0.0074 $  & $  1.669 \pm0.138 $  &  $ 1480.7 /( 1515 - 25 ),  p= 0.563$ \\


\noalign{\smallskip} \hline \hline \noalign{\smallskip}
Ly$\alpha$(Ly$\alpha$) $\times$ quasar: &  &  & & &\\ 
\noalign{\smallskip}

 $\mathrm{{Ly\alpha\;only}}$ &  $1.052 \pm 0.040$ & $0.938 \pm 0.050$ & $-0.0942 \pm 0.0051$ & $0.604 \pm 0.045$ & $3989.12/(3180-5), p=0$  \\
 $\mathrm{{\,\,\,\,+z_{q}}}$ &  $1.054 \pm 0.041$ & $0.931 \pm 0.053$ & $-0.294 \pm 0.012$ & $2.51 \pm 0.17$ & $3258.81/(3180-6), p=0.14$  \\
 $\mathrm{{\,\,\,\,+Metals }}$ &  $1.058 \pm 0.042$ & $0.929 \pm 0.053$ & $-0.295 \pm 0.012$ & $2.48 \pm 0.17$ & $3232.82/(3180-10), p=0.21$  \\
baseline   & $ 1.059\pm 0.032 $  & $  0.932\pm 0.039 $  & $ -0.2249\pm 0.0102 $  & $  1.946\pm 0.142 $  & $  3238.96 /( 3180 - 10 ),  p= 0.193$ \\
BBphysical   & $ 1.058\pm 0.031 $  &  $ 0.931\pm 0.038 $ & $ -0.2306\pm 0.013 $  & $  1.824\pm 0.166 $  & $  3041.11 /( 3030 - 22 ),  p= 0.332$ \\
BB   & $ 1.058\pm 0.031 $  &  $ 0.932\pm 0.038 $ & $ -0.2303\pm 0.0187 $  & $  1.762\pm 0.211 $  & $  3040.73 /( 3030 - 22 ),  p= 0.334$ \\

\noalign{\smallskip} \hline \hline \noalign{\smallskip}
 all\,combined: &  &  & & & \\  
\noalign{\smallskip}

 $\mathrm{{Ly\alpha\;only}}$ &  $1.026 \pm 0.020$ & $0.975 \pm 0.031$ & $-0.1872 \pm 0.0023$ & $1.107 \pm 0.024$ & $11666.49/(9540-6), p=0$  \\
 $\mathrm{{\,\,\,\,+Sky}}$ &  $1.027 \pm 0.020$ & $0.974 \pm 0.031$ & $-0.1674 \pm 0.0025$ & $0.965 \pm 0.023$ & $11113.07/(9540-10), p=0$  \\
 $\mathrm{{\,\,\,\,+z_{q}}}$ &  $1.027 \pm 0.020$ & $0.974 \pm 0.030$ & $-0.1892 \pm 0.0027$ & $1.190 \pm 0.030$ & $10255.42/(9540-11), p=1.4\times10^{-7}$  \\
 $\mathrm{{\,\,\,\,+Metals}}$ &  $1.042 \pm 0.023$ & $0.967 \pm 0.031$ & $-0.191 \pm 0.0030$ & $1.300 \pm 0.061$ & $9999.77/(9540-16), p=3.4\times10^{-4}$  \\
baseline   & $ 1.045 \pm 0.022 $  & $  0.956 \pm 0.028 $ & $ -0.2014 \pm 0.0032 $  &  $ 1.669 \pm 0.071 $  &  $ 9654.56 /( 9540 - 19 ),  p= 0.166$ \\

    \end{tabular}
    
\tablecomments{
    Uncertainties correspond to $\Delta\chi^2=1$.
      As discussed in Sect. \ref{section::The_correlation_functions},
      the only significant change to the BAO parameters occur
      with the addition of the metals which, because Si(1260) absorption
      interferes with the BAO peak near $\rperp=0$,
      changes $\apar$ by $\approx0.5\sigma$.
    }
    \label{table::best_fit_BAO_data_different_model}
    
\end{table*}

\begin{table*}
    \caption{
      Results of non-standard fits as explained in the text.
    }
    \centering
    
    \begin{tabular}{llll}
    
    Analysis
    &       $\alpha_{\parallel}$ 
    &       $\alpha_{\perp}$
    &       $\chi^{2}_{\mathrm{\min{}}}/DOF, \, proba$
    \\
    \noalign{\smallskip}
    \hline \hline
    \noalign{\smallskip}
    
    Ly$\alpha$(Ly$\alpha$) $\times$ Ly$\alpha$(Ly$\alpha$): &  &  & \\
    $\mathrm{baseline}$ &  $1.043 \pm 0.035$ & $0.986 \pm 0.051$ & $1599.55/(1590-14), p=0.33$  \\
    \noalign{\smallskip}

 $\mathrm{no\,SHP\,pairs,\,r_{\parallel}<4}$ &  $1.042 \pm 0.034$ & $0.985 \pm 0.051$ &  $1639.66/(1590-12), p=0.14$  \\
 $\mathrm{no\,DLA\,mask}$ &  $1.037 \pm 0.034$ & $0.978 \pm 0.053$ &  $1625.09/(1590-14), p=0.19$  \\
 $\mathrm{Z}$ &  $1.049 \pm 0.034$ & $0.971 \pm 0.054$ &  $1623.63/(1590-14), p=0.20$  \\
 $\mathrm{Z\_PCA(no\,Ly\alpha\,corr.)}$ &  $1.046 \pm 0.035$ & $0.980 \pm 0.055$  & $1618.25/(1590-14), p=0.22$  \\
 $\mathrm{Z\_CIV}$ &  $1.058 \pm 0.036$ & $0.967 \pm 0.052$ &  $1571.92/(1590-14), p=0.52$  \\
 $\mathrm{Z\_CIII}$ &  $1.051 \pm 0.039$ & $0.973 \pm 0.060$ &  $1628.38/(1590-14), p=0.17$  \\
 $\mathrm{dmat\,up\,300}$ &  $1.043 \pm 0.034$ & $0.984 \pm 0.051$ &  $1590.51/(1590-14), p=0.39$  \\
 $\mathrm{dmat\,rect.}$ &  $1.040 \pm 0.035$ & $0.986 \pm 0.053$ &  $1609.73/(1590-14), p=0.27$  \\
 $\mathrm{r_{\min} = 40}$ &  $1.038 \pm 0.033$ & $0.994 \pm 0.049$ &  $1490.83/(1515-14), p=0.57$  \\
 $\mathrm{r_{\max} = 150}$ &  $1.042 \pm 0.035$ & $0.988 \pm 0.051$  & $1089.86/(1097-14), p=0.44$  \\
 $\mathrm{low\,z}$ &  $1.024 \pm 0.052$ & $0.84 \pm 0.12$ &  $1542.74/(1590-14), p=0.72$  \\
 $\mathrm{high\,z}$ &  $1.066 \pm 0.048$ & $0.982 \pm 0.055$ &  $1624.99/(1590-14), p=0.19$  \\

    \noalign{\smallskip} \hline \hline \noalign{\smallskip}
    Ly$\alpha$(Ly$\alpha$) $\times$ quasar: &  &   & \\ 
    \noalign{\smallskip}
    
 $\mathrm{baseline}$ &  $1.058 \pm 0.042$ & $0.929 \pm 0.053$ &  $3232.82/(3180-10), p=0.21$  \\
    \noalign{\smallskip}

 $\mathrm{no\,DLA\,mask}$ &  $1.052 \pm 0.044$ & $0.943 \pm 0.056$ &  $3225.07/(3180-10), p=0.24$  \\
 $\mathrm{Z}$ &  $1.079 \pm 0.039$ & $0.913 \pm 0.047$ &  $3187.32/(3180-10), p=0.41$  \\
 $\mathrm{Z\_PCA(no\,Ly\alpha\,corr.)}$ &  $1.066 \pm 0.046$ & $0.930 \pm 0.052$ &  $3268.92/(3180-10), p=0.11$  \\
 $\mathrm{Z\_CIV}$ &  $1.062 \pm 0.046$ & $0.948 \pm 0.053$ &  $3298.39/(3180-10), p=0.055$  \\
 $\mathrm{Z\_CIII}$ &  $1.069 \pm 0.048$ & $0.929 \pm 0.064$ &  $3259.71/(3180-10), p=0.13$  \\
 $\mathrm{dmat\,up\,300}$ &  $1.059 \pm 0.042$ & $0.927 \pm 0.052$ &  $3220.68/(3180-10), p=0.26$  \\
 $\mathrm{dmat\,rect.}$ &  $1.059 \pm 0.042$ & $0.928 \pm 0.053$ &  $3226.78/(3180-10), p=0.24$  \\
 $\mathrm{r_{\min} = 40}$ &  $1.054 \pm 0.040$ & $0.936 \pm 0.052$ &  $3042.47/(3030-10), p=0.38$  \\
 $\mathrm{r_{\max} = 150}$ &  $1.054 \pm 0.043$ & $0.929 \pm 0.054$ &  $2288.26/(2194-10), p=0.059$  \\
 $\mathrm{low\,z}$ &  $1.021 \pm 0.065$ & $0.989 \pm 0.072$ &  $3191.71/(3180-10), p=0.39$  \\
 $\mathrm{high\,z}$ &  $1.092 \pm 0.056$ & $0.879 \pm 0.064$ &  $3267.25/(3180-10), p=0.11$  \\
    
    \noalign{\smallskip} \hline \hline \noalign{\smallskip}
    all\,combined: &  &  &  \\  
    \noalign{\smallskip}
 $\mathrm{baseline}$ &  $1.043 \pm 0.022$ & $0.960 \pm 0.029$ & $9626.45/(9540-20), p=0.22$  \\
 \noalign{\smallskip}

 $\mathrm{no\,DLA\,mask}$ &  $1.038 \pm 0.023$ & $0.958 \pm 0.030$ & $9705.05/(9540-20), p=0.091$  \\
 $\mathrm{Z}$ &  $1.053 \pm 0.022$ & $0.940 \pm 0.027$ & $9730.85/(9540-20), p=0.064$  \\
 $\mathrm{Z\_PCA(no\,Ly\alpha\,corr.)}$ &  $1.049 \pm 0.024$ & $0.954 \pm 0.030$ & $9734.68/(9540-20), p=0.061$  \\
 $\mathrm{Z\_CIV}$ &  $1.047 \pm 0.024$ & $0.962 \pm 0.029$ & $9746.02/(9540-20), p=0.052$  \\
 $\mathrm{Z\_CIII}$ &  $1.048 \pm 0.026$ & $0.957 \pm 0.031$ & $9786.63/(9540-20), p=0.027$  \\
 $\mathrm{dmat\,up\,300}$ &  $1.043 \pm 0.022$ & $0.958 \pm 0.029$ & $9587.65/(9540-20), p=0.31$  \\
 $\mathrm{dmat\,rect.}$ &  $1.041 \pm 0.023$ & $0.960 \pm 0.030$ & $9631.07/(9540-20), p=0.21$  \\
 $\mathrm{r_{\min} = 40}$ &  $1.039 \pm 0.021$ & $0.968 \pm 0.028$ & $9069.50/(9090-20), p=0.50$  \\
 $\mathrm{r_{\max} = 150}$ &  $1.041 \pm 0.023$ & $0.961 \pm 0.030$ & $6706.63/(6582-20), p=0.10$  \\
 $\mathrm{low\,z}$ &  $1.019 \pm 0.035$ & $0.979 \pm 0.061$ & $4751.87/(4770-18), p=0.50$  \\
 $\mathrm{high\,z}$ &  $1.076 \pm 0.031$ & $0.941 \pm 0.040$ & $4905.41/(4770-18), p=0.059$  \\

    \end{tabular}
    
\tablecomments{
    Uncertainties correspond to $\Delta\chi^2=1$.
    All fits use the HCD model of \Bautista~which has one more free
    parameter that the HCD model used in this paper,
      but has no significant
      impact on the BAO parameters.
    }
    \label{table::best_fit_BAO_data_systematic_tests}
    
\end{table*}

\section{Alternative analyses of the data}
\label{section::Systematic_tests_on_BAO}

In this appendix we present, in Tables
\ref{table::best_fit_BAO_data_different_model} and
\ref{table::best_fit_BAO_data_systematic_tests}
the results of alternative analyses.
Table~\ref{table::best_fit_BAO_data_different_model}
and Fig.~\ref{figure::four_models} show the evolution
of the BAO and bias parameters as components of the model are added.
We note that after the addition of metals, the BAO parameters
are very stable.  This is even true with the addition of
polynomial broadband (BB) terms to simulate possible unknown
components in the smooth part of the power spectrum.
As can be expected, the bias parameters are less stable,
changing significantly when the model includes HCDs and
polynomial broadbands.

  Table~\ref{table::best_fit_BAO_data_systematic_tests} presents
  the results of fits using the baseline model but under
  different data-selection criteria or parameter choices.
  With the exceptions discussed below, these alternative analyses
  have no significant effects on the BAO parameters.
  After the nominal baseline model, the table shows, first, the effect of
  not masking DLAs, thereby treating the intra-DLA flux-flux transmission
  field in the same way as the field due to the IGM.
  The next four entries in the table show the effect of using
  other quasar-redshift estimators.
    The variations in the BAO parameters for the cross-correlation
    are at the level of $\leq0.5\sigma$.
    As discussed in Appendix \ref{section::quasar_redshifts}, these variations are
    due to the statistical fluctuations in the number of pairs in
    $(\rpar,\rperp)$ bins because of random fluctuations in $\rpar$
    separation of quasars and pixels.
  
  The entry labeled ``dmat up 300'' uses a distortion matrix
  calculated up to separations of $300~\hMpc$, i.e. beyond
  the nominal maximum distance of $200~\hMpc$.
  This allows us to show the correlation function at large
  distance does not have an important impact on the distorted
  correlation function in the region of the fit.
  The entry labeled ``dmat rect.'' uses a distortion matrix that samples
  the model on a $(\rpar,\rperp)$ grid with spacings half the size
  of the grid used to measure the correlation functions.
  The next two entries change the range over which the correlation
  functions are fitted.

  The final two entries in
  Table~\ref{table::best_fit_BAO_data_systematic_tests}
  give the results found by splitting the sample into two
  independent redshift bins.
  These results are the only ones in the table that differ
  at the level of $1\sigma$ from the baseline model, as expected
  because of the independence of the two samples.

\section{MCMC sampler}
\label{section::mcmc_sampler}

The analysis presented in this work, as well as in several previous
\lya\ BAO analyses, have used a frequentist interpretation of probability.
In this appendix we present an alternative analysis using a Bayesian framework,
and discuss the differences between the two. 

The current approach, referred here as the $\chi^2$ scan, uses the frequentist
method of profiling the parameters of interest (in our case $\apar$ and
$\aperp$).
A grid is made over these parameters and, at each point in the grid, the
$\chi^2$ is minimized over all the other (nuisance) parameters.
This is not equivalent to the Bayesian concept of marginalized likelihood,
where the full likelihood is integrated over the nuisance parameters.
However, in some special cases, i.e., when the likelihood is Gaussian, the
profile likelihood is proportional to the marginalized one.
Assuming flat priors on the BAO parameters, the $\chi^2$ scan can be used to
construct a proxy for the marginalized posterior distribution of $\apar$ and
$\aperp$, allowing us to use the $\chi^2$ scans in popular packages like
CosmoMC \citep{2002PhRvD..66j3511L} or MontePython \citep{2013JCAP...02..001A},
where different cosmological results are combined within a Bayesian framework.

The $\chi^2$ scan can also be used to compute frequentist confidence levels
(CL).
In order to do that, a set of Monte Carlo (MC) simulations of the correlation
function is used to compute the values of $\Delta \chi^2$ that will correspond
to each confidence level, as described in appendix C of \duMasdesBourboux.

In Figure \ref{figure::sampler_contours} we compare the CLs obtained
using the $\chi^2$ scan (red) with Bayesian results obtained through
marginalization over the full posterior distribution (blue). 
This analysis assumed flat priors for all nuisance parameters, and used the
nested sampler Polychord \citep{2015MNRAS.450L..61H,2015MNRAS.453.4384H}.

In the same figure we also include results computed assuming the full
likelihood is Gaussian (green), where instead of computing a $\chi^2$ scan we
have found the maximum of the likelihood and computed its second derivatives.
The CLs were then computed using the same values of $\Delta \chi^2$ obtained
from the MC simulations described above.
The three methods give very similar results, which indicates that the
profile likelihood works well in this case, and the $\chi^2$ scan
can be interpreted as a posterior and safely used in Bayesian packages.
A more detailed description and comparison of the different methods is
presented in \citet{2020arXiv200402761C}.

\begin{figure*}
  \centering
  \includegraphics[width=.45\textwidth]{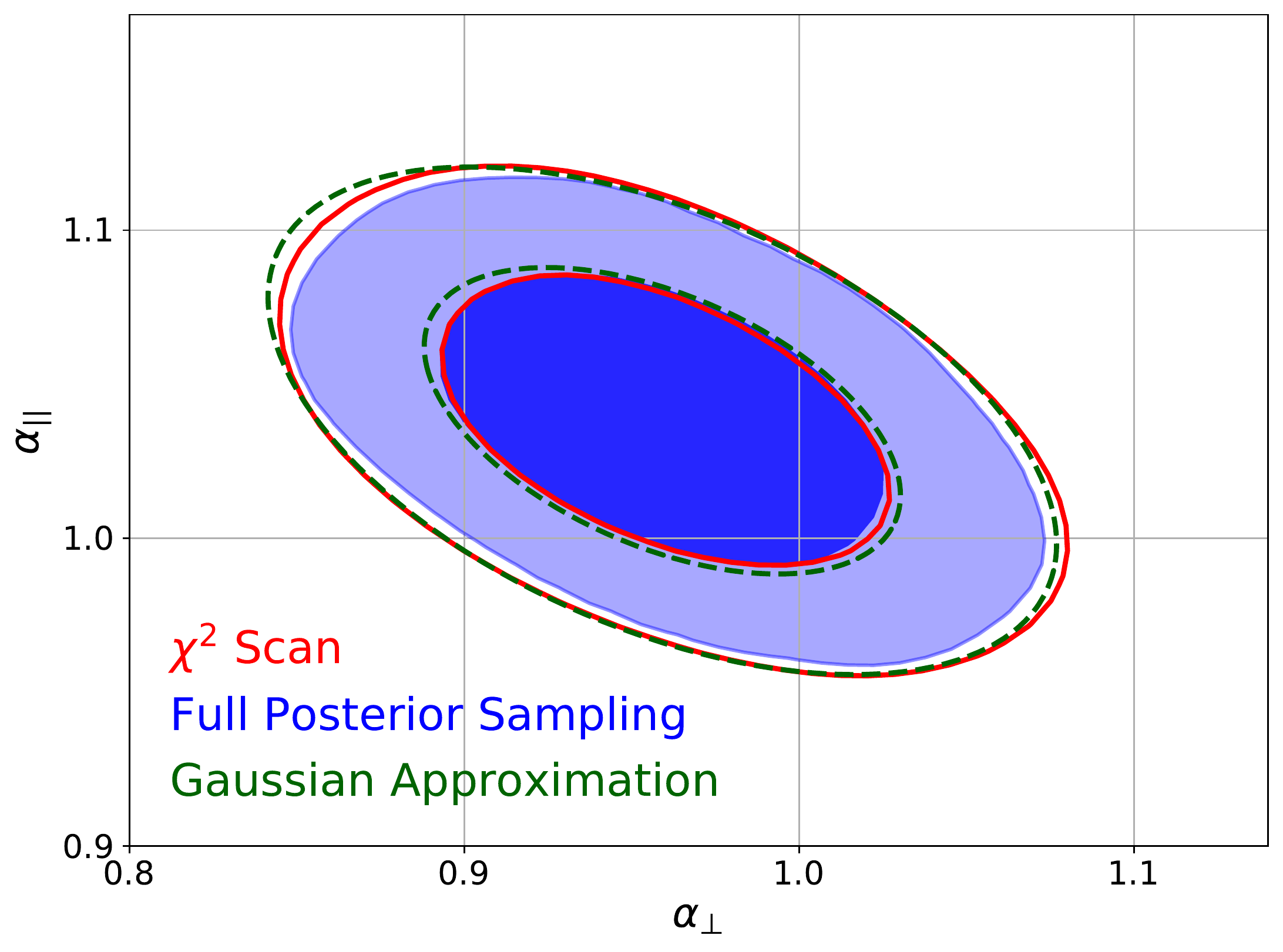}
  \includegraphics[width=.45\textwidth]{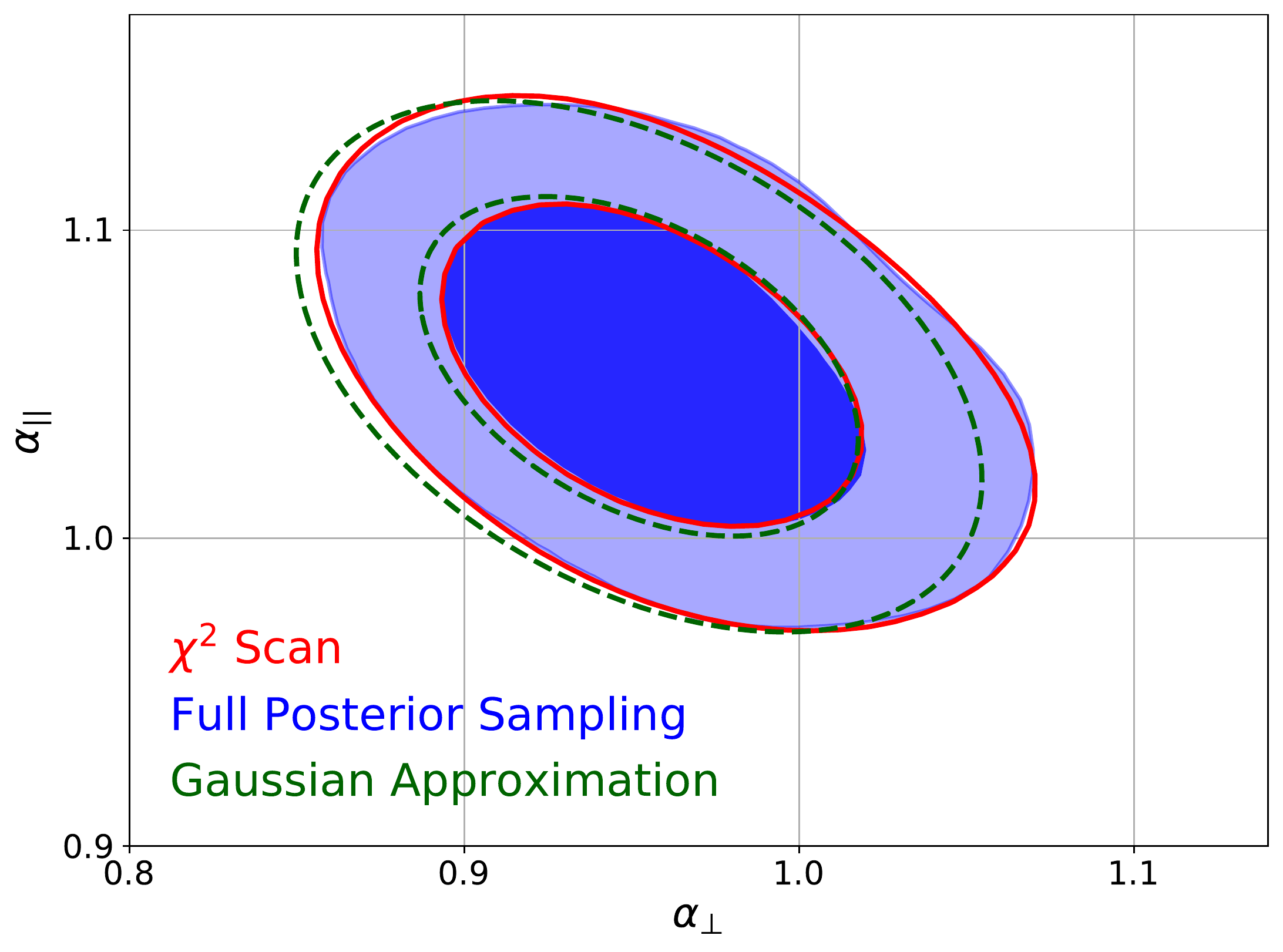}
  \caption{
    BAO results for the auto-correlation (left) and cross-correlation (right)
    computed using different methodologies:
    68\% and 95\% Confidence Levels (CL) from the $\chi^2$ scan, computed by
    minimizing the $\chi^2$ over all nuisance parameters (red);
    68\% and 95\% posterior probability from the full posterior sampling (blue);
    68\% and 95\% CLs from the Gaussian approximation of the likelihood (green).
    In both frequentists measurements, the CLs were computed using the 
    $\Delta \chi^2$ obtained from fastMC simulations (Table \ref{table::confidence}).
  }
  \label{figure::sampler_contours}
\end{figure*}

\section{Supplementary figures}
\label{section::more_plots}

\begin{figure*}
    \centering
    \includegraphics[width=.95\textwidth]{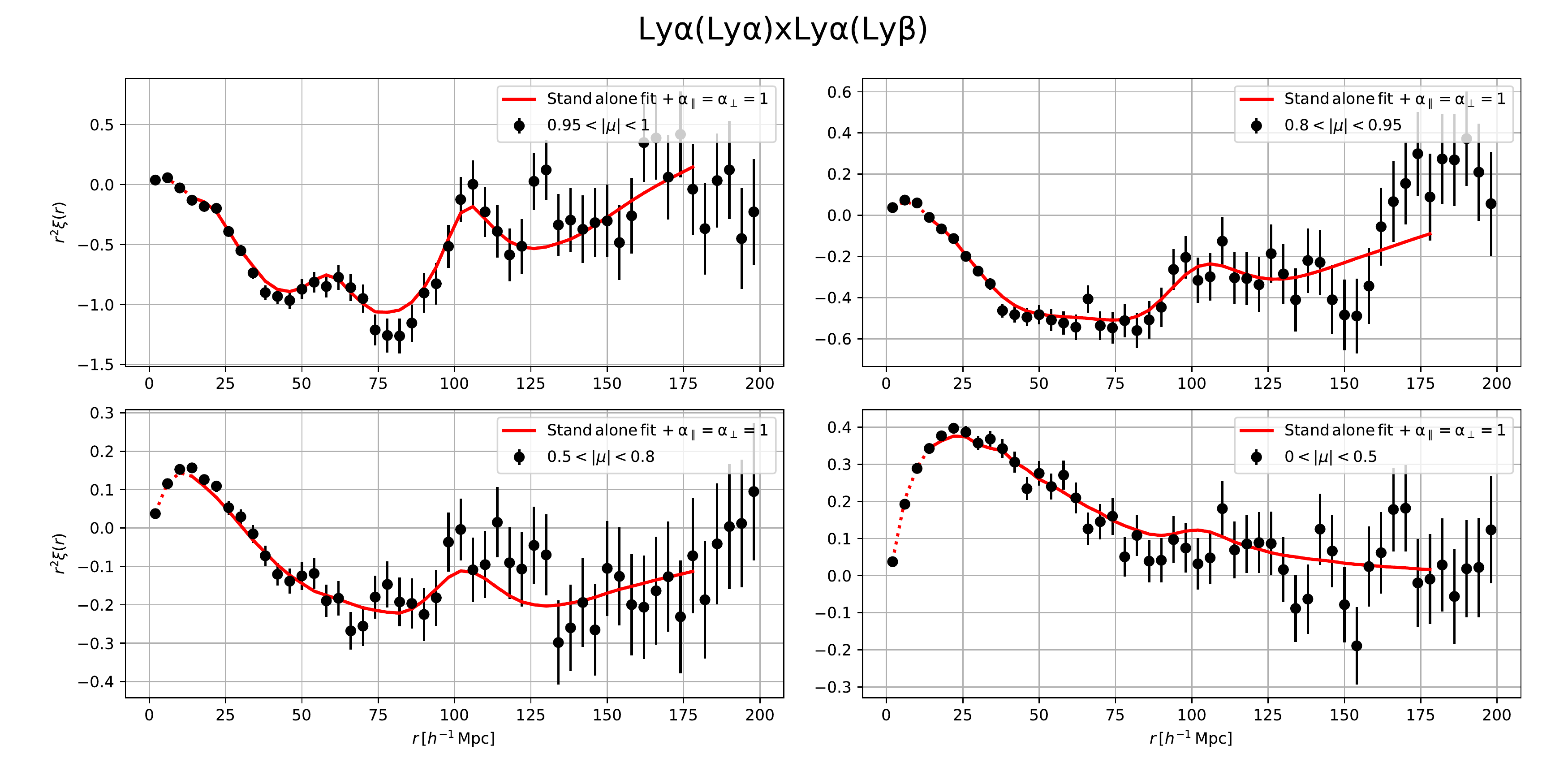} \\
    \includegraphics[width=.95\textwidth]{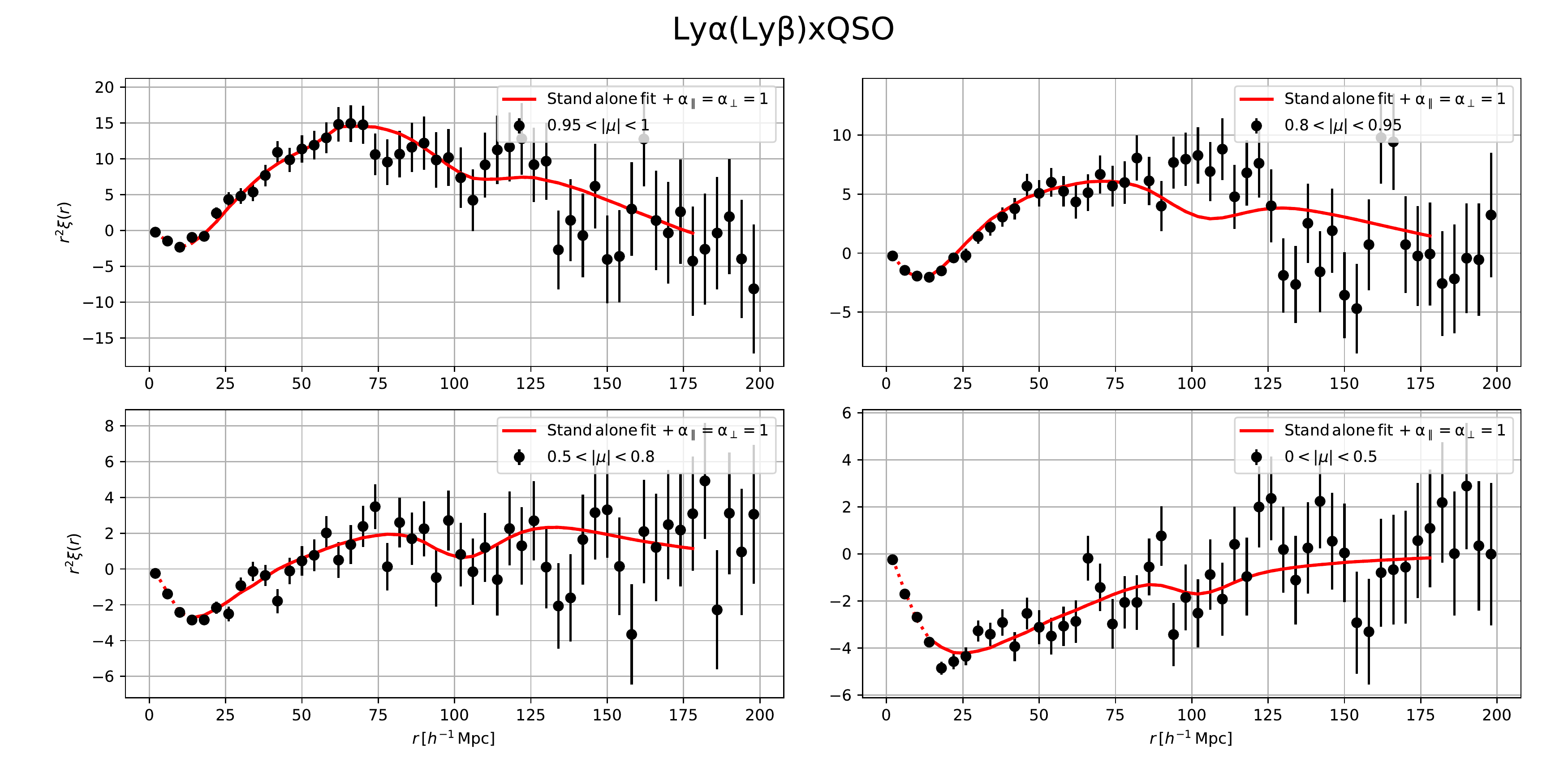}
    \caption{
      Same as Figure~\ref{figure::auto_4_wedges__cross_4_wedges}, but using
      forest pixels in the \lyb~region: \lyalyalyalyb~ (top four panels)
      and \lyalybq~ (bottom four panels.
              Because of the low significance of
              the BAO signal for this sample,
              the fits impose $\apar=\aperp=1$.
    }
    \label{figure::autoLyaLyainLyb_4_wedges__crossLyainLyb_4_wedges}
\end{figure*}

\begin{figure*}
    \centering
    \includegraphics[width=.95\textwidth]{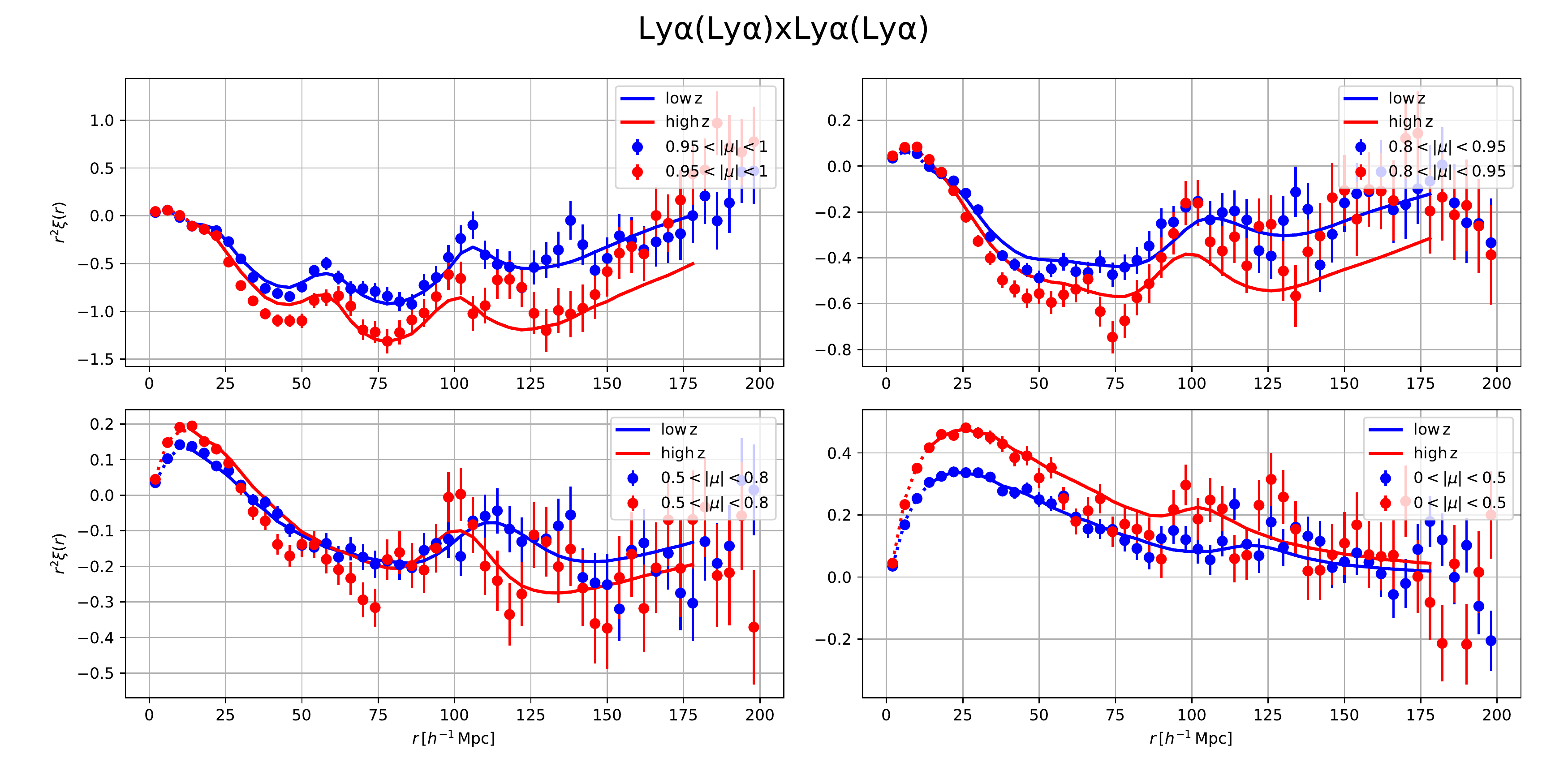} \\
    \includegraphics[width=.95\textwidth]{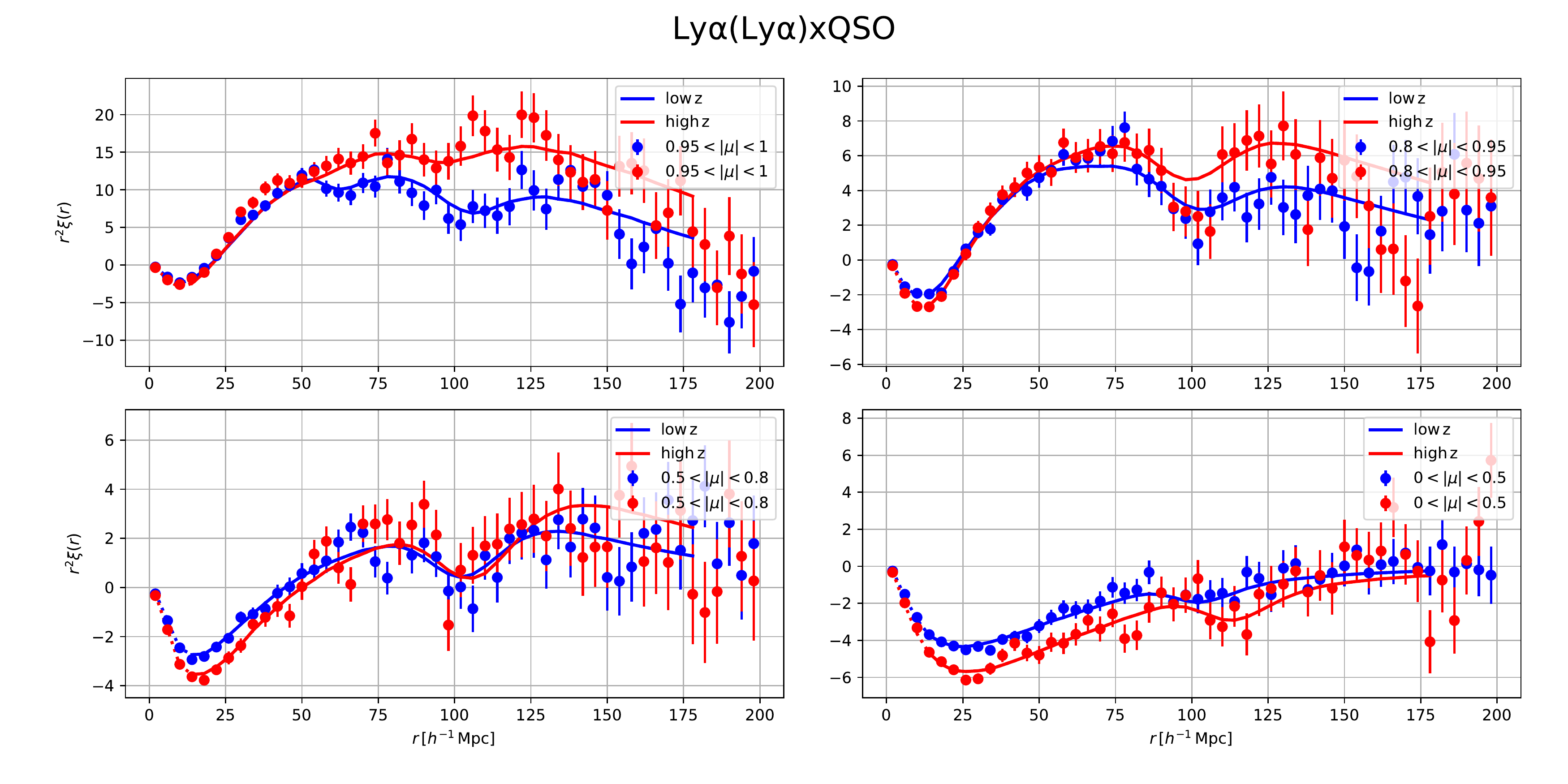}
    \caption{
      The correlation functions in two redshift bins: 
    \lyalyalyalya~ (top four panels) and \lyalyaq~
    (bottom four panels).
    Same as Figure~\ref{figure::auto_4_wedges__cross_4_wedges} but
    with two redshift bins, split at $z_{\mathrm{cut}} = 2.5$
    for the auto-correlation,
    and $z_{\mathrm{cut}} = 2.57$ for the cross-correlation.
    }
    \label{figure::auto_4_wedges_zBins__cross_4_wedges_zBins}
\end{figure*}

This appendix presents figures, that are referenced in the core of the analysis,
but that are not essential to the comprehension of our work.
The two correlations presented in
Figure~\ref{figure::autoLyaLyainLyb_4_wedges__crossLyainLyb_4_wedges}
and Figure~\ref{figure::auto_4_wedges_zBins__cross_4_wedges_zBins} are displayed
in a very similar manner as Figure~\ref{figure::auto_4_wedges__cross_4_wedges}.
They display the auto-correlation in their four top panels
and the cross-correlation in their four bottom panels, for
four wedges of $|\mu| = |r_{\parallel}/r|$.
Figure~\ref{figure::autoLyaLyainLyb_4_wedges__crossLyainLyb_4_wedges}
shows these two correlation functions when using one pixel in the Ly$\beta$
spectral region, instead of all in the Ly$\alpha$ spectral region.
The auto-correlation is then called \lyalyalyalyb\
and the cross-correlation called \lyalybq.
Figure~\ref{figure::auto_4_wedges_zBins__cross_4_wedges_zBins} displays
the two correlation functions, using pixels in the Ly$\alpha$ spectral region,
split into two redshift bins.

\begin{figure*}
    \centering
    \includegraphics[width=.95\textwidth]{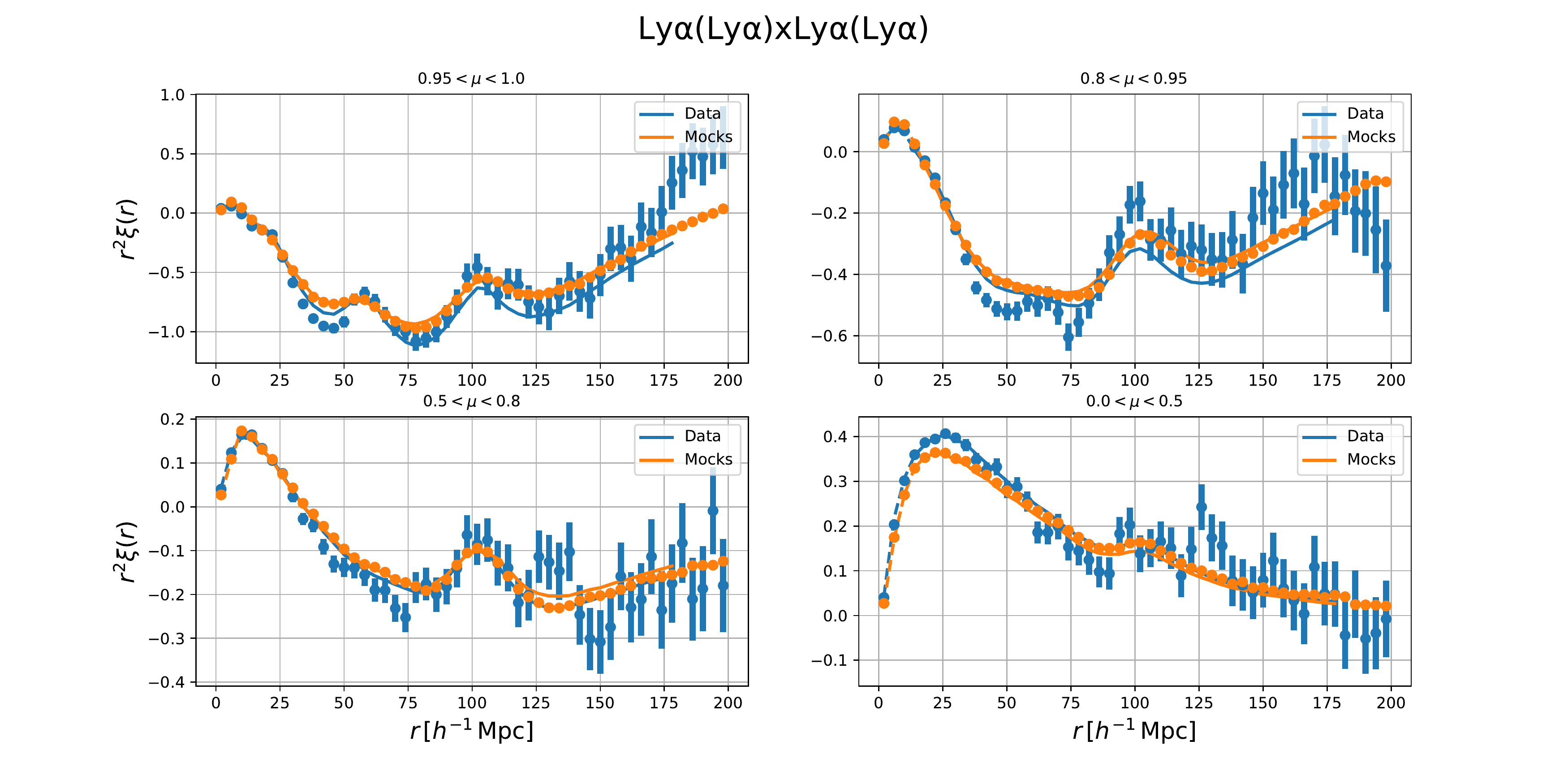} \\
    \includegraphics[width=.95\textwidth]{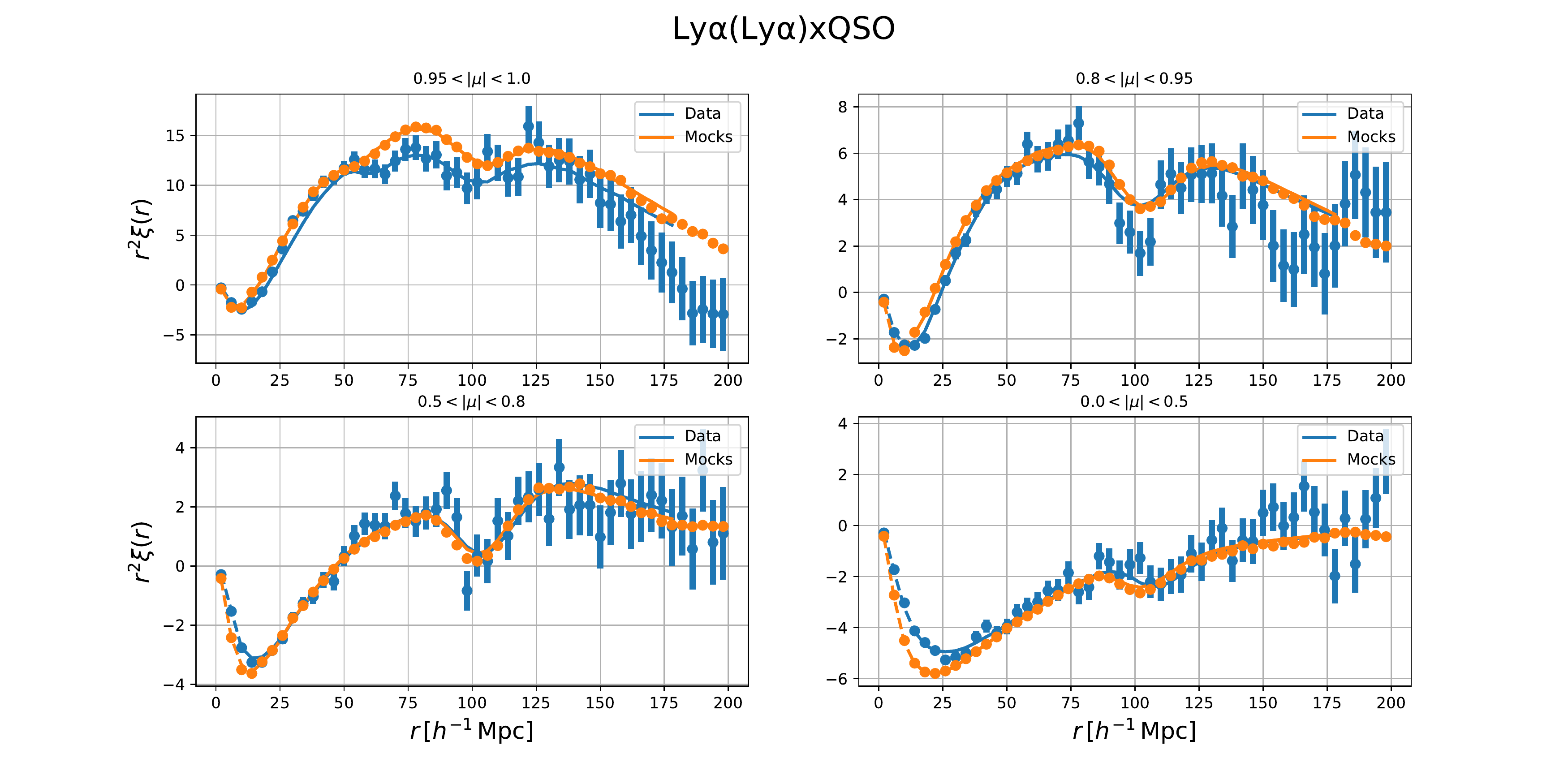}
    \caption{Comparison of the auto-correlation and cross-correlation measured
  in mocks and in the data, when using the \lya\ absorption in the 
  \lya\ region.}
    \label{figure::comparison_data_mocks_auto__comparison_data_mocks_cross}
\end{figure*}

\begin{figure*}
    \centering
    \includegraphics[width=.95\textwidth]{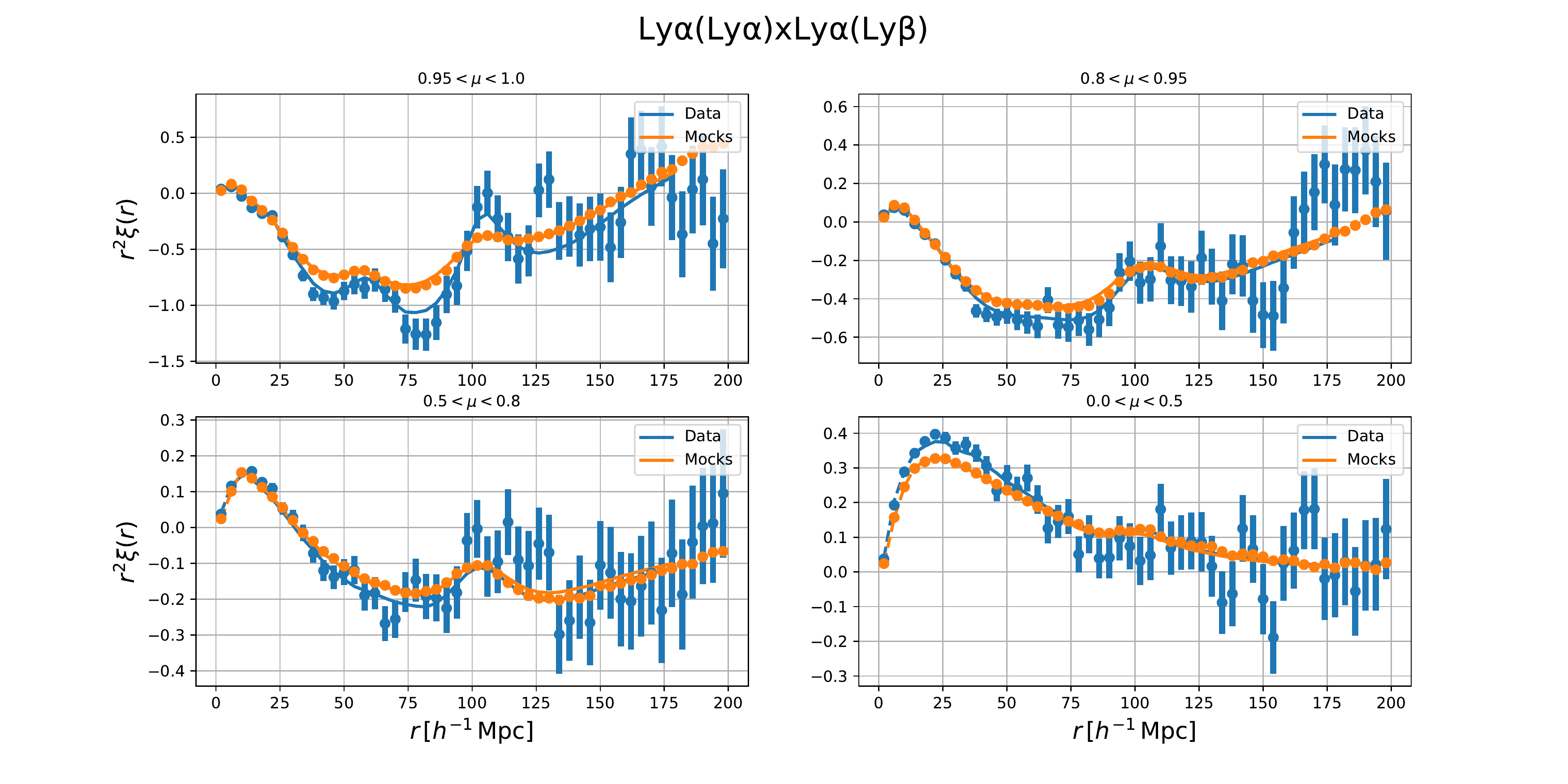} \\
    \includegraphics[width=.95\textwidth]{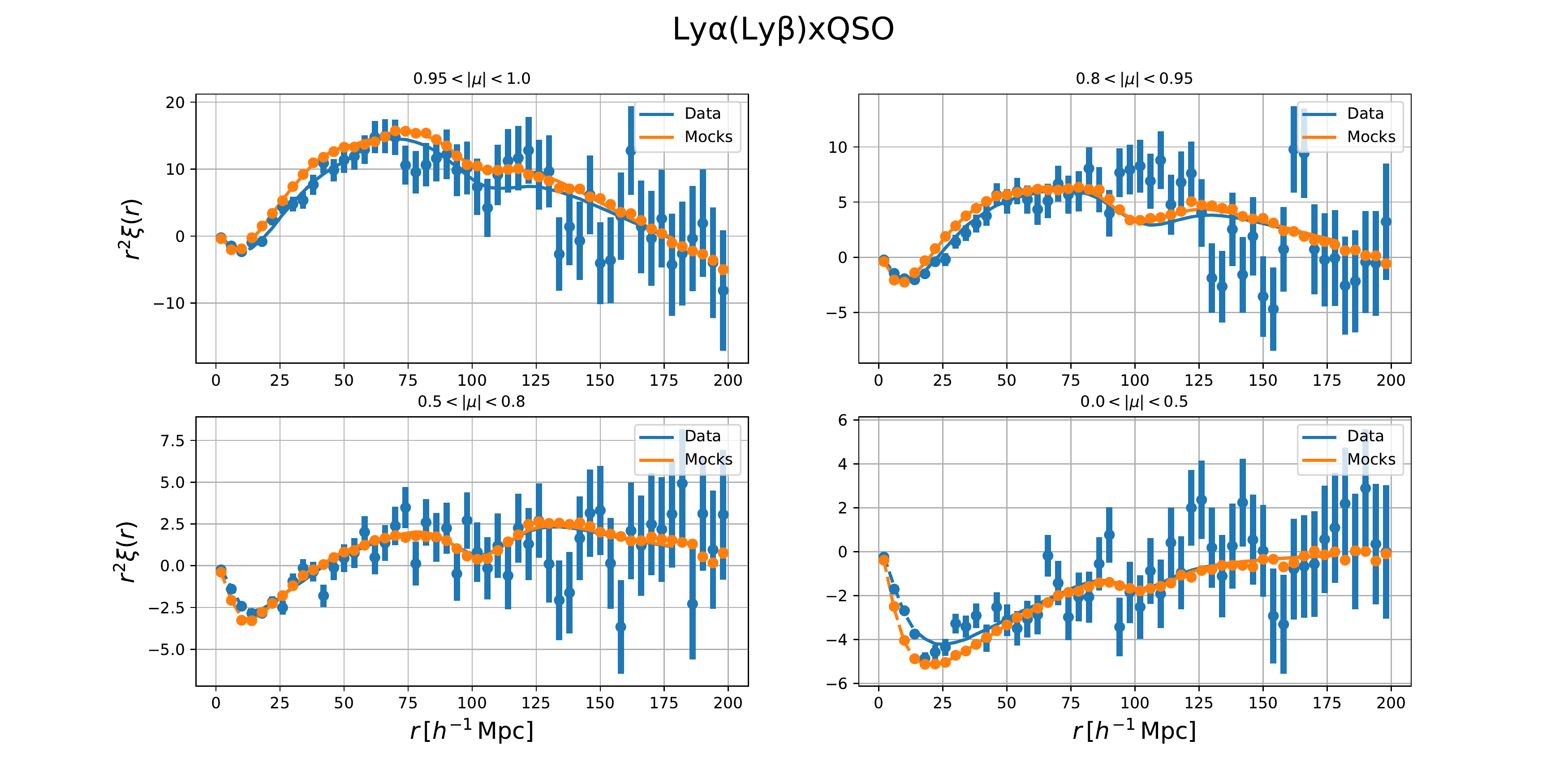}
    \caption{Comparison of the auto-correlation and cross-correlation measured
  in mocks and in the data, when using the \lya\ absorption in the 
  \lyb\ region.}
    \label{figure::comparison_data_mocks_autoLyb__comparison_data_mocks_crossLyb}
\end{figure*}

In Figures \ref{figure::comparison_data_mocks_auto__comparison_data_mocks_cross}
and \ref{figure::comparison_data_mocks_autoLyb__comparison_data_mocks_crossLyb}
we present a comparison of the correlations of the data and mocks.
The differences reflect the different values of the bias parameters
for the data (Table \ref{table::best_fit_parameters_data})
and mocks (Table~\ref{table::best_fit_BAO_mocks} ).

\section{Catalog of fluctuations of transmitted flux fraction}
\label{section::catalog_of_fluctuations_of_transmitted_flux_fraction}

This appendix presents the data format of the public release
of the flux-transmission field,
$\delta_q(\lambda)$, 
in the Ly$\alpha$ and Ly$\beta$ spectral regions.
These two catalogs can be found
on \url{https://dr16.sdss.org/sas/dr16/eboss/lya/} and were produced
using the ``Package for Igm Cosmological-Correlations Analyses'',
\picca, as presented in
section~\ref{subsection::Measurement_of_the_fluctuations_of_the_transmission_field}.
Documentation is at
\url{https://www.sdss.org/dr16/spectro/lyman-alpha-forest}.

The two catalogs are produced independently.
They are used in this analysis to extract the fluctuations of Ly$\alpha$
absorption, but were produced without any assumption on the nature of the
absorption.
To optimize reading time and simplify the structure, the sky is split into
HEALPix pixels \citep{2005ApJ...622..759G}, with the parameters
\texttt{nside=8} (\texttt{nside=4} for the Ly$\beta$ spectral region),
with the \texttt{RING} ordering.
All line-of-sights of a given HEALpix are written in a \texttt{FITS}
file, and are given the following name: \texttt{delta-<healpix>.fits.gz}, where 
\texttt{<healpix>} is the index of the HEALpix.
For the Ly$\alpha$ spectral region, the $247$ files with the
fluctuations of transmitted flux fraction are given in the repository
\texttt{Delta\_LYA}, for the Ly$\beta$ spectral region,
the $76$ files in the repository
\texttt{Delta\_LYB}.

Each file is composed of one \texttt{HDU} per line-of-sight, starting by
\texttt{HDU=1}.
The header section and the data section are given in Table~\ref{table::delta_data_format},
and presented here.
The external name of this \texttt{HDU}, is defined to be
given by the DR16Q identification integer: \texttt{EXTNAME=THING\_ID}.
In the file header, the different quantities \texttt{RA, DEC, Z, THING\_ID, PLATE,
MJD, FIBERID} are propagated from DR16Q. Where \texttt{Z} is the
redshift given in  DR16Q as \zlyawg, as discussed in
section~\ref{section::Quasar_samples_and_data_reduction}.
In this same header, \texttt{PMF} is given by \texttt{PLATE-MJD-FIBERID} and \texttt{ORDER}
is the order of the $\log_{10}(\lambda)$ polynomial for the continuum fit
(eqn.~\ref{equation::definition_quasar_continuum}).
The data columns of the \texttt{HDU} give:
\texttt{LOGLAM}, the common logarithm of the observed wavelength, $\log_{10}(\lambda)$,
\texttt{DELTA}, the measured fluctuations of transmitted flux fraction, $\delta$
(eqn~\ref{equation::definition_delta}), and its associated weight \texttt{WEIGHT}
(eqn~\ref{equation::definition_weight}), and finally \texttt{CONT}, the estimated continuum of the
quasar spectrum, $\overline{F}(z)C_{q}(\lrf)$ (eqn~\ref{equation::definition_delta}).

\begin{table}
    \caption{
      Description of data format of the flux-transmission field files,
      for each line-of-sight,
    containing
    the measured $\delta_q(\lambda)$ in the
    Ly$\alpha$ region and in the Ly$\beta$ region.
    }
    \centering
    \begin{tabular}{l l l}
        Name & Format & Description \\

        \noalign{\smallskip}
        \hline \hline
        \noalign{\smallskip}

        & Header & \\

        \noalign{\smallskip}
        \hline \hline
        \noalign{\smallskip}

        RA & DOUBLE & Right Ascension (from DR16Q) [rad]\\
        DEC & DOUBLE & Declination (from DR16Q) [rad]\\
        Z & DOUBLE & Redshift (\zlyawg~from DR16Q) \\
        PMF & STR &  PLATE-MJD-FIBERID\\
        THING\_ID & INT & Thing\_ID, eBOSS  source identifier (from DR16Q)\\
        PLATE & INT & Spectroscopic Plate number of the best spectrum (from DR16Q) \\
        MJD & INT & Spectroscopic Modified Julian Date of the best spectrum (from DR16Q) \\
        FIBERID & INT & Spectroscopic Fiber number of the best spectrum (from DR16Q) \\
        ORDER & INT & Order of the $\log_{10}(\lambda)$ polynomial for the continuum fit (eqn.~\ref{equation::definition_quasar_continuum})\\

        \noalign{\smallskip}
        \hline \hline
        \noalign{\smallskip}

        & Data & \\
        
        \noalign{\smallskip}
        \hline \hline
        \noalign{\smallskip}

        LOGLAM & DOUBLE & $\log_{10}(\lambda)$, where $\lambda$ is the observed wavelength $[\mathrm{Angstrom}]$ \\
        DELTA & DOUBLE & Flux-transmission field $\delta_q$ (eqn.~\ref{equation::definition_delta})\\
        WEIGHT & DOUBLE & $1/\sigma_q(\lambda)^2$ (eqn.~\ref{equation::definition_weight})\\
        CONT & DOUBLE & Best fit continuum, $\overline{F}(z)C_{q}(\lrf)$,
            of the quasar spectrum (eqn.~\ref{equation::definition_delta})\\

    \end{tabular}
    \label{table::delta_data_format}
\end{table}

\end{appendix}

\end{document}